\documentclass[journal]{IEEEtran}
\usepackage{epsbox}
\usepackage[dvips]{graphicx,color}
\usepackage{latexsym}
\usepackage{array}
\begin{document}
\arraycolsep 0.5mm

\newcommand{\bfig}{\begin{figure}[t]}
\newcommand{\efig}{\end{figure}}
\setcounter{page}{1}
\newenvironment{indention}[1]{\par
\addtolength{\leftskip}{#1}\begingroup}{\endgroup\par}
%form: \begin{indention}{2.3cm}
%      \end{indention}
%
%namelist environment
%form: \begin{namelist}{width}
\newcommand{\namelistlabel}[1]{\mbox{#1}\hfill} 
\newenvironment{namelist}[1]{%
\begin{list}{}
{\let\makelabel\namelistlabel
\settowidth{\labelwidth}{#1}
\setlength{\leftmargin}{1.1\labelwidth}}
}{%
\end{list}}
%
%\input{ieeecom2e.tex}
%
%---------------------------- ieeecom2e.tex--------------------------
%
%
\newcommand{\bc}{\begin{center}}  %
\newcommand{\ec}{\end{center}}
\newcommand{\befi}{\begin{figure}[h]}  %
\newcommand{\enfi}{\end{figure}}
\newcommand{\bsb}{\begin{shadebox}\begin{center}}   %
\newcommand{\esb}{\end{center}\end{shadebox}}
\newcommand{\bs}{\begin{screen}}     %
\newcommand{\es}{\end{screen}}
\newcommand{\bib}{\begin{itembox}}   %
\newcommand{\eib}{\end{itembox}}
\newcommand{\bit}{\begin{itemize}}   %
\newcommand{\eit}{\end{itemize}}
\newcommand{\defeq}{\stackrel{\triangle}{=}}
\newcommand{\qed}{\hbox{\rule[-2pt]{3pt}{6pt}}}
\newcommand{\beq}{\begin{equation}}
\newcommand{\eeq}{\end{equation}}
\newcommand{\beqa}{\begin{eqnarray}}
\newcommand{\eeqa}{\end{eqnarray}}
\newcommand{\beqno}{\begin{eqnarray*}}
\newcommand{\eeqno}{\end{eqnarray*}}
\newcommand{\ba}{\begin{array}}
\newcommand{\ea}{\end{array}}
\newcommand{\vc}[1]{\mbox{\boldmath $#1$}}
\newcommand{\lvc}[1]{\mbox{\scriptsize \boldmath $#1$}}
\newcommand{\svc}[1]{\mbox{\scriptsize\boldmath $#1$}}

\newcommand{\wh}{\widehat}
\newcommand{\wt}{\widetilde}
\newcommand{\ts}{\textstyle}
\newcommand{\ds}{\displaystyle}
\newcommand{\scs}{\scriptstyle}
\newcommand{\vep}{\varepsilon}
\newcommand{\rhp}{\rightharpoonup}
\newcommand{\cl}{\circ\!\!\!\!\!-}
\newcommand{\bcs}{\dot{\,}.\dot{\,}}
\newcommand{\eqv}{\Leftrightarrow}
\newcommand{\leqv}{\Longleftrightarrow}
\newtheorem{co}{Corollary} 
\newtheorem{lm}{Lemma} 
\newtheorem{Ex}{Example} 
\newtheorem{Th}{Theorem}
\newtheorem{df}{Definition} 
\newtheorem{pr}{Property} 
\newtheorem{pro}{Proposition} 
\newtheorem{rem}{Remark} 

\newcommand{\lcv}{convex } 

\newcommand{\hugel}{{\arraycolsep 0mm
                    \left\{\ba{l}{\,}\\{\,}\ea\right.\!\!}}
\newcommand{\Hugel}{{\arraycolsep 0mm
                    \left\{\ba{l}{\,}\\{\,}\\{\,}\ea\right.\!\!}}
\newcommand{\HUgel}{{\arraycolsep 0mm
                    \left\{\ba{l}{\,}\\{\,}\\{\,}\vspace{-1mm}
                    \\{\,}\ea\right.\!\!}}
\newcommand{\huger}{{\arraycolsep 0mm
                    \left.\ba{l}{\,}\\{\,}\ea\!\!\right\}}}
\newcommand{\Huger}{{\arraycolsep 0mm
                    \left.\ba{l}{\,}\\{\,}\\{\,}\ea\!\!\right\}}}
\newcommand{\HUger}{{\arraycolsep 0mm
                    \left.\ba{l}{\,}\\{\,}\\{\,}\vspace{-1mm}
                    \\{\,}\ea\!\!\right\}}}

\newcommand{\hugebl}{{\arraycolsep 0mm
                    \left[\ba{l}{\,}\\{\,}\ea\right.\!\!}}
\newcommand{\Hugebl}{{\arraycolsep 0mm
                    \left[\ba{l}{\,}\\{\,}\\{\,}\ea\right.\!\!}}
\newcommand{\HUgebl}{{\arraycolsep 0mm
                    \left[\ba{l}{\,}\\{\,}\\{\,}\vspace{-1mm}
                    \\{\,}\ea\right.\!\!}}
\newcommand{\hugebr}{{\arraycolsep 0mm
                    \left.\ba{l}{\,}\\{\,}\ea\!\!\right]}}
\newcommand{\Hugebr}{{\arraycolsep 0mm
                    \left.\ba{l}{\,}\\{\,}\\{\,}\ea\!\!\right]}}
\newcommand{\HUgebr}{{\arraycolsep 0mm
                    \left.\ba{l}{\,}\\{\,}\\{\,}\vspace{-1mm}
                    \\{\,}\ea\!\!\right]}}

\newcommand{\hugecl}{{\arraycolsep 0mm
                    \left(\ba{l}{\,}\\{\,}\ea\right.\!\!}}
\newcommand{\Hugecl}{{\arraycolsep 0mm
                    \left(\ba{l}{\,}\\{\,}\\{\,}\ea\right.\!\!}}
\newcommand{\hugecr}{{\arraycolsep 0mm
                    \left.\ba{l}{\,}\\{\,}\ea\!\!\right)}}
\newcommand{\Hugecr}{{\arraycolsep 0mm
                    \left.\ba{l}{\,}\\{\,}\\{\,}\ea\!\!\right)}}

\newcommand{\hugepl}{{\arraycolsep 0mm
                    \left|\ba{l}{\,}\\{\,}\ea\right.\!\!}}
\newcommand{\Hugepl}{{\arraycolsep 0mm
                    \left|\ba{l}{\,}\\{\,}\\{\,}\ea\right.\!\!}}
\newcommand{\hugepr}{{\arraycolsep 0mm
                    \left.\ba{l}{\,}\\{\,}\ea\!\!\right|}}
\newcommand{\Hugepr}{{\arraycolsep 0mm
                    \left.\ba{l}{\,}\\{\,}\\{\,}\ea\!\!\right|}}

\newenvironment{jenumerate}
	{\begin{enumerate}\itemsep=-0.25em \parindent=1zw}{\end{enumerate}}
\newenvironment{jdescription}
	{\begin{description}\itemsep=-0.25em \parindent=1zw}{\end{description}}
\newenvironment{jitemize}
	{\begin{itemize}\itemsep=-0.25em \parindent=1zw}{\end{itemize}}
\renewcommand{\labelitemii}{$\cdot$}

\newcommand{\iro}[2]{{\color[named]{#1}#2\normalcolor}}
\newcommand{\irr}[1]{{\color[named]{Red}#1\normalcolor}}
\newcommand{\irg}[1]{{\color[named]{Green}#1\normalcolor}}
\newcommand{\irb}[1]{{\color[named]{Blue}#1\normalcolor}}
\newcommand{\irBl}[1]{{\color[named]{Black}#1\normalcolor}}
\newcommand{\irWh}[1]{{\color[named]{White}#1\normalcolor}}

\newcommand{\irY}[1]{{\color[named]{Yellow}#1\normalcolor}}
\newcommand{\irO}[1]{{\color[named]{Orange}#1\normalcolor}}
\newcommand{\irBr}[1]{{\color[named]{Purple}#1\normalcolor}}
\newcommand{\IrBr}[1]{{\color[named]{Purple}#1\normalcolor}}
\newcommand{\irBw}[1]{{\color[named]{Brown}#1\normalcolor}}
\newcommand{\irPk}[1]{{\color[named]{Magenta}#1\normalcolor}}
\newcommand{\irCb}[1]{{\color[named]{CadetBlue}#1\normalcolor}}
%\newcommand{\irDg}[1]{{\color[named]{DarkSlateGray}#1\normalcolor}}
%
%-----------------------notaion2.tex -----------------------------%
%
\newcommand{\lcov}{%lower
                     convex }

\newcommand{\cars}{s}
\newcommand{\sigNi}{\sigma_{N_i}^2}
\newcommand{\sigN}{\sigma_{N}^2}

\newcommand{\cef}{c_i}
\newcommand{\cefsq}{c_i^2}

\newcommand{\Iset}{L}
\newcommand{\Ntn}{\mbox{\boldmath $N$}}
\newcommand{\lNtn}{\mbox{\scriptsize\boldmath $N$}}
\newcommand{\Nitn}{\mbox{\boldmath $N$}_i}
\newcommand{\lNitn}{\mbox{\scriptsize\boldmath $N$}_i}
\newcommand{\Nost}{N_{0,t}}
\newcommand{\Nast}{N_{1,t}}
\newcommand{\Nlst}{N_{L,t}}

\newcommand{\tinbNs}{\tilde{N}%_S
                    }
\newcommand{\tiNs}{\tilde{\mbox{\boldmath $N$}}%_S
                  }
\newcommand{\tilNs}{\tilde{\mbox{\scriptsize\boldmath$N$}}%_S
                   }
\newcommand{\Xotn}{\mbox{\boldmath $X$}_0}
\newcommand{\lXotn}{\mbox{\scriptsize\boldmath $X$}_0}
\newcommand{\Xost}{X_{0,t}}

\newcommand{\Xatn}{\mbox{\boldmath $X$}_1}
\newcommand{\Xast}{X_{1}(t)}

\newcommand{\Xbtn}{\mbox{\boldmath $X$}_2}
\newcommand{\Xbst}{X_{2}(t)}

\newcommand{\Xltn}{\mbox{\boldmath $X$}_L}
\newcommand{\Xlst}{X_{L}(t)}
\newcommand{\Xlsn}{X_{L}(n)}

\newcommand{\Xitn}{\mbox{\boldmath $X$}_i}
\newcommand{\lXitn}{\mbox{\scriptsize\boldmath $X$}_i}
\newcommand{\Xisa}{X_{i}(1)}
\newcommand{\Xisb}{X_{i}(2)}
\newcommand{\Xist}{X_{i}(t)}
\newcommand{\Xisn}{X_{i}(n)}
\newcommand{\Xtn}{\mbox{\boldmath $X$}}

\newcommand{\hatXotn}{\hat{\mbox{\boldmath $X$}}_0}
\newcommand{\hatXost}{\hat{X}_{0,t}}

\newcommand{\tiXs}{%\tilde
                  {\mbox{\boldmath $Y$}}%_S
                  }
\newcommand{\tilXs}{%\tilde
                   {\mbox{\scriptsize\boldmath $Y$}}%_S
                   }

\newcommand{\Zatn}{\mbox{\boldmath $Z$}_1}
\newcommand{\lZatn}{\mbox{\scriptsize\boldmath $Z$}_1}
\newcommand{\Zbtn}{\mbox{\boldmath $Z$}_2}
\newcommand{\lZbtn}{\mbox{\scriptsize\boldmath $Z$}_2}

\newcommand{\rdf}{J}  

\newcommand{\Dff}{-}  
\newcommand{\coS}{S^{\rm c}} 

\newcommand{\piso}{\pi(k_1)} %{\pi_S(1)}
\newcommand{\pisi}{\pi(k_i)} %{\pi_S(i)}
\newcommand{\pisiadd}{\pi(k_{i+1})} %{\pi_S(i+1)}
\newcommand{\pisimo}{\pi(k_{i-1})} %{\pi_S(i-1)}
\newcommand{\pisj}{\pi(k_j)} %{\pi_S(j)}
\newcommand{\pisjmo}{\pi(k_{j-1})} %{\pi_S(j-1)}
\newcommand{\pisjadd}{\pi(k_{j+1})} %{\pi_S(j+1)}
\newcommand{\piss}{\pi(k_s)} %{\pi_S(s)}
\newcommand{\pissmo}{\pi(k_{s-1})} %{\pi_S(s-1)}
\newcommand{\pios}{\pi(S)} %{\pi_S}

\newcommand{\pibi}{\pi(B_i)} 
\newcommand{\picobi}{\pi(S\Dff B_{i})}
\newcommand{\pibimo}{\pi(B_{i-1})} 
\newcommand{\picobimo}{\pi(S\Dff B_{i-1})}
\newcommand{\pibj}{\pi(B_j)} 
\newcommand{\picobj}{\pi(S\Dff B_{j})}
\newcommand{\pibs}{\pi(B_s)} 
\newcommand{\pibsmo}{\pi(B_{s-1})} 

\newcommand{\maho}{many-help-one } %{Many-Help-One}
\newcommand{\Mho}{Many-help-one } %{Many-Help-One}
\newcommand{\oho}{one-helps-one } %{One-Helps-One}

\newcommand{\D}{\mbox{\rm d}} %
\newcommand{\E}{\mbox{\rm E}} %

\newcommand{\conv}{\mbox{\rm conv}} %
\newcommand{\rsub}{\empty}

\newcommand{\DisT}{\Sigma_d}
\newcommand{\DisTi}{\Sigma_{\tilde{d}}}
\newcommand{\Npre}{\preceq{\!\!\!\!\!|\:\:}}
\newcommand{\EP}[1]
{
\ts \frac{1}{2\pi{\rm e}}{\rm e}^{\scriptstyle \frac{2}{n}h(#1)}
}
\newcommand{\CdEP}[2]
{
\ts \frac{1}{2\pi{\rm e}}{\rm e}^{\frac{2}{n}
h(\scriptstyle #1|\scriptstyle #2)}
}
\newcommand{\MEq}[1]{\stackrel{%\mbox
{\rm (#1)}}{=}}

\newcommand{\MLeq}[1]{\stackrel{%\mbox
{\rm (#1)}}{\leq}}

\newcommand{\ML}[1]{\stackrel{%\mbox
{\rm (#1)}}{<}}

\newcommand{\MGeq}[1]{\stackrel{%\mbox
{\rm (#1)}}{\geq}}

\newcommand{\MG}[1]{\stackrel{%\mbox
{\rm (#1)}}{>}}

\newcommand{\MPreq}[1]{\stackrel{%\mbox
{\rm (#1)}}{\preceq}}

\newcommand{\MSueq}[1]{\stackrel{%\mbox
{\rm (#1)}}{\succeq}}
%                                                           %
%-----------------------------------------------------------%
%                                                           %
%\date{}
\title{
%Rate Distortion Region for the 
Distributed Source Coding 
for Correlated Memoryless Gaussian Sources
}
%Separate Coding 
%Rate Distortion Region for the Vector Gaussian CEO Problem
%
\author{Yasutada~Oohama%,~\IEEEmembership{Member,~IEEE,}
\thanks{Manuscript received xxx, 20XX; revised xxx, 20XX.}% 
\thanks{Y. Oohama is with the Department of Information Science 
        and Intelligent Systems, 
        University of Tokushima,
        2-1 Minami Josanjima-Cho, Tokushima 
        770-8506, Japan.}
}
\markboth{
%IEEE Transactions on Information Theory,~Vol.~XX,No.~Y, 
%~Month~20XX
}
{Oohama: Separate Source Coding of Correlated Gaussian Remote Sources
}
% *** Note that you probably will NOT want to include the author's name in ***
% *** the headers of peer review papers.                                   ***

% If you want to put a publisher's ID mark on the page
% (can leave text blank if you just want to see how the
% text height on the first page will be reduced by IEEE)
%\pubid{0000--0000/00\$00.00~\copyright~2002 IEEE}

%\specialpapernotice{(Invited Paper)}

\maketitle

\begin{abstract}
We consider a distributed source coding problem of $L$ correlated
Gaussian observations $Y_i, i=1,2,\cdots,L$. We assume that the
random vector 
$Y^{L}={}^{\rm t}
(Y_1,Y_2,$ $\cdots,Y_L)$ is an observation of the Gaussian 
random vector  
$X^K={}^{\rm t}
(X_1,X_2,\cdots,X_K)$, 
having the form $Y^L=AX^K+N^L\,,$ where $A$ is a $L\times K$ 
matrix and $N^L={}^{\rm t}
(N_1,N_2,\cdots,N_L)$ is a vector of $L$ 
independent Gaussian random variables also independent 
of $X^K$. The estimation error on $X^K$ is measured by 
the distortion covariance matrix. The rate distortion 
region is defined by a set of all rate vectors for which
the estimation error is upper bounded by an arbitrary prescribed
covariance matrix in the meaning of positive semi definite.
In this paper we derive explicit outer and inner bounds 
of the rate distortion region. This result provides 
a useful tool to study the direct and indirect source coding 
problems on this Gaussian distributed source coding 
system, which remain open in general. 
%Our result may provide a useful 
%tool to study two other open problems on this Gaussian 
%distributed source coding system. 
\end{abstract}
%
%
%for the vector Gaussian CEO problem. 
%We show that the above inner and outer region coincides with each 
%other when Gaussian correlated sources satisfy some condition.
%Estimation error is measured by the distortion covariance matrix. 
%This coding problem is called the vector Gaussian CEO problem since 
%it can be considered as a vector version of the Gaussian 
%CEO problem, where $X_i,i=1,2,\cdots, L$ are identical. 
%For a given 
%
%Furthermore, we extend the results 
%to the case of $L$ correlated Gaussian observations.
%

\begin{keywords}
Multiterminal source coding, %\maho problem, 
rate-distortion region, CEO problem.   
\end{keywords}
% Will be ignored for other modes.
\IEEEpeerreviewmaketitle

\section{Introduction}

Distributed source coding of correlated information sources are a form of 
communication system which is significant from both theoretical and 
practical points of view in multi-user source networks. The first 
fundamental theory in those coding systems was established 
by Slepian and Wolf \cite{sw}. They considered a distributed 
source coding system of two correlated information sources. 
Those two sources are separately encoded and sent to a single 
destination, where the decoder reconstruct the original sources. 

%In this system, Slepian and Wolf \cite{sw} 
%determined the admissible rate region, the set that consists of a pair 
%of transmission rates for which two sources can be decoded with an 
%arbitrary small error probability. 

In the above distributed source coding systems we can consider the 
case where the source outputs should be reconstructed with average 
distortions smaller than prescribed levels. Such a situation 
suggests the multiterminal rate distortion theory.

%The rate-distortion theory for the separate coding system formulated by
%Slepian and Wolf has been studied by Wyner and Ziv \cite{wz}, Wyner
%\cite{w}, Berger \cite{bt}, Tung \cite{syt}, Berger et. al \cite{bh},
%Kaspi and Berger \cite{kb}, and Berger and Yeung \cite{by}. This problem,
%in general, remains an open problem and characterization of the rate
%distortion region has been unknown yet except for special cases. 

The rate distortion theory for the above distributed source 
coding system formulated by Slepian and Wolf has been studied by
\cite{wz}-\cite{oh1}. Wagner {\it et al.} \cite{wg3} gave 
a complete solution to this problem in the case of Gaussian
information sources and quadratic distortion by proving 
that sum rate part of the inner bound of Berger \cite{bt} and Tung \cite{syt} 
is tight. Wang {\it et al.} \cite{wa} gave a new alternative proof. 

%This situation
%was first studied by Yamamoto and Ito \cite{yam0} and subsequently,
%investigated by Flynn and R. M. Gray \cite{fg}.

As a practical situation of distributed source coding systems, we can
consider a case where the distributed encoders can not directly access
to the source outputs but can access to their noisy observations. This
situation was first studied by Yamamoto and Ito \cite{yam0}. They call
the investigated coding system the communication system with a remote
source. Subsequently, a similar distributed source coding 
system %or with a incomplete source output 
was studied by Flynn and R. M. Gray \cite{fg}.

%by Yamamoto \cite{yam1}
%We consider the 
%case where $L$ distributed encoders can not directly access the
%source vector $X^K={}^{\rm t}(X_1,X_2,\cdots, X^K)$, but can obtain
%$L$ correlated Gaussian sources $Y_i, i=1,2,\cdots,L$ which are noisy
%observations of $X^K$. 

In this paper we consider a distributed source coding problem of $L$
correlated Gaussian sources $Y_i,i=1,2,\cdots,L$ which are noisy
observations of $X_i, i=1,2,\cdots,K$. We assume that 
$Y^{L}={}^{\rm t}(Y_1,Y_2,$ $\cdots, Y_L)$ is an observation of the
source vector $X^K={}^{\rm t}(X_1,X_2,\cdots, X_K)$, having 
the form $Y^L=AX^K+N^L\,,$ where $A$ is a $L\times K$ matrix 
and $N^L={}^{\rm t}(N_1,N_2,\cdots,N_L)$ is a vector of $L$ 
independent Gaussian random variables also independent of $X^K$.

We consider two distortion criterions based on the covariance matrix of
the estimation error on $X^K$.  One is the criterion called the vector
distortion criterion distortion region where each of the the diagonal
elements of the covariance matrix is upper bounded by a prescribed
level. The other is the criterion called the sum distortion criterion
where the trace of the covariance matrix is upper bounded by a
prescribed level.  For each of the above two distirion criterions we
derive explicit inner and outer bounds of the rate distiron region. 
We also derive an explicit matching condition in the case of the sum
distortion criterion.

When $K=1$, the source coding system becomes that of the quadratic 
Gaussian CEO problem investigated by \cite{wa}, \cite{vb}-\cite{oh4}. 
The system in the case of $K=L$ and sum distortion criterion was studied 
by Pandya {\it et al.} \cite{pdya}. They derived lower and upper bounds 
of the minimum sum rate in the rate distortion region. Several partial 
solutions in the case of $K=L$, $A=I_L$ and sum distortion criterion are 
obtained by \cite{oh5}-\cite{oh9a}. The case of $K=L$, $A=I_L$ 
and vector distortion criterion is studied by \cite{oh8}.

The remote source coding problem treated in this paper is also referred 
to as the indirect distributed source coding problem. On the other hand, the 
multiterminal rate distortion  problem in the frame work of distributed 
source coding is called the direct distributed source coding problem. 
As shown in the paper of Wagner {\it et al.} \cite{wg3} and 
in the recent work by Wang {\it et al.} \cite{wa}, 
we have a strong connection between the direct and indirect 
distributed source coding problems. 

In this paper we also consider the multiterminal rate distortion
problem, i.e., the direct distributed source coding problem for the
Gaussian information source specified with $Y^L=X^L+N^L$, which
corresponds to the case of $K=L$ and $A=I_L$.  We shall derive a
result which implies a strong connection between the remote source
coding problem and the multiterminal rate distortion problem. This
result states that all results on the rate distortion region of the
remote source coding problem can be converted into those on the rate
distortion region of the multiterminal source coding problem. Using
this result, we drive several new partial solutions to the Gaussian
multiterminal rate distortion problem.

\section{Problem Statement and Previous Results}

\bfig
\setlength{\unitlength}{1.00mm}
\begin{picture}(80,48)(2,0)
%(100,47)(-30,4)

%Information sources:
\put(5,33){\framebox(6,6){$X_1$}}
\put(11,36){\vector(1,0){7.5}}

\put(5,19){\framebox(6,6){$X_2$}}
\put(11,22){\vector(1,0){7.5}}

\put(7,12){$\vdots$}

\put(5,2){\framebox(6,6){$X_K$}}
\put(11,5){\vector(1,0){7.5}}

\put(18.5,1.5){\framebox(4,38){$A$}}

%Random vectors X_1 and Y_1:
\put(12.5,38.6){${\vc X}_1$}
\put(30,40.6){${\vc Y}_1$}
\put(26,45){${\vc N}_1$}

%Random vectors X_2 and Y_2:
\put(12.5,24.6){${\vc X}_2$}
\put(30,25.6){${\vc Y}_2$}
\put(26,30){${\vc N}_2$}

%Random vectors X_L and Y_L:
\put(12.5,7.6){${\vc X}_K$}
\put(30,5.6){${\vc Y}_L$}
\put(26,10){${\vc N}_L$}

%Additive Noise 1:
\put(22.5,38){\vector(1,0){4}}
\put(29.5,38){\vector(1,0){7.5}}
\put(28,44){\vector(0,-1){4.5}}
\put(28,37){\line(0,1){2}}
\put(27,38){\line(1,0){2}}
\put(28,38){\circle{3.0}}

%Additive Noise 2:
\put(22.5,23){\vector(1,0){4}}
\put(29.5,23){\vector(1,0){7.5}}
\put(28,29){\vector(0,-1){4.5}}
\put(28,22){\line(0,1){2}}
\put(27,23){\line(1,0){2}}
\put(28,23){\circle{3.0}}

%Additive Noise l:
\put(22.5,3){\vector(1,0){4}}
\put(29.5,3){\vector(1,0){7.5}}
\put(28,9){\vector(0,-1){4.5}}
\put(28,2){\line(0,1){2}}
\put(27,3){\line(1,0){2}}
\put(28,3){\circle{3.0}}

%Encoder 1: 
\put(37,34.5){\framebox(7,7){$\varphi_1^{(n)}$}}
\put(45,40.6){$\varphi_1^{(n)}({\vc Y}_1)$}

%Encoder 2: 
\put(37,19.5){\framebox(7,7){$\varphi_2^{(n)}$}}
\put(45,25.6){$\varphi_2^{(n)}({\vc Y}_2)$}
\put(39.5,11){$\vdots$}
%Encoder L: 
\put(37,-0.5){\framebox(7,7){$\varphi_L^{(n)}$}}
\put(45,5.6){$\varphi_L^{(n)}({\vc Y}_L)$}

%Transmission line 1:
\put(44,38){\line(1,0){15}}
\put(59,38){\vector(1,-3){5}}

%Transmission line 2:
\put(44,23){\line(1,0){15}}
\put(59,23){\vector(1,0){5}}

%Transmission line l:
\put(44,3){\line(1,0){15}}
\put(59,3){\vector(1,4){5}}

%Decoder: 
\put(64,19.5){\framebox(7,7){$\psi^{(n)}$}}
\put(71,23){\vector(1,0){3}}
\put(74,22){$\left[
             \ba[c]{c}
              \hat{\vc X}_1\\
              \hat{\vc X}_2\\
              \vdots\\
             \hat{\vc X}_K
             \ea\right]
             $}
\end{picture}
\begin{flushleft}
{\small Fig. 1. Distributed source coding system for $L$ correlated 
        Gaussian observations}
\end{flushleft}
\efig
%on ${\cal R}_L(\DisT)$ 
%Our goal is determine the admissible 
%rate distortion region ${\cal R}_{\Iset}(\DisT)$ 
%in an explicit form.
%
\subsection{
Formal Statement of Problem
}

\newcommand{\baseN}{\rm e}

In this subsection we present a formal statement 
of problem. Throughout this paper all logarithms 
are taken to the base natural. 
Let $X_i, i=1,2,\cdots, K$ be correlated zero 
mean Gaussian random variable. 
For each $i=1,2,\cdots,K$, $X_i$ takes values 
in the real line ${\cal X}_i$. 
We write a $k$ dimensional random vector as 
$X^K=$ ${}^{\rm t}(X_1,X_2,$ $\cdots, X_K)$. 
We denote the covariance matrix of $X^K$ by $\Sigma_{X^K}$. 
Let $Y^{L}\defeq {}^{\rm t}(Y_1,Y_2,$ 
$\cdots, Y_L)$ 
be an observation of the source vector $X^K$, 
having the form $Y^L=AX^K+N^L$, where $A$ is 
a $L\times K$ %attenuation 
matrix and 
$N^L={}^{\rm t}(N_1,N_2,\cdots,N_L)$ 
is a vector of $L$ independent zero mean Gaussian random 
variables also independent of $X^K$. 
For $i=1,2,\cdots,L$, $\sigma_{N_i}^2$ stands for 
the variance of $N_i$. 
Let $\{(\Xast,$ $\Xbst, \cdots, X_K(t))\}_{t=1}^{\infty}$
be a stationary memoryless multiple Gaussian source. 
For each $t=1,$$2,\cdots,$ $X^K(t)\defeq $ 
${}^{\rm t}(X_{1}(t),X_{2}(t),\cdots,$ $\!X_{k}(t))\,$ 
has the same distribution as $X^K$. 
A random vector consisting of $n$ independent copies of 
the random variable $X_i$ is denoted by 
$$
{\vc X}_i\defeq (X_{i}(1),X_{i}(2),\cdots, X_{i}(n)).
$$
For each $t=1,2,\cdots$, $Y_i(t),i=1,2,\cdots,L$ 
is a vector of $L$ correlated observations of 
$X^K(t)$, having the form 
$
Y^L(t)=AX^K(t)+N^L(t),
$
where $N^L(t),t=1,2,\cdots,$ are independent 
identically distributed (i.i.d.) Gaussian random 
vector having the same distribution as $N^L$. 
We have no assumption on the number of observations $L$, 
which may be $L\geq K$ or $L<K$.   

%$L$ distributed encoders can only access 
%noisy observations 
%$Y_i$, $i=1,2, \cdots, L$ of $X^K$.

The distributed source coding system for $L$ 
correlated Gaussian observations treated in this 
paper is shown in Fig. 1. In this coding system 
the distributed encoder functions $\varphi_i, i=1,2,\cdots,L$ 
are defined by
$$
\varphi_i^{(n)}: {\cal X}_i^n \to {\cal M}_i 
\defeq \left\{1,2,\cdots, M_i\right\}\,.  
\label{eqn:enc}
$$
For each $i=1,2,\cdots,L$, set 
$
R_i^{(n)}\defeq \frac{1}{n}\log M_i\,, 
$
which stands for the transmission rate of 
the encoder function $\varphi_i^{(n)}$.
The joint decoder function $\psi^{(n)}=$ 
$(\psi_1^{(n)},$ $\psi_2^{(n)},$ $\cdots,\psi_K^{(n)})$ 
is defined by  
$$
\psi_i^{(n)}: {\cal M}_1 \times \cdots \times {\cal M}_L 
\to \hat{\cal X}_i^n\,,i=1,2,\cdots,K,
$$
where $\hat{\cal X}_i$ is the real line in which 
a reconstructed random variable of $X_i$ takes values. 
For ${\vc X}^K$ $=({\vc X}_1,$ ${\vc X}_2,$ $\cdots,$ 
${\vc X}_K)$, set
\beqno
\varphi^{(n)}({\vc Y}^{L}) & \defeq &
             \varphi_1^{(n)}({\vc Y}_1),
             \varphi_2^{(n)}({\vc Y}_2),
      \cdots,\varphi_L^{(n)}({\vc Y}_L)),
\\
   \hat{\vc X}^{K}
   &=&\left[
\ba{c}
\hat{\vc X}_1\\
\hat{\vc X}_2\\
      \vdots\\
\hat{\vc X}_K\\
\ea
\right]
\defeq 
\left[
\ba{c}
\psi_1^{(n)}(\varphi^{(n)}({\vc Y}^L))\\
\psi_2^{(n)}(\varphi^{(n)}({\vc Y}^L))\\
\vdots\\ 
\psi_K^{(n)}(\varphi^{(n)}({\vc Y}^L))\\
\ea
\right]\,,
\\
d_{ii}%({\vc X}_i-\hat{\vc X}_i)
& \defeq & {\rm E}||{\vc X}_i-\hat{\vc X}_i||^2 
\,,
\nonumber\\
d_{ij}%({\vc X}_i-\hat{\vc X}_i,{\vc X}_j-\hat{\vc X}_j) 
& \defeq &
{\rm E} \langle {\vc X}_i-\hat{\vc X}_i,
        {\vc X}_j-\hat{\vc X}_j\rangle\,, 1 \leq i\ne j \leq K,
\nonumber
\eeqno
where $||{\vc a}||$ stands for the Euclid norm of $n$ 
dimensional vector ${\vc a}$ and $\langle {\vc a},{\vc b}\rangle$
stands for the inner product between 
%the $n$ dimensional vectors
${\vc a}$ and ${\vc b}$. Let $\Sigma_{{\lvc X}^K-\hat{\lvc X}^K}$ 
be a covariance matrix with $d_{ij}$ %({\vc X}_i,\hat{\vc X}_j)$ 
in its $(i,j)$ entry. Let $\DisT$ be a given $L\times L$ 
covariance matrix which serves as a distortion criterion. 
We call this matrix a distortion matrix. 

For a given distortion matrix $\DisT$, the rate 
vector $(R_1,$ $R_2,\cdots, R_L)$ is $\DisT$-{\it admissible} 
if there exists a sequence 
$\{(\varphi_1^{(n)},$
   $\varphi_2^{(n)}, \cdots,$ 
   $\varphi_L^{(n)},$ 
   $\psi^{(n)})\}_{n=1}^{\infty}$ 
such that
\beqno
& &\limsup_{n\to\infty}R_i^{(n)}\leq R_i, 
   \mbox{ for }i=1,2,\cdots, L\,,
\\
& &\limsup_{n\to\infty}{\ts \frac{1}{n}}
   \Sigma_{{\lvc X}^K-\hat{\lvc X}^K} \preceq \DisT \,, 
\eeqno
where $A_1\preceq A_2$ means that $A_2-A_1$ 
is positive semi-definite matrix. 
Let ${\cal R}_{\Iset}(\DisT|\Sigma_{X^KY^L})$ 
denote the set of all $\DisT$-admissible 
rate vectors. We often have a particular interest 
in the minimum sum rate part of the rate distortion 
region. To examine this quantity, we set  
$$ 
R_{{\rm sum}, L}(\DisT|\Sigma_{X^KY^L})
\defeq \min_{\scs (R_1,R_2,\cdots,R_L)
\atop{\scs \in {\cal R}_{\Iset}(\Gamma,D^K|\Sigma_{X^KY^L})}}
\left\{\sum_{i=1}^{L}R_i\right\}\,.
$$
%Our goal is determine the admissible 
%rate distortion region ${\cal R}_{\Iset}(\DisT)$ 
%in an explicit form.
%By the rate distortion theory for single Gaussian sources, when
%$\Sigma_{X^L}\preceq $ $\DisT,$ 
%$R_1=R_2=$$\cdots=R_L=0$ is admissible. 
%In this case, we have  
%$$
%{\cal R}_L(\DisT)=
%\left\{(R_1,\cdots,R_L):R_i\geq 0, i\in \Lambda \right\}.
%$$
%In the subsequent arguments we focus on the case of 
%$\Sigma_{X^{L}}$ $\Npre\DisT$.
%
%In this Next,
% 
We consider two types of distortion criterion. For each 
distortion criterion we define the determination problem 
of the rate distortion region. 

{\it Problem 1. Vector Distortion Criterion: }  
Fix $K\times K$ invertible matrix $\Gamma$ and 
positive vector ${D}^K=$ $(D_1,$ $D_2,\cdots$ $, D_K)$. 
For given $\Gamma$ and $D^K$, the rate vector $(R_1,R_2,\cdots, R_L)$ 
is $(\Gamma,D^K)$-{\it admissible} if there exists a sequence 
$\{(\varphi_1^{(n)},$ 
   $\varphi_2^{(n)}, \cdots,$ 
   $\varphi_L^{(n)},$ $\psi^{(n)})\}_{n=1}^{\infty}$ 
such that
\beqno
& &\limsup_{n\to\infty}R^{(n)}\leq R_i,\mbox{ for }i=1,2,\cdots,L,  
\\
& &\limsup_{n\to\infty}
\left[\Gamma\left({\ts \frac{1}{n}}
\Sigma_{{\lvc X}^K-\hat{\lvc X}^K}\right){}^{\rm t}\Gamma\right]_{ii} 
\leq D_i\,,\mbox{ for }i=1,2,\cdots,K, 
\eeqno
where $[C]_{ij}$ stands for the $(i,j)$ entry of the matrix $C$. 
Let ${\cal R}_{\Iset}(\Gamma,D^K|\Sigma_{X^KY^L})$ denote the set 
of all $(\Gamma,D^K)$-admissible rate vectors. When $\Gamma$ 
is equal to the $K\times K$ identity matrix $I_K$, 
we omit $\Gamma$ in 
${\cal R}_{\Iset}(\Gamma,D|\Sigma_{X^KY^L})$ 
to simply write ${\cal R}_{\Iset}(D|\Sigma_{X^KY^L})$. 
Similar notations are used for other sets or quantities. 
To examine the sum rate part of 
${\cal R}_{\Iset}(\Gamma,D^K|\Sigma_{X^KY^L})$, define 
$$ 
R_{{\rm sum}, L}(\Gamma,D^K|\Sigma_{X^KY^L})
\defeq \min_{\scs (R_1,R_2,\cdots,R_L)
\atop{\scs \in {\cal R}_{\Iset}(\Gamma,D^K|\Sigma_{X^KY^L})}}
\left\{\sum_{i=1}^{L}R_i\right\}\,.
$$

{\it Problem 2. Sum Distortion Criterion:}
Fix $K\times K$ positive definite invertible matrix $\Gamma$ and 
positive $D$. For given $\Gamma$ and $D$, 
the rate vector $(R_1,R_2,\cdots, R_L)$ is 
$(\Gamma,D)$-{\it admissible} 
if there exists a sequence 
$\{(\varphi_1^{(n)},$ 
   $\varphi_2^{(n)}, \cdots,$ 
   $\varphi_L^{(n)},$ $\psi^{(n)})\}_{n=1}^{\infty}$ 
such that
\beqno
&&\limsup_{n\to\infty}R^{(n)}\leq R_i,
\mbox{ for }i=1,2,\cdots,L, 
\\
&&\limsup_{n\to\infty}
{\rm tr}\left[
\Gamma\left({\ts \frac{1}{n}}\Sigma_{{\lvc X}^K-\hat{\lvc X}^K}\right)
{}^{\rm t}\Gamma\right] 
\leq D. 
\eeqno
To examine the sum rate part of 
${\cal R}_{\Iset}(\Gamma,D|\Sigma_{X^KY^L})$, define
$$ 
R_{{\rm sum},L}(\Gamma,D|\Sigma_{X^KY^L})
\defeq \min_{\scs (R_1,R_2,\cdots,R_L)
\atop{\scs \in{\cal R}_{\Iset}(\Gamma,D|\Sigma_{X^KY^L})}}
\left\{\sum_{i=1}^{L}R_i\right\}\,.
$$
Let ${\cal S}_K(D^K)$ be a set of all $K\times K$ 
covariance matrices whose $(i,i)$ entry do not 
exceed $D_i$ for $i=1,2,\cdots,K$. Then we have
\beqa
& &{\cal R}_L(\Gamma,D^K|\Sigma_{X^KY^L})
   =\bigcup_{\Gamma \DisT {}^{\rm t}\Gamma \in {\cal S}_K(D^K)}
   \hspace*{-2mm}{\cal R}_L({\DisT}|\Sigma_{X^KY^L}),
\label{eqn:char1z}
\\
& &{\cal R}_L(\Gamma,D|\Sigma_{X^KY^L})
  =\bigcup_{{\rm tr}[\Gamma\DisT {}^{\rm t}\Gamma] \leq D}
   \hspace*{-2mm}{\cal R}_L({\DisT}|\Sigma_{X^KY^L}).
\label{eqn:char2z}
\eeqa
Furthermore, we have 
\beq
{\cal R}_L(\Gamma,D|\Sigma_{X^KY^L})
=\bigcup_{\sum_{i=1}^K D_i\leq D}{\cal R}_L(\Gamma,D^K|\Sigma_{X^KY^L})\,.
\eeq
In this paper we establish explicit inner and outer bounds 
of ${\cal R}_L({\DisT}|\Sigma_{X^KY^L})$. 
Using the above bounds and equations (\ref{eqn:char1z}) 
and (\ref{eqn:char2z}), we give new outer 
bounds of ${\cal R}_L(\Gamma,D|\Sigma_{X^KY^L})$ and 
${\cal R}_L(\Gamma,D^K|\Sigma_{X^KY^L})$.

\subsection{
Inner Bounds 
%of the Rate Distortion Region
and Previous Results
}

In this subsection we present inner bounds of 
${\cal R}_L(\DisT$ $|\Sigma_{X^KY^L})$,
${\cal R}_L(\Gamma,D^L$ $|\Sigma_{X^KY^L})$, and
${\cal R}_L(\Gamma,D$ $|\Sigma_{X^KY^L})$.
Those inner bounds can be obtained by a standard technique 
developed in the field of multiterminal source coding.

Set $\Lambda\defeq\{1,2,\cdots,L\}$. For $i\in \Lambda$, let ${U}_i$ 
be a random variable taking values in 
the real line ${\cal U}_i$. 
For any subset 
$S\subseteq \Lambda$, 
we introduce the notation 
$U_S=(U_i)_{i\in S}$. In particular $U_\Lambda=$ $U^L=$ 
$(U_1,$ $U_2,$ $\cdots, U_L)$.   
Define 
\beqno
%& &
{\cal G}(\DisT)
%\\
&\defeq& \ba[t]{l}
 \left\{U^L \right.:
  \ba[t]{l} 
  U^L\mbox{ is a Gaussian }
  \vspace{1mm}\\
  \mbox{random vector that satisfies}
  \vspace{1mm}\\
  U_S\to Y_S \to X^K \to Y_{S^{\rm c}} \to U_{S^{\rm c}}\,, 
  \vspace{1mm}\\
  U^L \to Y^L \to X^K
  \vspace{1mm}\\
  \mbox{for any $S\subseteq \Lambda$ and }\\ 
  \Sigma_{X^K-{\psi}(U^L)} \preceq \DisT
  \vspace{1mm}\\
  \mbox{for some linear mapping }
  \vspace{1mm}\\
  {\psi}: {\cal U}^L\to \hat{\cal X}^K\,. 
  \left. \right\}
  \ea
\ea
\eeqno
and set 
\beqno
& &
\hat{\cal R}_{L}^{({\rm in})}(\DisT|\Sigma_{X^KY^L})
\nonumber\\
&\defeq&{\rm conv}\ba[t]{l}
\left\{R^L \right. : 
%\vspace{1mm}\\
  \ba[t]{l}
  \mbox{There exists a random vector}
  \vspace{1mm}\\ 
  U^L\in {\cal G}(\DisT) \mbox{ such that }
  \vspace{1mm}\\
  \ds \sum_{i \in S} R_i \geq I(U_S;Y_S|U_{S^{\rm c}})
  \vspace{1mm}\\
  \mbox{ for any } S\subseteq \Lambda\,.
  \left. \right\}\,,
  \ea
\ea
\eeqno 
where $\conv\{A\}$ stands for the convex hull of the set $A$. 
Set 
\beqno
& &\hat{\cal R}_L^{\rm (in)}(\Gamma,D^K|\Sigma_{X^KY^L})
\\
&\defeq& \conv\left\{
    \bigcup_{\Gamma \DisT {}^{\rm t}\Gamma \in {\cal S}_K(D^K)}
    {\cal R}_L({\DisT}|\Sigma_{X^KY^L})
    \right\},
\\
& &\hat{\cal R}_L^{\rm (in)}(\Gamma,D|\Sigma_{X^KY^L})
\\
&\defeq&\conv\left\{
    \bigcup_{{\rm tr}[\Gamma\DisT {}^{\rm t}\Gamma] \leq D}
   {\cal R}_L({\DisT}|\Sigma_{X^KY^L})
   \right\}.
\eeqno
Define 
\beqno
\Sigma_{X^K|Y^L}
\defeq (\Sigma_{X^K}^{-1}+{}^{\rm t}A\Sigma_{N^L}^{-1}A)^{-1}
\eeqno
and set
\beqno
d^K(\Gamma\Sigma_{X^K|Y^L}{}^{\rm t}\Gamma)
&\defeq&\left([\Gamma\Sigma_{X^K|Y^L}{}^{\rm t}\Gamma]_{11},
         [\Gamma\Sigma_{X^K|Y^L}{}^{\rm t}\Gamma]_{22},
\right.\\
& &\left.
\:\cdots,[\Gamma\Sigma_{X^K|Y^L}{}^{\rm t}\Gamma]_{LL}\right)\,.
\eeqno
We can show that  
$\hat{\cal R}_{L}^{({\rm in})}(\DisT|\Sigma_{X^KY^L})$,
$\hat{\cal R}_{L}^{({\rm in})}(\Gamma,D^L|\Sigma_{X^KY^L})$,
and $\hat{\cal R}_{L}^{({\rm in})}(\Gamma,D|\Sigma_{X^KY^L})$
satisfy the following property.
\begin{pr}
\label{pr:prz001z} $\quad$\\
\begin{itemize}
\item[{\rm a)}] The set 
$\hat{\cal R}_{L}^{({\rm in})}(\DisT|\Sigma_{X^KY^L})$ is not void 
if and only if $\DisT \succ \Sigma_{X^K|Y^L}$.
\item[{\rm b)}] The set 
$\hat{\cal R}_{L}^{({\rm in})}(\Gamma,D^K|\Sigma_{X^KY^L})$ is not void 
if and only if $D^K > d^K(\Gamma$ 
$\Sigma_{X^K|Y^L}{}^{\rm t}\Gamma)$.
\item[{\rm c)}] The set 
$\hat{\cal R}_{L}^{({\rm in})}(\Gamma,D|\Sigma_{X^KY^L})$ is not void 
if and only if $D > {\rm tr }[\Gamma\Sigma_{X^K|Y^L}{}^{\rm t}\Gamma]$.
\end{itemize}
\end{pr}
%Details of our arguments will be stated in the subsequent 
%sections. 

On inner bounds of 
${\cal R}_{L}(\DisT|\Sigma_{X^KY^L})$, 
${\cal R}_{L}(\Gamma,D^L|\Sigma_{X^KY^L}$ $)$,
and $\hat{\cal R}_{L}(\Gamma,D|\Sigma_{X^KY^L})$,  
we have the following result.
\begin{Th}[Berger \cite{bt} and Tung \cite{syt}]\label{th:direct}
For any $\DisT$ $\succ$ \\$\Sigma_{X^K|Y^L}$, we have
\beqno
& &\hat{\cal R}_{L}^{({\rm in})}(\DisT|\Sigma_{X^KY^L})
   \subseteq {\cal R}_{L}(\DisT|\Sigma_{X^KY^L})\,.
\eeqno
For any $\Gamma$ and any $D^K$ $>$ 
$d^K(\Gamma\Sigma_{X^K|Y^L}{}^{\rm t}\Gamma)$, 
we have 
\beqno
& &\hat{\cal R}_{L}^{({\rm in})}(\Gamma,D^K|\Sigma_{X^KY^L})
   \subseteq {\cal R}_{L}(\Gamma,D^K|\Sigma_{X^KY^L})\,.
\eeqno
For any $\Gamma$ and any 
$D$ $>{\rm tr}[\Gamma\Sigma_{X^K|Y^L}{}^{\rm t}\Gamma]$, 
we have 
\beqno
& &\hat{\cal R}_{L}^{({\rm in})}(\Gamma,D|\Sigma_{X^KY^L})
   \subseteq {\cal R}_{L}(\Gamma,D|\Sigma_{X^KY^L})\,.
\eeqno
\end{Th}

The above three inner bounds can be regarded as variants 
of the inner bound which is well known as that of 
Berger \cite{bt} and Tung \cite{syt}.

When $K=1$ and $L\times 1$ column vector $A$ has the form 
$
A={}^{\rm t}[{11\cdots 1}],$ 
the system considered here becomes the quadratic Gaussian CEO problem. 
This problem was first posed and investigated by Viswanathan and Berger 
\cite{vb}. They further assumed $\Sigma_{N^L}=\sigma^2I_L$. 
Set $\sigma_X^2\defeq\Sigma_X$ and 
$$ 
R_{\rm sum}(D|\sigma_X^2,\sigma^2)
\defeq \liminf_{L\to\infty} R_{{\rm sum},L}(D|\Sigma_{XY^L})\,.
$$ 
Viswanathan and Berger \cite{vb} studied an asymptotic form 
of $R_{\rm sum}(D|\sigma_X^2,\sigma^2)$ for small $D$. 
Subsequently, Oohama \cite{oh2} determined an exact form 
of $R_{\rm sum}(D|\sigma_X^2,\sigma^2)$. 
The region ${\cal R}_L(D|\Sigma_{XY^L})$ 
was determined by Oohama \cite{oh4}.
%
%They dealt with a case where $k=l$ and  
%$A$ is $l \times L$ a positive definite authentication matrix. 
%Subsequently, Zhang and Wicker \cite{zh} studied the 
%same coding problem in the case where $A=I_L$ 
%and derived an explicit inner bound of 
%${\cal R}_{L}(D^K|\Sigma_{X^LY^L})$. 

%The source coding system in the case where 
%$l=km$ and $km\times k$ matrix $A$ has the form 
%$$
%A={}^{\rm t}[\underbrace{I_KI_K\cdots I_K}_{m}],
%$$ 
%was investigated Zhang and Wicker \cite{zh}. 
%They called the determination problem of 
%${\cal R}_L(D,$ 
%$\Sigma_{X^KY^{km}})$ the vector Gaussian CEO problem.  
%They derived inner bounds of ${\cal R}_L(D,$  
%$\Sigma_{X^KY^{km}})$.  
%Characterization of ${R}_{{\rm sum},L}(D|\Sigma_{X^LY^L})$
%was investigated by Pandya {\it et al.} \cite{pdya}. 
%They derived upper and lower bounds of 
%$R_{{\rm sum},L}(D|\Sigma_{X^LY^L})$. 

In the case where $K=L$ and $\Gamma=A=I_L$, 
Oohama \cite{oh5}-\cite{oh8} 
derived inner and outer bounds of ${\cal R}_{L}(D|\Sigma_{X^LY^L})$. 
Oohama %\cite{oh6},
\cite{oh7} also derived explicit 
sufficient conditions for inner and outer bounds to 
match and found examples of information sources for 
which rate distortion region are explicitly determined.
In \cite{oh8}, Oohama derived explicit outer bounds of 
${\cal R}_{L}(\DisT$ $|\Sigma_{X^LY^L}),$
${\cal R}_{L}(D^L$ $|\Sigma_{X^LY^L}),$ and 
${\cal R}_{L}(D$ $|\Sigma_{X^LY^L}).$

Recently, Wagner {\it et al.} \cite{wg3}
have determined ${\cal R}_2(D^2|$ $\Sigma_{X^2Y^2})$. 
Their result is as follows. 
\begin{Th}[Wagner {\it et al.} \cite{wg3}]
For any $D^2>d^2([\Sigma_{X^K|}$ ${}_{Y^L}])$, we have  
$$
{\cal R}_{2}(D^2|\Sigma_{X^2Y^2})
=\hat{\cal R}_{2}^{({\rm in})}(D^2|\Sigma_{X^2Y^2})\,.
$$
\end{Th}

Their method for the proof depends heavily on the specific 
property of $L=2$. It is hard to generalize it to the 
case of $L\geq 3$.

\section{Main Results}

\subsection{Inner and Outer Bounds of the Rate Distortion Region}
In this subsection we state our result on 
the characterizations of 
${\cal R}_L(\DisT$ $|\Sigma_{X^KY^L})$, 
${\cal R}_L(\Gamma, D^K$ $|\Sigma_{X^KY^L})$, and 
${\cal R}_L(\Gamma, D$ $|\Sigma_{X^KY^L})$. 
%Let $\Lambda\defeq \{1,2,\cdots, L\}$.
To describe those results we define several functions 
and sets. For $r_i\geq 0, i\in \Lambda$, 
let $N_{i}(r_i),$ $i\in \Lambda$ be $L$ independent 
Gaussian random variables with mean 0 and variance 
$\sigma_{N_i}^2/(1-{\baseN}^{-2r_i})$. 
Let $\Sigma_{N^L(r^L)}$ be a covariance matrix for 
the random vector $N^L(r^L)$. 
%For $A=$ $[a_{ij}]_{i\in \Lambda, 1\leq j\leq K}$ and $S\subset \Lambda$, set 
%$A_S=$ $[a_{ij}]_{i\in S, 1\leq j\leq K}$.
%we introduce the notation 
%$r_S=(r_i)_{i\in S}$. In particular $r_\Lambda=$ $r^L=$ 
%$(r_1,$ $r_2,$ $\cdots, r_L)$. 
Fix nonnegative vector $r^L$. For $\theta >0$ and 
for $S \subseteq  \Lambda$, define
\beqno
\Sigma_{N_{S^{\rm c}}(r_{S^{\rm c}})}^{-1} 
&\defeq& 
\left. \Sigma_{N^L(r^L)}^{-1} \right|_{r_{S}={\lvc 0}} \,,
\\
\underline{J}_{S}(\theta, r_S|r_{\coS})
&\defeq &\frac{1}{2}\log^{+}
   \left[\ts 
   \frac{\ds \prod_{i\in S} {\baseN}^{2r_i} }
        {\ds  \theta \left|\Sigma_{X^K}^{-1}
                 +{}^{\rm t}A\Sigma_{N_{S^{\rm c}}(r_{S^{\rm c}})}^{-1}A
                           \right|
        }
  \right],
\\
{J}_{S}\left(r_S|r_{\coS}\right)
&\defeq &\frac{1}{2}\log
   \left[\ts 
   \frac{\ds \left|\Sigma_{X^K}^{-1}
    +{}^{\rm t}A\Sigma_{N^L(r^L)}^{-1}A\right|
             %\left\{
             \prod_{i\in S} {\baseN}^{2r_i}
             %\right\}
        }
        {\ds  \left|\Sigma_{X^K}^{-1}
                 +{}^{\rm t}A
                 \Sigma_{N_{S^{\rm c}}(r_{S^{\rm c}})}^{-1}
                            A\right|
        }
  \right],
\eeqno
where $S^{\rm c}=\Lambda-S$ 
and $\log^{+}x\defeq\max\{\log x,0\}\,.$
Set
$$
{\cal A}_L(\DisT)
\defeq 
\left\{ r^L\geq 0:
\left[\Sigma_{X^K}^{-1}+
{}^{\rm t}A\Sigma_{N^L(r^L)}^{-1}A\right]^{-1}
             \preceq \DisT\right\}\,.
$$
We can show that for $S\subseteq \Lambda$, 
$\underline{J}_S(|\DisT|,$ $r_S|r_{\coS})$ and $J_S(r_S|r_{\coS})$ 
satisfy the following two properties.
\begin{pr}{
\label{pr:prz01z}
$\quad$
\begin{itemize}
\item[{\rm a)}] If $r^L\in {\cal A}_L(\DisT)$, then for any 
$S\subseteq \Lambda$, 
$$
\underline{J}_S(|\DisT|,r_S|r_{\coS})\leq J_S(r_S|r_{\coS})\,.
$$
\item[{\rm b)}] Suppose that $r^L\in {\cal A}_L(\DisT)$. 
If $\left. r^L\right|_{r_S={\lvc 0}}$ still belongs to 
${\cal A}_L(\DisT)$, then 
\beqno
& &\left. \underline{J}_S(|\DisT|, r_S|r_{\coS})
 \right|_{r_S={\lvc 0}}
 =\left. J_S(r_S|r_{\coS})\right|_{r_S={\lvc 0}}
\\
& &=0\,.
\eeqno
\end{itemize}
}\end{pr}

\begin{pr}\label{pr:matroid}{\rm 
Fix $r^L\in {\cal A}_L(\DisT)$. For $S \subseteq \Lambda$, set 
\beqno
{f}_S&=&{f}_S(r_S|r_{\coS})\defeq \underline{J}_S(|\DisT|,r_S|r_{\coS})\,.
\eeqno
By definition, it is obvious that ${f}_S, S \subseteq \Lambda$ 
are nonnegative. We can show that
$f\defeq \{{f}_S\}_{S \subseteq \Lambda}$ 
satisfies the followings:
\begin{itemize}
\item[{\rm a)}] ${f}_{\emptyset}=0$. 
\item[{\rm b)}] 
${f}_A\leq {f}_B$ 
for $A\subseteq B\subseteq \Lambda$.  
\item[{\rm c)}] ${f}_A+{f}_B \leq {f}_{A \cap B}+{f}_{A\cup B}\,.$
\end{itemize}
In general $(\Lambda,f)$ is called a {\it co-polymatroid} 
if the nonnegative function $\rho$ on $2^{\Lambda}$ satisfies 
the above three properties. Similarly, we set
\beqno
\tilde{f}_S&=&\tilde{f}_S(r_S|r_{\coS})\defeq J_S(r_S|r_{\coS})\,,
\quad \tilde{f}=\left\{\tilde{f}_S\right\}_{S \subseteq \Lambda}\,.
\eeqno
Then $(\Lambda,\tilde{f})$ also has the same three properties 
as those of $(\Lambda,f)$ and becomes a co-polymatroid. 
}\end{pr}

To describe our result on ${\cal R}_L(\DisT|\Sigma_{X^KY^L})$, set 
\beqno
& &
{\cal R}_L^{({\rm out})}(\theta,r^L|\Sigma_{X^KY^L})
\\
&\defeq&
\ba[t]{l}
  \left\{R^L \right. : 
  \ba[t]{l}
  \ds \sum_{i \in S} R_i 
  \geq \underline{J}_{S}\left(\theta,r_S|r_{\coS}\right)
  \vspace{1mm}\\
  \mbox{ for any }S \subseteq \Lambda\,. 
  \left. \right\}\,,
  \ea
\ea
\nonumber\\
& &{\cal R}_{L}^{({\rm out})}(\DisT|\Sigma_{X^KY^L})
\\
&\defeq& 
\bigcup_{r^L \in {\cal A}_L(\DisT)}
{\cal R}_L^{({\rm out})}(|\DisT|, r^L|\Sigma_{X^KY^L})\,, 
\nonumber\\
& &
{\cal R}_L^{({\rm in})}(r^L)
\\
&\defeq&
\ba[t]{l}
  \left\{R^L \right.: 
  \ba[t]{l}
  \ds \sum_{i \in S} R_i 
  \geq {J}_{S}\left(r_S|r_{\coS}\right)
  \vspace{1mm}\\
  \mbox{ for any }S \subseteq \Lambda\,. 
  \left. \right\}\,,
  \ea
\ea
\nonumber\\
& &{\cal R}_L^{({\rm in})}(\DisT|\Sigma_{X^KY^L})
\nonumber\\
&\defeq&{\rm conv}
        \left\{
        \bigcup_{r^L \in {\cal A}_L(\DisT)}
        {\cal R}_L^{({\rm in})}(r^L|\Sigma_{X^KY^L})
        \right\}\,. 
\eeqno
We can show that 
${\cal R}_L^{({\rm in})}(\DisT|\Sigma_{X^KY^L})$
and 
${\cal R}_L^{({\rm out})}(\DisT|\Sigma_{X^KY^L})$
satisfy the following property.
%We can show that  
%$\hat{\cal R}_{L}^{({\rm in})}(\DisT|\Sigma_{X^KY^L})$,
%$\hat{\cal R}_{L}^{({\rm in})}(\Gamma,D^L|\Sigma_{X^KY^L})$,
%and $\hat{\cal R}_{L}^{({\rm in})}(\Gamma,D|\Sigma_{X^KY^L})$
%satisfy the following property.
\begin{pr}
%\label{pr:prz001z} $\quad$\\
%\begin{itemize}
%\item[{\rm a)}] 
The sets 
${\cal R}_{L}^{({\rm in})}(\DisT|\Sigma_{X^KY^L})$ 
and ${\cal R}_{L}^{({\rm out})}(\DisT$ $|\Sigma_{X^KY^L})$ 
are not void if and only if $\DisT \succ \Sigma_{X^K|Y^L}$.
\end{pr}

Our result on inner and outer bounds of 
${\cal R}_{L}(\DisT|\Sigma_{X^KY^L})$ is as follows.
\begin{Th}\label{th:conv2}
For any $\DisT$$\succ$ $\Sigma_{X^K|Y^L}$, 
we have 
\beqno
& &{\cal R}_{L}^{({\rm in})}(\DisT|\Sigma_{X^KY^L})
\subseteq \hat{\cal R}_{L}^{({\rm in})}(\DisT|\Sigma_{X^KY^L})
\\
&\subseteq& {\cal R}_{L}(\DisT|\Sigma_{X^KY^L})
 \subseteq {\cal R}_{L}^{({\rm out})}(\DisT|\Sigma_{X^KY^L})\,.
\eeqno
\end{Th}

Proof of this theorem is given in Section V. This result 
includes the result of Oohama \cite{oh8} as a special 
case by letting $K=L$ and $\Gamma=A=I_L$. From this 
theorem we can derive outer and inner bounds of   
${\cal R}_{L}(\Gamma,D^K|$ $\Sigma_{X^KY^L})$ 
and 
${\cal R}_{L}(\Gamma,$$D|\Sigma_{X^KY^L})\,.$
To describe those bounds, set
\beqno
& &{\cal R}_L^{\rm (out)}(\Gamma,D^K|\Sigma_{X^KY^L})
\\
&\defeq& 
    \bigcup_{\Gamma \DisT {}^{\rm t}\Gamma \in {\cal S}_K(D^K)}
    {\cal R}_L^{\rm (out)}({\DisT}|\Sigma_{X^KY^L}),
\\
& &{\cal R}_L^{\rm (in)}(\Gamma,D^K|\Sigma_{X^KY^L})
\\
&\defeq& \conv\left\{
    \bigcup_{\Gamma \DisT {}^{\rm t}\Gamma \in {\cal S}_K(D^K)}
    {\cal R}_L^{\rm (in)}({\DisT}|\Sigma_{X^KY^L})
    \right\},
\\
& &{\cal R}_L^{\rm (out)}(\Gamma,D|\Sigma_{X^KY^L})
\\
&\defeq&
    \bigcup_{{\rm tr}[\Gamma\DisT {}^{\rm t}\Gamma] \leq D}
   {\cal R}_L^{\rm (out)}({\DisT}|\Sigma_{X^KY^L}),
\\
& &{\cal R}_L^{\rm (in)}(\Gamma,D|\Sigma_{X^KY^L})
\\
&\defeq&\conv\left\{
    \bigcup_{{\rm tr}[\Gamma\DisT {}^{\rm t}\Gamma] \leq D}
   {\cal R}_L^{\rm (in)}({\DisT}|\Sigma_{X^KY^L})
   \right\}.
\eeqno
Set 
\beqno
{\cal A}(r^L)
&\defeq & 
    \left\{\DisT:
    \DisT 
    \succeq (\Sigma_{X^K}^{-1}
+{}^{\rm t}A\Sigma_{N^L(r^L)}^{-1}A)^{-1}\right\}\,,
\\
\theta(\Gamma,D^K,r^L)
&\defeq &  
\max_{\scs \DisT:\DisT \in {\cal A}_L(r^L), 
      \atop{\scs
      \Gamma\DisT {}^{\rm t}\Gamma\in{\cal S}_K(D^K)}
     }
\left|\DisT \right|\,,
\\
\theta(\Gamma,D,r^L)
&\defeq & 
\max_{\scs \DisT: \DisT \in {\cal A}_L({r^L}),
      \atop{\scs 
       {\rm tr}[\Gamma\DisT {}^{\rm t}\Gamma]\leq D}
      }
\left|\DisT \right|\,.
\eeqno
Furthermore, set 
\beqno
& &{\cal B}_L(\Gamma,D^K)
\\
&\defeq&
\left\{r^L\geq 0:
\Gamma(\Sigma_{X^K}^{-1}+{}^{\rm t}
A\Sigma_{N^L(r^L)}^{-1}A)^{-1}{}^{\rm t}\Gamma\in{\cal S}_K(D^K)
\right\}\,,
\\
& &{\cal B}_L(\Gamma,D)
\\
&\defeq &
\left\{r^L\geq 0:
{\rm tr}[\Gamma(\Sigma_{X^K}^{-1}
+{}^{\rm t}A\Sigma_{N^L(r^L)}^{-1}A)^{-1}{}^{\rm t}\Gamma]
\leq D 
\right\}\,.
\eeqno
It can easily be verified that 
${\cal R}_{L}^{({\rm out})}(\Gamma,$ $D^K|\Sigma_{X^KY^L})$,
${\cal R}_{L}^{({\rm in})}($ $\Gamma,$ $D^K|\Sigma_{X^KY^L})$,
${\cal R}_{L}^{({\rm out})}(\Gamma,$ $D|\Sigma_{X^KY^L})$, and 
${\cal R}_{L}^{({\rm in})}(\Gamma,$ $D|$ $\Sigma_{X^KY^L})$ 
satisfies the following property.  
\begin{pr} $\quad$
\begin{itemize}
\item[{\rm a)}] 
The sets ${\cal R}_{L}^{({\rm in})}(\Gamma,D^K|\Sigma_{X^KY^L})$ 
and  ${\cal R}_{L}^{({\rm out})}(\Gamma,D^K|\Sigma_{X^K}$ ${}_{Y^L})$ 
are not void 
if and only if $D^K > d^K(\Gamma\Sigma_{X^K|Y^L}{}^{\rm t}\Gamma)$.
\item[{\rm b)}] 
The sets ${\cal R}_{L}^{({\rm in})}(\Gamma,D|\Sigma_{X^KY^L})$ 
and ${\cal R}_{L}^{({\rm out})}(\Gamma, D |\Sigma_{X^K}$ ${}_{Y^L})$ 
are not void 
if and only if $D>{\rm tr}[\Gamma \Sigma_{X^K|Y^L}$ ${}^{\rm t}\Gamma]$.
\item[{\rm c)}] 
\beqno
& &{\cal R}_{L}^{({\rm out})}(\Gamma,D^K|\Sigma_{X^KY^L})
\\
&=& 
\bigcup_{r^L \in {\cal B}_L(\Gamma,D^K)}
{\cal R}_L^{({\rm out})}(\theta(\Gamma,D^K,r^L),r^L|\Sigma_{X^KY^L})\,, 
\nonumber\\
& &  {\cal R}_L^{({\rm in})}(\Gamma,D^K|\Sigma_{X^KY^L})
\\
&=&
        \conv\left\{\bigcup_{r^L \in {\cal B}_L(\Gamma,D^K)}
        {\cal R}_L^{({\rm in})}(r^L|\Sigma_{X^KY^L})\right\}\,,
\\
& &{\cal R}_{L}^{({\rm out})}(\Gamma,D|\Sigma_{X^KY^L})
\\
&=& 
\bigcup_{r^L \in {\cal B}_L(\Gamma,D)}
{\cal R}_L^{({\rm out})}(\theta(\Gamma,D,r^L), r^L|\Sigma_{X^KY^L})\,, 
\\
& &
{\cal R}_L^{({\rm in})}(\Gamma,D|\Sigma_{X^KY^L})
\\
&=&\conv\left\{
        \bigcup_{r^L \in {\cal B}_L(\Gamma,D)}
        {\cal R}_L^{({\rm in})}(r^L)\right\}\,.
\eeqno
\end{itemize}
\end{pr}

The following result is obtained as a 
simple corollary from Theorem \ref{th:conv2}. 
\begin{co}\label{co:conv2z}
%(Oohama \cite{oh5}) 
For any $\Gamma$ and any $D^K>$ 
$d^K(\Gamma\Sigma_{X^K|Y^L}{}^{\rm t}\Gamma)$, 
we have 
\beqno
& &{\cal R}_{L}^{({\rm in})}(\Gamma,D^K|\Sigma_{X^KY^L})
\subseteq \hat{\cal R}_{L}^{({\rm in})}(\Gamma,D^K|\Sigma_{X^KY^L})
\\
&\subseteq& {\cal R}_{L}(\Gamma,D^K|\Sigma_{X^KY^L})
 \subseteq {\cal R}_{L}^{({\rm out})}(\Gamma,D^K|\Sigma_{X^KY^L})\,.
\eeqno
For any $\Gamma$ and any 
$D>{\rm tr}[\Gamma\Sigma_{X^K|Y^L}{}^{\rm t}\Gamma]$, 
we have 
\beqno
& &{\cal R}_{L}^{({\rm in})}(\Gamma,D|\Sigma_{X^KY^L})
\subseteq \hat{\cal R}_{L}^{({\rm in})}(\Gamma,D|\Sigma_{X^KY^L})
\\
&\subseteq& {\cal R}_{L}(\Gamma,D|\Sigma_{X^KY^L})
 \subseteq {\cal R}_{L}^{({\rm out})}(\Gamma,D|\Sigma_{X^KY^L})\,.
\eeqno
\end{co}

Those result includes the result of Oohama \cite{oh8} as a special 
case by letting $K=L$ and $\Gamma=A=I_L$. 
Next we compute $\theta(\Gamma,D,r^L)$ to 
derive a more explicit expression of ${\cal R}_{L}^{({\rm out})}
(\Gamma$ $,D|\Sigma_{X^KY^L})$. This expression 
will be quite useful for finding a sufficient 
condition for the outer bound 
${\cal R}_{L}^{({\rm out})}
(\Gamma$ $,D|\Sigma_{X^KY^L})$
to be tight.   
Let $\alpha_i=\alpha_i(r^L), i=1,2,\cdots,K$ 
be $K$ eigen values of the matrix 
$$
\Gamma^{-1}
\left(\Sigma_{X^K}^{-1}
+{}^{\rm t}A\Sigma_{N^L(r^L)}^{-1}A\right){}^{\rm t}\Gamma^{-1}\,.
$$
%Let $\partial{\cal B}_L(D)$ be the boundary of ${\cal B}_L(D)$, 
%that is, the set of all nonnegative vectors $r^L$ that satisfy
%$$
%{\rm tr}\left[\left(
%        \Sigma_{X^L}^{-1}+\Sigma_{N(r^L)}^{-1}
%        \right)^{-1}\right]=D\,. 
%$$
Let $\xi$ be a nonnegative number that satisfy 
$$
\sum_{i=1}^K\left\{[\xi- \alpha_i^{-1}]^{+}
+\alpha_i^{-1}\right\}=D.
$$    
Define  
$$
\omega(\Gamma,D,r^L)\defeq |\Gamma|^{-2}
\prod_{i=1}^K\left\{[\xi-\alpha_i^{-1}]^{+}+\alpha_i^{-1}\right\}.
$$
The function ${\omega}(\Gamma, D,r^L)$ has 
an expression of the so-called water filling 
solution to the following optimization problem:
\beqa
{\omega}(\Gamma, D,r^L)
=|\Gamma|^{-2}
  \max_{\scs \xi_i\alpha_{i}\geq 1,i\in\Lambda\,, 
  \atop{\scs
          \sum_{i=1}^K\xi_i\leq D
       }   
      }\prod_{i=1}^K\xi_{i}\,. 
\eeqa
Then we have the following theorem.
\begin{Th}\label{th:conv2a}
For any ${\Gamma}$ and any positive $D$, we have 
$$
\theta(\Gamma,D,r^L)=\omega(\Gamma,D,r^L)\,.
$$
A more explicit expression of 
${\cal R}_{L}^{({\rm out})}(\Gamma,D|\Sigma_{X^KY^L})$ 
using $\omega(\Gamma,D,r^L)$ is given by
\beqno
& &{\cal R}_{L}^{({\rm out})}(\Gamma,D|\Sigma_{X^KY^L})
\\
&\defeq& 
\bigcup_{r^L \in {\cal B}_L(\Gamma, D)}
{\cal R}_L^{({\rm out})}(\omega
(\Gamma,D,r^L),r^L|\Sigma_{X^KY^L})\,. 
\eeqno
\end{Th}

Proof of this theorem will be given in Section V. 
The above expression of the outer bound includes the result of 
Oohama \cite{oh8} as a special case by 
letting $K=L$ and $\Gamma=A=I_L$. 

\subsection{Matching Condition Analysis}

For $L\geq 3$, we present a sufficient 
condition for 
${\cal R}^{{(\rm out)}}_L(\Gamma,$ $D|$ $\Sigma_{X^KY^L})$ 
$\subseteq$ ${\cal R}_L^{({\rm in})}($$D|\Sigma_{X^KY^L})\,.$ 
We consider the following condition on 
$\theta(\Gamma, D,r^L)$.

{\it Condition: } For any $i\in \Lambda$, 
${\baseN}^{-2r_i}\theta(\Gamma, D,r^L)$ is a monotone 
decreasing function of $r_i\geq 0$.

We call this condition the MD condition. The following 
is a key lemma to derive the matching condition. 
This lemma is due to 
Oohama \cite{oh7}, \cite{oh7b}.
\begin{lm}[Oohama \cite{oh7},\cite{oh7b}] 
\label{lm:lem1}
If $\theta(\Gamma,D,r^L)$ satisfies 
the MD condition on ${\cal B}_L($ $\Gamma,D)$, then 
\beqno
 {\cal R}_L^{({\rm in})}(\Gamma,D|\Sigma_{X^KY^L})
&=&{\cal R}_L(\Gamma,D|\Sigma_{X^KY^L})
\\
&=&{\cal R}_L^{({\rm out})}(\Gamma,D|\Sigma_{X^KY^L}).
\eeqno 
\end{lm}

Based on Lemma \ref{lm:lem1}, we derive a sufficient 
condition for $\theta(\Gamma,D,r^L)$ to satisfy 
the MD condition. This sufficient condition is closely related 
to the distribution of eigen values of 
$$
{}^{\rm t}\Gamma^{-1}
(\Sigma_{X^K}^{-1}+{}^{\rm t}A\Sigma_{N^L(r^L)}^{-1}A)\Gamma^{-1}.
$$
Define
\beq
u_i \defeq \ts\frac{1}{\sigma_{N_i}^2}(1-{\baseN}^{-2r_i})
\,, \mbox{ for }i=1,2,\cdots,L. 
\label{eqn:trz}
\eeq
From (\ref{eqn:trz}), we have 
$$
2r_i=\log\frac{\frac{1}{\sigma_{N_i}^2}}{\frac{1}{\sigma_{N_i}^2}-u_i}\,. 
$$
By the above transformation we regard 
$$
{}^{\rm t}\Gamma^{-1}
(\Sigma_{X^K}^{-1}+{}^{\rm t}A\Sigma_{N^L(r^L)}^{-1}A)\Gamma^{-1}
$$
and $\theta(\Gamma,D,r^L)$ as functions of $u^L$, that is, 
\beqno
& &{}^{\rm t}\Gamma^{-1}
   (\Sigma_{X^K}^{-1}+{}^{\rm t}A\Sigma_{N^L(r^L)}^{-1}A)\Gamma^{-1}
\\
&=&{}^{\rm t}\Gamma^{-1}
   (\Sigma_{X^K}^{-1}+{}^{\rm t}A\Sigma_{N^L(u^L)}^{-1}A)\Gamma^{-1},
\eeqno
and $\theta(\Gamma,D,r^L)=\theta(\Gamma,D,u^L)$. 
Let $\hat{a}_{ij}$ be the $(i,j)$ entry of $A\Gamma^{-1}$. 
Set 
$
\hat{\vc a}_i\defeq [
\hat{a}_{i1}\hat{a}_{i2} \cdots \hat{a}_{iK}]\,.
$
Let $Q$ be a $K\times K$ unitary matrix. We consider the following 
matrix:
\beqno
& & {}^{\rm t}Q{}^{\rm t}\Gamma^{-1}
    (\Sigma_{X^K}^{-1}+{}^{\rm t}A\Sigma_{N^L(u^L)}^{-1}A)\Gamma^{-1}Q
\\
&=& {}^{\rm t}Q{}^{\rm t}\Gamma^{-1}\Sigma_{X^K}^{-1}\Gamma^{-1}Q
    +\sum_{j=1}^Lu_j{}^{\rm t}(\hat{\vc a}_jQ)(\hat{\vc a}_jQ)\,.
\eeqno
For each $i=1,2,\cdots,L$, choose the $K\times K$ 
unitary matrix $Q=Q_i$ so that 
$
\hat{\vc a}_iQ_i=[||\hat{\vc a}_i|| 0 \cdots 0 ]\,.
$ 
For this choice of $Q=Q_i$, set
\beqno
& &\eta_{i}=\eta_{i}(u^L_{[i]})
\\
&\defeq& 
\left[
{}^{\rm t}Q_i{}^{\rm t}\Gamma^{-1}\Sigma_{X^K}^{-1}\Gamma^{-1}Q_i
\right]_{11}
+\sum_{j\ne i}u_j
\left[{}^{\rm t}(\hat{\vc a}_jQ_i)(\hat{\vc a}_jQ_i)\right]_{11},  
\eeqno
where 
$
u^L_{[i]} \defeq u_{1}\cdots u_{i-1} u_{i+1}\cdots u_L\,.
$
Similar notations are used for other variables or random variables.
Then we have 
\beqno
& &[{}^{\rm t}Q_i{}^{\rm t}\Gamma^{-1}\left(\Sigma_{X^K}^{-1}
      +{}^{\rm t}A\Sigma_{N^L(u^L)}^{-1}A
  \right)\Gamma^{-1}Q_i]_{11}
\\
&=&
||\hat{\vc a}_i||^2u_i+\eta_i\,. 
\eeqno
If $(i^{\prime},i^{\prime \prime})\ne (1,1)$, then the value of 
\beqno
& &[{}^{\rm t}Q_i{}^{\rm t}\Gamma^{-1}
  (\Sigma_{X^K}^{-1}+{}^{\rm t}A\Sigma_{N^L(u^L)}^{-1}A)
\Gamma^{-1}Q_i]_{i^{\prime}i^{\prime \prime}}
\\
&=&[{}^{\rm t}Q_i{}^{\rm t}\Gamma^{-1}\Sigma_{X^K}^{-1}
   \Gamma^{-1}Q_i]_{i^{\prime}i^{\prime \prime}}
   +\sum_{j=1}^Lu_j
   \left[
   {}^{\rm t}(\hat{\vc a}_jQ_i)(\hat{\vc a}_jQ_i)
   \right]_{i^{\prime}i^{\prime \prime}}  
\eeqno
does not depend on $u_i$. Note that the matrix
$$
{}^{\rm t}Q_i{}^{\rm t}\Gamma^{-1}
(\Sigma_{X^K}^{-1}+{}^{\rm t}A\Sigma_{N^L(u^L)}^{-1}A)\Gamma^{-1}Q_i
$$
has the same eigen values as those of
$$
{}^{\rm t}\Gamma^{-1}
(\Sigma_{X^K}^{-1}+{}^{\rm t}A\Sigma_{N^L(u^L)}^{-1}A)\Gamma^{-1}\,.
$$
We recall here that $\alpha_i=\alpha_j(u^L), j=1,2,\cdots,K$ 
are $K$ eigen values of the above two matrices. 
Let $\alpha_{\min}=\alpha_{\min}(u^L)$ and 
$\alpha_{\max}=\alpha_{\max}(u^L)$ be the minimum 
and maximum eigen values among $\alpha_j,j=1,2,\cdots,K$. 
According to Oohama \cite{oh7}, \cite{oh7b}, we have the following 
lemma on those eigen values. 
\begin{lm}[Oohama\cite{oh7},\cite{oh7b}] 
\label{lm:Egn1}$\:$For each $i=1,2,\cdots,L$, we have
\beqno
& &\alpha_{\min}(u^L)\leq 
   ||\hat{\vc a}_i||^2u_i+\eta_i(u_{[i]}^L) 
   \leq \alpha_{\max}(u^L),
\\
& &\frac{\partial \alpha_j}{\partial u_i}
   \geq 0, \mbox{ for }j=1,2,\cdots,K,\:\:
\sum_{j=1}^K\frac{\partial \alpha_j}{\partial u_i}
   =||\hat{\vc a}_i||^2\,.
\eeqno
\end{lm}

The following is a key lemma to derive a sufficient 
condition for the MD condition to hold. 
\begin{lm}\label{lm:pro3}
If 
$\alpha_{\min}(u^L)$ and $\alpha_{\max}(u^L)$ satisfy 
\beqno
& &\frac{1}{\alpha_{\min}(u^L)}-\frac{1}{\alpha_{\max}(u^L)}
\leq \frac{1}{||\hat{\vc a}_i||^2
     \frac{1}{\sigma^2_{N_i}}+\eta_i(u^L_{[i]})}\,,
\\
& &\mbox{ for }i=1,2,\cdots,L
\eeqno
on ${\cal B}_L(\Gamma,D)$, then $\theta(\Gamma, D,u^L)$ 
satisfies the MD condition on ${\cal B}_L(\Gamma,D)$.
\end{lm}
%
%Set
%\beqno 
%{\cal C}
%&\defeq & 
%\{(\Gamma, D,\Sigma_{X^KY^L}): 
%\ba[t]{l} r^L\in {\cal B}_L(\Gamma, D)\\ 
%\mbox{ for some nonnegative }r^L. \}.   
%\ea
%\eeqno
%When $r^L\geq s^L$, we have 
%\beqno
%& & 
%{}^{\rm t}\Gamma^{-1}
%(\Sigma_{X^K}^{-1}+{}^{\rm t}A\Sigma_{N^L(r^L)}^{-1}A)
%\Gamma^{-1}
%\\
%& &\succeq 
%{}^{\rm t}\Gamma^{-1}
%(\Sigma_{X^K}^{-1}+{}^{\rm t}A\Sigma_{N^L(s^L)}^{-1})
%\Gamma^{-1}
%\\
%&\Rightarrow &
%\Gamma
%\left(\Sigma_{X^K}^{-1}+\Sigma_{N^L(r^L)}^{-1}\right)^{-1}
%{}^{\rm t}\Gamma
%\preceq 
%\Gamma
%\left(\Sigma_{X^K}^{-1}+\Sigma_{N^L(s^L)}^{-1}\right)^{-1}
%{}^{\rm t}\Gamma
%\\
%&\Rightarrow &
%{\rm tr}\left[
%\Gamma
%\left(\Sigma_{X^K}^{-1}+{}^{\rm t}
%A\Sigma_{N^L(r^L)}^{-1}A
%\right)^{-1}{}^{\rm t}\Gamma
%\right]
%\\& & \leq 
%{\rm tr}\left[
%\Gamma
%\left(\Sigma_{X^K}^{-1}+{}^{\rm t}
%A\Sigma_{N^L(s^L)}^{-1}A
%\right)^{-1}{}^{\rm t}\Gamma
%\right]
%\eeqno
%which implies that 
%$
%{\rm tr}\left[
%\Gamma
%\left(\Sigma_{X^K}^{-1}+{}^{\rm t}
%A\Sigma_{N^L(r^L)}^{-1}A
%\right)^{-1}{}^{\rm t}\Gamma
%\right]
%$
%is a monotone decreasing function of $r^L$. Hence, 
%we have 
%$$
%{\cal C}
%=
%\ba[t]{l}
%\Bigl\{(\Gamma,D,\Sigma_{X^KY^L}):\Bigr.
%\\
%\Bigl.
%D>{\rm tr}\left[
%\Gamma\left(
%  \Sigma_{X^K}^{-1}+{}^{\rm t}A\Sigma_{N^L}^{-1}A
%\right)^{-1}{}^{\rm t}\Gamma\right]\Bigr\}\,.
%\ea
%$$

Proof of Lemma \ref{lm:pro3} will be stated in Section V. 
Let $\alpha_{\max}^{*}$ be the maximum eigen value of 
$$
{}^{\rm t}\Gamma^{-1}
(\Sigma_{X^K}^{-1}+{}^{\rm t}A\Sigma_{N^L}^{-1}A)\Gamma^{-1}.
$$
From Lemmas \ref{lm:lem1}-\ref{lm:pro3} and an elementary 
computation we obtain the following. 
\begin{Th}\label{th:matchTh}
If we have 
\beqno
& &
{\rm tr}[\Gamma\Sigma_{X^K|Y^L}{}^{\rm t}\Gamma] 
< D \leq \ts \frac{K+1}{\alpha_{\max}^{*}}\,,
\eeqno
then 
\beqno
  & &{\cal R}_L^{({\rm in})}(\Gamma, D|\Sigma_{X^K})
=\hat{\cal R}_L^{({\rm in})}(\Gamma, D|\Sigma_{X^K})
\\
&=&{\cal R}_L(\Gamma,D|\Sigma_{X^KY^L})
 ={\cal R}_L^{({\rm out})}(\Gamma,D|\Sigma_{X^KY^L}).
\eeqno
In particular,
\beqno
& &R_{{\rm sum},L}(\Gamma, D)
\nonumber\\
&=& \min_{r^L\in {\cal B}_L(\Gamma, D)}
\left\{
\sum_{i=1}^Lr_i
+\frac{1}{2}\log
      %\left[ 
      \frac{\left|\Sigma_{X^K}^{-1}
      +{}^{\rm t}A\Sigma_{N^L(r^L)}^{-1}A\right|}
      {\left|\Sigma_{X^K}^{-1}\right|}
      %\right] 
\right\}.
%\label{eqn:miniz}
\eeqno
\end{Th}

Proof of Theorem \ref{th:matchTh} 
will be stated in Section V. From this theorem, we can see 
that if the value of $D$ is very close to 
${\rm tr}[\Gamma\Sigma_{X^K|Y^L}{}^{\rm t}\Gamma]$,  
${\cal R}_L^{(\rm in)}(\Gamma,$  $D|\Sigma_{X^KY^L})$ 
and ${\cal R}_L^{({\rm out})}(\Gamma,$ $D|\Sigma_{X^KY^L})$ 
match. 

\section{
Application to the Multiterminal 
Rate Distortion Problem
}

\bfig
\setlength{\unitlength}{1.00mm}
\begin{picture}(79,58)(0,0)
%(100,47)(-30,4)
\put(2,-1){\dashbox{0.5}(15,50)}
%Information sources:
\put(3,35){\framebox(6,6){$X_1$}}
\put(19,35){\framebox(6,6){$Y_1$}}

\put(3,20){\framebox(6,6){$X_2$}}
\put(19,20){\framebox(6,6){$Y_2$}}

\put(5,11){$\vdots$}
\put(3,0){\framebox(6,6){$X_L$}}
\put(19,0){\framebox(6,6){$Y_L$}}

%Random vectors X_1 and Y_1:
%\put(10.5,40.6){${\vc X}_1$}
\put(26,40.6){${\vc Y}_1$}
\put(12,45){${N}_1$}

%Random vectors X_2 and Y_2:
%\put(10.5,25.6){${\vc X}_2$}
\put(26,25.6){${\vc Y}_2$}
\put(12,30){${N}_2$}

%Random vectors X_L and Y_L:
%\put(10.5,5.6){${\vc X}_L$}
\put(26,5.6){${\vc Y}_L$}
\put(12,10){${N}_L$}

%Additive Noise 1:
\put(9,38){\vector(1,0){3.5}}
\put(14,44){\vector(0,-1){4.5}}
\put(14,37){\line(0,1){2}}
\put(13,38){\line(1,0){2}}
\put(14,38){\circle{3.0}}

%Additive Noise 2:
\put(9,23){\vector(1,0){3.5}}
\put(14,29){\vector(0,-1){4.5}}
\put(14,22){\line(0,1){2}}
\put(13,23){\line(1,0){2}}
\put(14,23){\circle{3.0}}

%Additive Noise l:
\put(9,3){\vector(1,0){3.5}}
\put(14,9){\vector(0,-1){4.5}}
\put(14,2){\line(0,1){2}}
\put(13,3){\line(1,0){2}}
\put(14,3){\circle{3.0}}

%Encoder 1: 
\put(32,34.5){\framebox(7,7){$\varphi_1^{(n)}$}}
\put(40,40.6){$\varphi_1^{(n)}({\vc Y}_1)$}

%Encoder 2: 
\put(32,19.5){\framebox(7,7){$\varphi_2^{(n)}$}}
\put(40,25.6){$\varphi_2^{(n)}({\vc Y}_2)$}

\put(34.5,11){$\vdots$}

%Encoder L: 
\put(32,-0.5){\framebox(7,7){$\varphi_L^{(n)}$}}
\put(40,5.6){$\varphi_L^{(n)}({\vc Y}_L)$}

%Transmission line 1:
\put(15.5,38){\vector(1,0){3.5}}
\put(25,38){\vector(1,0){7}}
\put(39,38){\line(1,0){15}}
\put(54,38){\vector(2,-3){10}}

%Transmission line 2:
\put(15.5,23){\vector(1,0){3.5}}
\put(25,23){\vector(1,0){7}}
\put(39,23){\line(1,0){15}}
\put(54,23){\vector(1,0){10}}

%Transmission line L:
\put(15.5,3){\vector(1,0){3.5}}
\put(25,3){\vector(1,0){7}}
\put(39,3){\line(1,0){15}}
\put(54,3){\vector(1,2){10}}

%Decoder: 
\put(64,19.5){\framebox(7,7){$\phi^{(n)}$}}
\put(71,23){\vector(1,0){3}}
\put(74,22){$\left[
             \ba{c}
             \hat{\vc Y}_1\\
             \hat{\vc Y}_2\\
             \vdots \\ 
             \hat{\vc Y}_L
             \ea
             \right]$}
\end{picture}
\begin{flushleft}
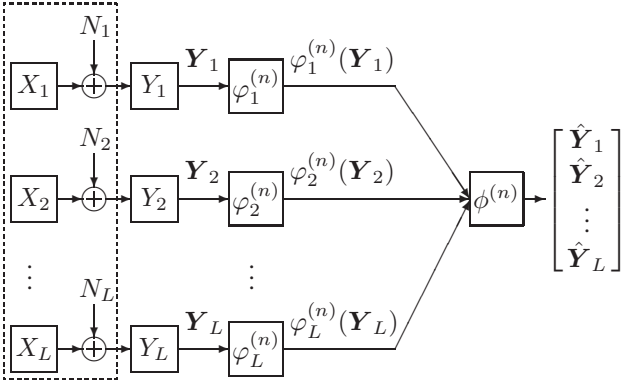
{\small Fig. 2. Distributed source coding system for $L$ correlated 
        Gaussian sources}
\end{flushleft}
\efig

In this section we consider the multiterminal rate distortion problem 
for Gaussian information source specified with $Y^L$. We consider the
case where $K=L$ and $A=I_L$. In this case we have $Y^L=X^L+N^L$. The
Gaussian random variables $Y_i$,$i=1,2,\cdots,L$ are $L$-noisy
components of random vector $X^L$. The Gaussian random vector $X^L$
can be regarded as a ``hidden'' information source of $Y^L$. Note that
$(X^L,Y^L)$ satisfies $Y_S \to X^L \to Y_{S^{\rm c}} \mbox{ for any }S
\subseteq \Lambda\,. 
$

\subsection{
Problem Formulation and Previous Results
}

The distributed source coding system for $L$ 
correlated Gaussian source treated here is 
shown in Fig. 2. Definitions of encoder 
functions $\varphi_i, i=1,2,\cdots, L$ 
are the same as the previous definitions.
The decoder function $\phi^{(n)}=$ 
$(\phi_1^{(n)},$ $\phi_2^{(n)},$ $\cdots,\phi_L^{(n)})$ 
is defined by  
$$
\phi_i^{(n)}: {\cal M}_1 \times \cdots \times {\cal M}_L 
\to \hat{\cal Y}_i^n\,,i=1,2,\cdots,K,
$$
where $\hat{\cal Y}_i$ is the real line in which 
estimations of $Y_i$ take values.
For ${\vc Y}^L$ $=({\vc Y}_1,$ ${\vc Y}_2,$ $\cdots,$ 
${\vc Y}_L)$, set
\beqno
   \hat{\vc Y}^{L}
   &=&\left[
\ba{c}
\hat{\vc Y}_1\\
\hat{\vc Y}_2\\
      \vdots\\
\hat{\vc Y}_L\\
\ea
\right]
\defeq 
\left[
\ba{c}
\phi_1^{(n)}(\varphi^{(n)}({\vc Y}^L))\\
\phi_2^{(n)}(\varphi^{(n)}({\vc Y}^L))\\
\vdots\\ 
\phi_L^{(n)}(\varphi^{(n)}({\vc Y}^L))\\
\ea
\right]\,,
\\
\tilde{d}_{ii}%({\vc X}_i-\hat{\vc X}_i)
& \defeq & {\rm E}||{\vc Y}_i-\hat{\vc Y}_i||^2 
\,,
\nonumber\\
\tilde{d}_{ij}%({\vc X}_i-\hat{\vc X}_i,{\vc X}_j-\hat{\vc X}_j) 
& \defeq &
{\rm E} \langle {\vc Y}_i-\hat{\vc Y}_i,
        {\vc Y}_j-\hat{\vc Y}_j\rangle\,, 1 \leq i\ne j \leq L\,.
\nonumber
\eeqno
Let $\Sigma_{{\lvc Y}^L-\hat{\lvc Y}^L}$ 
be a covariance matrix with $\tilde{d}_{ij}$ 
in its $(i,j)$ entry. 

For a given $\DisT$, the rate vector $(R_1,R_2,\cdots, R_L)$ 
is $\DisT$-{\it admissible} if there exists a sequence 
$\{(\varphi_1^{(n)},$
   $\varphi_2^{(n)}, \cdots,$ 
   $\varphi_L^{(n)},$ 
   $\psi^{(n)})\}_{n=1}^{\infty}$ 
such that
\beqno
& &\limsup_{n\to\infty}R^{(n)}\leq R_i, 
   \mbox{ for }i=1,2,\cdots, L\,,
\\
& &\limsup_{n\to\infty}{\ts \frac{1}{n}}
   \Sigma_{{\lvc Y}^L-\hat{\lvc Y}^L} \preceq \DisT\,. 
\eeqno
Let ${\cal R}_{\Iset}(\DisT|\Sigma_{Y^L})$ 
denote the set of all $\DisT$-admissible 
rate vectors. 

We consider two types of distortion criterion. For each distortion 
criterion we define the determination 
problem of the rate distortion region. 

{\it Problem 3. Vector Distortion Criterion:}  
For given $L\times L$ invertible matrix $\Gamma$ and 
$D^L>0$, the rate vector $(R_1,R_2,$ $\cdots,R_L)$ is 
$(\Gamma,D^L)$-{\it admissible} if there exists a sequence 
$\{(\varphi_1^{(n)},$ 
   $\varphi_2^{(n)}, \cdots,$ 
   $\varphi_L^{(n)},$ $\phi^{(n)})\}_{n=1}^{\infty}$ 
such that
\beqno
& &\limsup_{n\to\infty}R^{(n)}\leq R_i\,,
\mbox{ for }i=1,2,\cdots,L\,,  
\\
& &\limsup_{n\to\infty}
\left[\Gamma\left(
{\ts \frac{1}{n}}
\Sigma_{{\lvc Y}^L-\hat{\lvc Y}^L}
\right){}^{\rm t}\Gamma\right]_{ii} 
\leq D_i\,,
\mbox{ for }i=1,2,\cdots,L\,. 
\eeqno
Let ${\cal R}_{\Iset}(\Gamma,D^L|\Sigma_{Y^L})$ denote the set 
of all $(\Gamma,D^L)$-admissible rate vectors. The sum rate 
part of the rate distortion region is defined by
$$ 
R_{{\rm sum}, L}(\Gamma,D^L|\Sigma_{Y^L})
\defeq \min_{\scs (R_1,R_2,\cdots,R_L)
\atop{\scs \in {\cal R}_{\Iset}(\Gamma,D^L|\Sigma_{Y^L})}}
\left\{\sum_{i=1}^{L}R_i\right\}\,.
$$

{\it Problem 4. Sum Distortion Criterion:}
For given $L\times L$ invertible matrix $\Gamma$ 
and $D>0$, the rate vector 
$(R_1,R_2,\cdots, R_L)$ is 
$(\Gamma,D)$-{\it admissible} 
if there exists a sequence 
$\{(\varphi_1^{(n)},$ 
   $\varphi_2^{(n)}, \cdots,$ 
   $\varphi_L^{(n)},$ $\phi^{(n)})\}_{n=1}^{\infty}$ 
such that
\beqno
&&\limsup_{n\to\infty}R^{(n)}\leq R_i,
\mbox{ for }i=1,2,\cdots,L, 
\\
&&\limsup_{n\to\infty}
{\rm tr}\left[
\Gamma
\left({\ts \frac{1}{n}}\Sigma_{{\lvc Y}^L-\hat{\lvc Y}^L}
\right){}^{\rm t}\Gamma\right] 
\leq D\,. 
\eeqno
Let ${\cal R}_{\Iset}(\Gamma,D|\Sigma_{Y^L})$ denote 
the set of all admissible rate vectors. The sum 
rate part of the rate distortion region 
is defined by
$$ 
R_{{\rm sum}, L}(\Gamma,D|\Sigma_{Y^L})
\defeq \min_{\scs (R_1,R_2,\cdots,R_L)
\atop{\scs \in {\cal R}_{\Iset}(\Gamma,D|\Sigma_{Y^L})}}
\left\{\sum_{i=1}^{L}R_i\right\}\,.
$$

Relations between 
${\cal R}_{\Iset}(\DisT|\Sigma_{Y^L}),$ 
${\cal R}_{\Iset}(\Gamma,D^L|\Sigma_{Y^L}),$ and  
${\cal R}_{\Iset}(\Gamma,$ $D|\Sigma_{Y^L})$ 
are as follows.
\beqa
& &{\cal R}_L(\Gamma,D^L|\Sigma_{Y^L})
  =\bigcup_{\Gamma \DisT {}^{\rm t}\Gamma\in {\cal S}_L(D^L)}
   {\cal R}_L({\DisT}|\Sigma_{Y^L}),
\label{eqn:char1}
\\
& &{\cal R}_L(\Gamma,D|\Sigma_{Y^L})
   =\bigcup_{{\rm tr}[\Gamma\DisT{}^{\rm t}\Gamma ] \leq D}
   {\cal R}_L({\DisT}|\Sigma_{Y^L}).
\label{eqn:char2}
\eeqa
Furthermore, we have 
\beq
{\cal R}_L(\Gamma,D|\Sigma_{Y^L})
=\bigcup_{\sum_{i=1}^L D_i\leq D}{\cal R}_L(\Gamma,D^L|\Sigma_{Y^L})\,.
\eeq

We first present inner bounds of 
${\cal R}_L(\DisT$ $|\Sigma_{Y^L})$,
${\cal R}_L(\Gamma,D^L$ $|\Sigma_{Y^L})$, and 
${\cal R}_L(\Gamma,$ $D|\Sigma_{Y^L})$.
Those inner bounds can be obtained by a standard technique 
of multiterminal source coding. Define 
\beqno
\tilde{\cal G}(\DisT)
&\defeq& \ba[t]{l}
 \left\{U^L \right.:
  \ba[t]{l} 
  U^L\mbox{ is a Gaussian }
  \vspace{1mm}\\
  \mbox{random vector that satisfies}
  \vspace{1mm}\\
  U_S\to Y_S \to X^L \to 
  Y_{S^{\rm c}} \to U_{S^{\rm c}} 
  \vspace{1mm}\\
  U^L \to Y^L \to X^L
  \vspace{1mm}\\
  \mbox{for any $S\subset \Lambda$ and }\\ 
  %\ds \sum_{i=1}^L{\rm E}\left
\Sigma_{Y^L-{\phi}(U^L)} \preceq \DisT
  \vspace{1mm}\\
  \mbox{for some linear mapping }
  \vspace{1mm}\\
  {\phi}: {\cal U}^L\to \hat{\cal Y}^L\,.  
  \left. \right\}
  \ea
\ea
\eeqno
and set 
\beqno
& &
\hat{\cal R}_{L}^{({\rm in})}(\DisT|\Sigma_{Y^L})
\nonumber\\
&\defeq&{\rm conv}\ba[t]{l}
\left\{R^L \right. : 
%\vspace{1mm}\\
  \ba[t]{l}
  \mbox{There exists a random vector}
  \vspace{1mm}\\ 
  U^L\in \tilde{\cal G}(\DisT) \mbox{ such that }
  \vspace{1mm}\\
  \ds \sum_{i \in S} R_i \geq I(U_S;Y_S|U_{S^{\rm c}})
  \vspace{1mm}\\
  \mbox{ for any } S\subseteq \Lambda\,.
  \left. \right\}\,,
  \ea
\ea
\eeqno
\beqno
& &\hat{\cal R}_{L}^{({\rm in})}(\Gamma,D^L|\Sigma_{Y^L}) 
\\
&\defeq& 
\conv 
\left\{
\bigcup_{\Gamma\DisT{}^{\rm t}\Gamma \in {\cal S}_L(D^L)}
\hat{\cal R}_{L}^{({\rm in})}(\DisT|\Sigma_{Y^L})
\right\}\,,
\\
& &\hat{\cal R}_{L}^{({\rm in})}(\Gamma,D|\Sigma_{Y^L}) 
\\
&\defeq &\conv 
\left\{
\bigcup_{{\rm tr}[\Gamma\DisT{}^{\rm t}\Gamma]\leq D}
\hat{\cal R}_{L}^{({\rm in})}(\DisT|\Sigma_{Y^L}) 
\right\}\,.
\eeqno
Then we have the following result.
\begin{Th}[Berger \cite{bt} and Tung \cite{syt}]\label{th:direct2}
For any positive \\ definite $\DisT$, we have
\beqno
& &\hat{\cal R}_{L}^{({\rm in})}(\DisT|\Sigma_{Y^L})
\subseteq {\cal R}_{L}(\DisT|\Sigma_{Y^L}).
\eeqno
For any invertible $\Gamma$ and any $D^L>0$, we have
\beqno
& &\hat{\cal R}_{L}^{({\rm in})}(\Gamma,D^L|\Sigma_{Y^L})
  \subseteq {\cal R}_{L}(\Gamma,D^L|\Sigma_{Y^L})\,.
\eeqno
For any invertible $\Gamma$ and any $D>0$, 
we have
\beqno
& &\hat{\cal R}_{L}^{({\rm in})}(\Gamma,D|\Sigma_{Y^L})
\subseteq {\cal R}_{L}(\Gamma,D|\Sigma_{Y^L})\,.
\eeqno
\end{Th}

The inner bound $\hat{\cal R}_{L}^{({\rm in})}(D^L|\Sigma_{Y^L})$
for $\Gamma=I_L$ is well known as the inner bound of 
Berger \cite{bt} and Tung \cite{syt}. The above 
three inner bounds are variants of this inner bound. 

Optimality of $\hat{\cal R}_{2}^{({\rm in})}(D^2|\Sigma_{Y^2})$
was first studied by Oohama \cite{oh1}. Without loss 
of generality we may assume that
$$
%\beq
{\arraycolsep 1mm
\Sigma_{Y^2}=
\left[
\ba{cc}
1 &\rho \\
\rho& 1 \\
\ea
\right]\quad \rho\in [0,1)\,.
}
%\label{eqn:zzaa}
%\eeq
$$
For $i=1,2$, set
$$
{\cal R}_{i,2}(D_i|\Sigma_{Y^2})
\defeq \bigcup_{D_{3-i}>0}{\cal R}_{2}(D^2|\Sigma_{Y^2})\,.
$$
Oohama \cite{oh1} obtained the following result.
\begin{Th}[Oohama \cite{oh1}] \label{th:Oh97} 
For $i=1,2$, we have 
$$
{\cal R}_{i,2}(D_i|\Sigma_{Y^2})={\cal R}_{i,2}^{*}(D_i|\Sigma_{Y^2}),
$$
where 
$$
\ba{l} 
{\cal R}_{i,2}^{*}(D_i|\Sigma_{Y^2})\defeq
\vspace{2mm}\\
\hspace*{-1.5mm}
        \ba[t]{rl}
        \Bigl\{(R_1,R_2):
        &\,\hspace*{-1mm}R_i\geq\ts\frac{1}{2}\log^{+}
                       \left[
                       (1-\rho^2)\frac{1}{D_i}
                       \left(
                       1+\frac{\rho^2}{1-\rho^2}\cdot s
                       \right)
                       \right],
\vspace{2mm}\\
         &\,\hspace*{-1mm}R_{3-i}\geq\ts
                        \frac{1}{2}\log \left[\frac{1}{s}\right]
 \vspace{2mm}\\
         &\,\mbox{ for some }0<s\leq 1\:\Bigl.\Bigr\}\,.
\ea
\ea
$$
\end{Th}

Since ${\cal R}^{*}_{i,2}(D_i|\Sigma_{Y^2}),$ $i=1,2$ serve
as outer bounds of ${\cal R}_{2}(D^2|\Sigma_{Y^2})$, we have 
\beq
{\cal R}_{2}(D^2|\Sigma_{Y^2})
\subseteq {\cal R}_{1,2}^*(D_1|\Sigma_{Y^2})
\cap{\cal R}_{2,2}^*(D_2|\Sigma_{Y^2}).
\label{eqn:outer0}
\eeq
Wagner {\it et al.} \cite{wg3} derived the condition where
the outer bound in the right hand side of 
(\ref{eqn:outer0}) is tight. To describe their result 
set
\beqno
{\cal D}&\defeq&\{(D_1,D_2): D_1,D_2>0,
\\
& &
\ba[t]{l}
\max\{D_1,D_2\}
\leq \min\{1,\rho^2\min\{D_1,D_2\}+1-\rho^2\}\}\,.
\ea
\eeqno
Wagner {\it et al.} \cite{wg3} showed that 
if $D^2\notin {\cal D}$, 
we have 
$$
{\cal R}_{2}(D^2|\Sigma_{Y^2})
={\cal R}_{1,2}^*(D_1|\Sigma_{Y^2})
\cap{\cal R}_{2,2}^*(D_2|\Sigma_{Y^2}).
$$
%which was first pointed out by 
%Wagner {\it et al.} \cite{wg3}. 
Next we consider the case of $D^2\in {\cal D}$. 
In this case by an elementary computation 
we can show that 
$\hat{\cal R}_2^{\rm(in)}(D^2|\Sigma_{Y^2})$ 
has the following form:
\beqno
& &\hat{\cal R}_2^{\rm(in)}(D^2|\Sigma_{Y^2})
\\
&=&{\cal R}_{1,2}^{*}(D_1|\Sigma_{Y^2})
\cap{\cal R}_{2,2}^{*}(D_2|\Sigma_{Y^2})
\cap{\cal R}^{*}_{3,2}(D^2|\Sigma_{Y^2})\,,
\eeqno
where
\beqno
& &{\cal R}_{3,2}^*(D^2|\Sigma_{Y^2})
\\
&\defeq&\Bigl\{
    (R_1,R_2):\:\Bigr.
    \ba[t]{l}
         R_1+R_2 \\
         \geq\ts
           \frac{1}{2}\log
           \Bigl[
           (1-\rho^2)\frac{\beta^{*}}{2}\cdot
           \frac{1}{D_1D_2}
           \Bigr] \Bigl.\Bigr\}\,,
    \ea 
\\
& &\beta^{*}\defeq
\ts 1+\sqrt{1+\frac{4\rho^2}{(1-\rho^2)^2}\cdot D_1D_2}\,.
\eeqno
%the determination problem of 
%${\cal R}_2(D^2|\Sigma_{Y^2})$ was solved by 
%Wagner et. al \cite{wg2}. In the case of $L\geq 3$, 
%the determination problems of 
%${\cal R}_L(D|\Sigma_{Y^L})$ and 
%${\cal R}_L(D^L|\Sigma_{Y^L})$ 
%still remain open. 
The boundary of $\hat{\cal R}_2^{\rm (in)}(D^2|\Sigma_{Y^2})$ 
consists of one straight line
segment defined by the boundary of ${\cal R}_{3,2}^*(D^2|\Sigma_{Y^2})$ 
and two curved portions defined by the boundaries of 
${\cal R}_{1,2}^{*}(D_1|\Sigma_{Y^2})$ and
${\cal R}_{2,2}^{*}(D_2|\Sigma_{Y^2})$. 
Accordingly, %by Theorem$\,$2 and Proposition$\,1$ 
the inner bound established by Berger \cite{bt} and Tung \cite{syt} partially
coincides with ${\cal R}_2(D^2|\Sigma_{Y^2})$ 
at two curved portions of its boundary.
Recently, Wagner {\it et al.} \cite{wg3} have completed the proof of the 
optimality of $\hat{\cal R}_2^{\rm (in)}(D^2|\Sigma_{Y^2})$ 
by determining the sum rate part ${R}_{{\rm sum}, 2}(D^2|\Sigma_{Y^2})$. 
Their result is as follows.
\begin{Th}[Wagner {\it et al.} \cite{wg3}]
For any $D^2\in {\cal D}$, we have 
\beqno
{R}_{{\rm sum}, 2}(D^2|\Sigma_{Y^2}) 
&=&\min_{
\scs (R_1,R_2)
\atop{\scs \in \hat{\cal R}_{2}^{({\rm in})}(D^2|\Sigma_{Y^2})
}}
(R_1+R_2)
\\
&=&
 \frac{1}{2}\log
           \Bigl[
           (1-\rho^2)\frac{\beta^{*}}{2}\cdot
           \frac{1}{D_1D_2}
           \Bigr]\,.
\eeqno
\end{Th}

According to Wagner {\it et al.} \cite{wg3}, the 
results of Oohama \cite{oh2} and \cite{oh4} play an 
essential role in deriving their result. Their method for the 
proof depends heavily on the specific property of $L=2$. 
It is hard to generalize it to the case of $L\geq 3$. 

%The source coding problem considered in this paper 
%A characterization of the sum rate part of 
%${\cal R}_L(D|\Sigma_{X^LY^L})$ 
%was investigated by Pandya {\it et al.} \cite{pdya}. 
%They dealt with a case where $k=l$ and  
%$
%Y^L=AX^L+N^L\,,
%$
%where 
%$A$ is $l \times L$ a positive definite authentication matrix. 
%They derived upper and lower 
%bounds of the minimum of the sum rate 
%$\sum_{i=1}^LR_i$ for 
%$R^L\in {\cal R}_{l}(D|\Sigma_{X^LY^L})$. 
%Subsequently, Zhang and Wicker \cite{zh} studied the 
%same coding problem in the case where $A=I_L$ 
%and derived an explicit inner bound of ${\cal R}_{L}(\DisT)$. 
%In the case where $k=l$ and $A=I_L$, Oohama \cite{oh5}, \cite{oh6} 
%derived inner and outer bounds of 
%${\cal R}_{L}(D|\Sigma_{X^LY^L})$. 
%Oohama \cite{oh6} also derived explicit sufficient conditions 
%for inner and outer bounds and found examples of 
%information sources for which rate distortion region are 
%explicitly determined. 
%
%Here, we state the previous results on the above problem. 

\subsection{New Partial Solutions} 

In this subsection we state our results on 
the characterizations of 
${\cal R}_L(\DisT|\Sigma_{Y^L})$,  
${\cal R}_L(\Gamma,D^L|\Sigma_{Y^L})$, and 
${\cal R}_L(\Gamma,$ $D|\Sigma_{Y^L})$. 
Before describing those results we derive an important 
relation between remote source coding problem and 
multiterminal rate distortion problem. We first 
observe that by an elementary computation 
we have 
\beq
X^L=\tilde{A}Y^L+\tilde{N}^L\,,
\label{eqn:assad}
\eeq
where 
$\tilde{A}=(\Sigma_{X^L}^{-1}$
$+\Sigma_{N^L}^{-1})^{-1}\Sigma_{N^L}^{-1}$
and $\tilde{N}^L$ is a zero mean Gaussian random vector 
with covariance matrix 
$\Sigma_{\tilde{N}^L}$
$ 
=(\Sigma_{X^L}^{-1}$ $+\Sigma_{N^L}^{-1})^{-1}\,.
$
The random vector $\tilde{N}^L$ is independent of $Y^L$. 
Set
\beqno
B&\defeq&\tilde{A}^{-1}
  \Sigma_{\tilde{N}^L}{}^{\rm t}\tilde{A}^{-1}
 =\Sigma_{N^L}+\Sigma_{N^L}\Sigma_{X^L}^{-1}\Sigma_{N^L}\,,
\\
b^L&\defeq& {}^{\rm t}([B]_{11},[B]_{22},\cdots,[B]_{LL})\,,
\\
\tilde{B}&\defeq&\Gamma B {}^{\rm t}\Gamma\,,
\\
\tilde{b}^L&\defeq&{}^{\rm t}
([\tilde{B}]_{11},[\tilde{B}]_{22},\cdots,[\tilde{B}]_{LL})\,.
\eeqno
From (\ref{eqn:assad}), we have the following 
relation between ${\vc X}^L$ and ${\vc Y}^L$:   
\beq
{\vc X}^L=\tilde{A}{\vc Y}^L+\tilde{\vc N}^L,
\label{eqn:relation1}
\eeq
where $\tilde{\vc N}^L$ is a sequence of $n$ 
independent copies of $\tilde{N}^L$ and is 
independent of ${\vc Y}^L$. Now, we fix  
$\{(\varphi_1^{(n)},$
   $\varphi_2^{(n)}, \cdots,$ 
   $\varphi_L^{(n)},$ 
   $\psi^{(n)})\}_{n=1}^{\infty}$, arbitrary. 
For each $n=1,2,\cdots$, the estimation $\hat{\vc X}^{L}$ 
of ${\vc X}^L$ is given by    
\beqno
   \hat{\vc X}^{L}
   &=&
\left[
\ba{c}
        \psi_1^{(n)}(\varphi^{(n)}({\vc Y}^L))\\
        \psi_2^{(n)}(\varphi^{(n)}({\vc Y}^L))\\
        \vdots\\ 
        \psi_L^{(n)}(\varphi^{(n)}({\vc Y}^L))\\
\ea
\right]\,.
\eeqno
Using this estimation, we construct an estimation 
$\hat{\vc Y}^{L}$ of ${\vc Y}^L$ by 
\beq
\hat{\vc Y}^L=\tilde{A}^{-1}\hat{\vc X}^L\,,
\label{eqn:relation2}
\eeq
which is equivalent to 
\beq
\hat{\vc X}^L=\tilde{A}\hat{\vc Y}^L.
\label{eqn:relation3}
\eeq
From (\ref{eqn:relation1}) 
 and (\ref{eqn:relation3}), we have
\beq
           {\vc X}^L-\hat{\vc X}^L
=\tilde{A}({\vc Y}^L-\hat{\vc Y}^L) + \tilde{{\vc N}^L}\,.
\label{eqn:relation4}
\eeq
Since $\hat{\vc Y}^L$ is a function of ${\vc Y}^L$, 
$\hat{\vc Y}^L-{\vc Y}^L$ is independent 
of $\tilde{\vc N}^L$. Computing 
${\ts \frac{1}{n}}\Sigma_{{\lvc X}^L-\hat{\lvc X}^L}$ 
based on (\ref{eqn:relation4}), we obtain
\beq
{\ts \frac{1}{n}}\Sigma_{{\lvc X}^L-\hat{\lvc X}^L}=
\tilde{A}
\left({\ts \frac{1}{n}}\Sigma_{{\lvc Y}^L-\hat{\lvc Y}^L}
\right){}^{\rm t}\tilde{A}+\Sigma_{\tilde{N}^L}\,.
\label{eqn:zaa}
\eeq
From (\ref{eqn:zaa}), we have 
\beqa
{\ts \frac{1}{n}}\Sigma_{{\lvc Y}^L-\hat{\lvc Y}^L}
&=& \tilde{A}^{-1}
\left({\ts \frac{1}{n}}\Sigma_{{\lvc X}^L-\hat{\lvc X}^L}
-\Sigma_{\tilde{N}^L}\right){}^{\rm t}\tilde{A}^{-1}
\nonumber\\
&=&\tilde{A}^{-1}
\left({\ts \frac{1}{n}}\Sigma_{{\lvc X}^L-\hat{\lvc X}^L}\right)
{}^{\rm t}\tilde{A}^{-1}-B\,.
\label{eqn:relation5}
\eeqa
Conversely, we fix  
$\{(\varphi_1^{(n)},$
   $\varphi_2^{(n)}, \cdots,$ 
   $\varphi_L^{(n)},$ 
   $\phi^{(n)})\}_{n=1}^{\infty}$, arbitrary. 
For each $n=1,2,\cdots$, using the estimation 
$\hat{\vc Y}^{L}$ of ${\vc Y}^L$ 
given by    
\beqno
   \hat{\vc Y}^{L}
   &=&
\left[
\ba{c}
\phi_1^{(n)}(\varphi^{(n)}({\vc Y}^L))\\
\phi_2^{(n)}(\varphi^{(n)}({\vc Y}^L))\\
\vdots\\ 
\phi_L^{(n)}(\varphi^{(n)}({\vc Y}^L))\\
\ea
\right]\,,
\eeqno
we construct an estimation $\hat{\vc X}^L$ of ${\vc X}^L$ by  
(\ref{eqn:relation3}). Then 
using (\ref{eqn:relation1}) 
  and (\ref{eqn:relation3}),
we obtain (\ref{eqn:relation4}). 
Hence we have the relation (\ref{eqn:zaa}).

The following proposition provides an important strong 
connection between remote source coding problem and 
multiterminal rate distortion problem.

\begin{pro}\label{pro:MainTh1}
For any positive definite $\DisT$, we have 
\beqno
{\cal R}_{L}(\DisT|\Sigma_{Y^L})
&=&
{\cal R}_{L}(\tilde{A}(\DisT+B){}^{\rm t}\tilde{A}|\Sigma_{X^LY^L})\,.
\eeqno
For any invertible $\Gamma$ and any $D^L>0$, we have
\beqno
{\cal R}_{L}(\Gamma,D^L|\Sigma_{Y^L})
&=&{\cal R}_{L}(\Gamma\tilde{A}^{-1},D^L+\tilde{b}^L|\Sigma_{X^LY^L})\,.
\eeqno
For any invertible $\Gamma$ and any $D>0$, we have
\beqno
{\cal R}_{L}(\Gamma,D|\Sigma_{Y^L})
&=&{\cal R}_{L}(\Gamma\tilde{A}^{-1},
D+{\rm tr}[\tilde{B}]|\Sigma_{X^LY^L})\,.
\eeqno
\end{pro}

{\it Proof:}
Suppose that 
$R^L$ $\in {\cal R}_L(\tilde{A}(\DisT+B){}^{\rm t}\tilde{A}
|\Sigma_{X^LY^L})$. 
Then there exists
$\{(\varphi_1^{(n)},$
   $\varphi_2^{(n)}, \cdots,$ 
   $\varphi_L^{(n)},$ 
   $\psi^{(n)})\}_{n=1}^{\infty}$ 
such that
\beqno
& &\limsup_{n\to\infty}R^{(n)}\leq R_i, 
   \mbox{ for }i=1,2,\cdots, L\,,
\\
& &\limsup_{n\to\infty}{\ts \frac{1}{n}}
   \Sigma_{{\lvc X}^L-\hat{\lvc X}^L} \preceq 
\tilde{A}(\DisT+B){}^{\rm t}\tilde{A}\,. 
\eeqno
Using $\hat{\vc X}^L$, we construct an 
estimation $\hat{\vc Y}^L$ of ${\vc Y}^L$ by 
$\hat{\vc Y}^L$$=\tilde{A}^{-1}\hat{\vc X}^L$. 
Then from (\ref{eqn:relation5}), we have
\beqno
& &\limsup_{n\to\infty}{\ts \frac{1}{n}}\Sigma_{{\lvc Y}^L-\hat{\lvc Y}^L}
\\
&=& \limsup_{n\to\infty}
\tilde{A}^{-1}
\left({\ts \frac{1}{n}}\Sigma_{{\lvc X}^L-\hat{\lvc X}^L}\right)
{}^{\rm t}\tilde{A}^{-1}-B
\nonumber\\
&\preceq&
\tilde{A}^{-1}
\tilde{A}(\DisT+B){}^{\rm t}
\tilde{A}{}^{\rm t}\tilde{A}^{-1}-B
=\DisT\,,
\eeqno
which implies that 
$R^L\in$
${\cal R}_{L}(\tilde{A}(\DisT+B){}^{\rm t}\tilde{A}|
 \Sigma_{X^LY^L})\,.$
Thus
$$
{\cal R}_{L}(\DisT|\Sigma_{Y^L})
\supseteq 
{\cal R}_{L}(\tilde{A}(\DisT+B){}^{\rm t}\tilde{A}|\Sigma_{X^LY^L})
$$
is proved. Next we prove the reverse inclusion. Suppose 
that $R^L$ $\in {\cal R}_L(\DisT|\Sigma_{Y^L})$. 
Then there exists
$\{(\varphi_1^{(n)},$
$\varphi_2^{(n)}, \cdots,$ 
$\varphi_L^{(n)},$ 
$\phi^{(n)})\}_{n=1}^{\infty}$ 
such that
\beqno
& &\limsup_{n\to\infty}R^{(n)}\leq R_i, 
   \mbox{ for }i=1,2,\cdots, L\,,
\\
& &\limsup_{n\to\infty}{\ts \frac{1}{n}}
   \Sigma_{{\lvc Y}^L-\hat{\lvc Y}^L} \preceq \DisT\,. 
\eeqno
Using $\hat{\vc Y}^L$, 
we construct an estimation $\hat{\vc X}^L$ of ${\vc X}^L$ by 
$\hat{\vc X}^L$$=\tilde{A}\hat{\vc Y}^L$. 
Then from (\ref{eqn:zaa}), we have
\beqno
& &\limsup_{n\to\infty}{\ts \frac{1}{n}}\Sigma_{{\lvc X}^L-\hat{\lvc X}^L}
\\
&=& \limsup_{n\to\infty}
\tilde{A}
\left({\ts \frac{1}{n}}\Sigma_{{\lvc Y}^L-\hat{\lvc Y}^L}\right)
{}^{\rm t}\tilde{A}+\Sigma_{\tilde{N}^L}
\nonumber\\
&\preceq&
\tilde{A}\DisT{}^{\rm t}\tilde{A}{}^{\rm t}+\Sigma_{\tilde{N}^L}
=\tilde{A}(\DisT+B){}^{\rm t}\tilde{A}{}^{\rm t}\,,
\eeqno
which implies that
$R^L\in$ 
${\cal R}_{L}(\tilde{A}(\DisT+B){}^{\rm t}\tilde{A}$ 
$|\Sigma_{X^LY^L})\,.$
Thus,
$$
{\cal R}_{L}(\DisT|\Sigma_{Y^L})
\subseteq 
{\cal R}_{L}(\tilde{A}(\DisT+B){}^{\rm t}\tilde{A}|\Sigma_{X^LY^L})
$$
is proved. Next we prove the second equality. We have the following
chain of equalities:
%${\cal R}_{L}(\Gamma,D^K|\Sigma_{Y^L})$
%${\cal R}_{L}(\tilde{A}^{-1},D^L+b^L|\Sigma_{X^LY^L})$
\beqa
& &{\cal R}_{L}(\Gamma,D^L|\Sigma_{Y^L})
\nonumber\\
&=&
\bigcup_{\Gamma\DisT{}^{\rm t}\Gamma \in {\cal S}_L(D^L)}
{\cal R}_L({\DisT}|\Sigma_{Y^L})
\nonumber\\
&=&
\bigcup_{\Gamma\DisT{}^{\rm t}\Gamma\in {\cal S}_L(D^L)}
{\cal R}_L(\Gamma\tilde{A}(\DisT+B){}^{\rm t}\tilde{A}|\Sigma_{X^L Y^L})
\nonumber\\
&=&
\bigcup_{
\scs
\Gamma
\tilde{A}^{-1}
\tilde{A}(\DisT+B){}^{\rm t}\tilde{A}{}^{\rm t}\tilde{A}^{-1}
{}^{\rm t}\Gamma
\atop{\scs -\Gamma B {}^{\rm t}\Gamma \in {\cal S}_L(D^L)}}
{\cal R}_L(\tilde{A}(\DisT+B){}^{\rm t}\tilde{A}|\Sigma_{X^LY^L})
\nonumber\\
&=&
\bigcup_{
\scs
\Gamma\tilde{A}^{-1}\tilde{A}(\DisT+B){}^{\rm t}\tilde{A}
{}^{\rm t}(\Gamma\tilde{A}^{-1})
\atop{\scs \in {\cal S}_L(D^L+\tilde{b}^L)}}
{\cal R}_L(\tilde{A}(\DisT+B){}^{\rm t}\tilde{A}|\Sigma_{X^L Y^L})
\nonumber\\
&=&
\bigcup_{
\scs \hat{\Sigma}_d=\tilde{A}(\DisT+B){}^{\rm t}\tilde{A}
\succ \Sigma_{X^L|Y^L}, 
\atop{\scs
\Gamma\tilde{A}^{-1}\hat{\Sigma}_d
{}^{\rm t}(\Gamma\tilde{A}^{-1})\in {\cal S}_L(D^L+\tilde{b}^L)}}
{\cal R}_L(\hat{\Sigma}_d|\Sigma_{X^L Y^L})
\nonumber\\
&=&{\cal R}_{L}(\Gamma\tilde{A}^{-1},D^L+\tilde{b}^L|\Sigma_{X^LY^L})\,.
\nonumber
\eeqa
Thus the second equality is proved. 
Finally we prove the third equality. 
We have the following
chain of equalities:
\beqa
& &{\cal R}_{L}(\Gamma,D|\Sigma_{Y^L})
\nonumber\\
&=&
\bigcup_{{\rm tr}[\Gamma\DisT{}^{\rm t}\Gamma]\leq D}
{\cal R}_L({\DisT}|\Sigma_{Y^L})
\nonumber\\
&=&
\bigcup_{{\rm tr}[\Gamma\DisT{}^{\rm t}\Gamma]\leq D}
{\cal R}_L(\Gamma\tilde{A}(\DisT+B){}^{\rm t}\tilde{A}|\Sigma_{X^LY^L})
\nonumber\\
&=&
\bigcup_{
\scs
{\rm tr}[\Gamma\tilde{A}^{-1}
\tilde{A}(\DisT+B){}^{\rm t}\tilde{A}{}^{\rm t}
\tilde{A}^{-1}{}^{\rm t}\Gamma]
\atop{\scs -
{\rm tr}[\Gamma B {}^{\rm t}\Gamma]\leq D}}
{\cal R}_L(\tilde{A}(\DisT+B){}^{\rm t}\tilde{A}|\Sigma_{X^LY^L})
\nonumber\\
&=&
\bigcup_{
\scs
{\rm tr}[
\Gamma\tilde{A}^{-1}\tilde{A}(\DisT+B){}^{\rm t}\tilde{A}
{}^{\rm t}(\Gamma\tilde{A}^{-1})]
\atop{\scs \leq D+{\rm tr}[\tilde{B}]}}
{\cal R}_L(\tilde{A}(\DisT+B){}^{\rm t}\tilde{A}|\Sigma_{X^LY^L})
\nonumber\\
&=&
\bigcup_{
\scs \hat{\Sigma}_d=\tilde{A}(\Sigma_d +B){}^{\rm t}\tilde{A} 
\succ \Sigma_{X^L|Y^L} ,
\atop{\scs
{\rm tr}[\Gamma\tilde{A}^{-1}
\hat{\Sigma}_d{}^{\rm t}(\Gamma \tilde{A}^{-1})]
\leq D+{\rm tr}[\tilde{B}]}}
{\cal R}_L(\hat{\Sigma}_d|\Sigma_{X^LY^L})
\nonumber\\
&=&  {\cal R}_{L}(\Gamma\tilde{A}^{-1},
   D+{\rm tr}[\tilde{B}]|\Sigma_{X^LY^L})\,.
\nonumber
\eeqa
Thus the third equality is proved. 
\hfill\IEEEQED

Proposition \ref{pro:MainTh1} implies that all results on the 
rate distortion regions for the remote source coding 
problems can be converted into those on the multiterminal 
source coding problems. 
In the following we derive inner and outer bounds of 
${\cal R}_{L}(\DisT|\Sigma_{Y^L})$, 
${\cal R}_{L}(\Gamma,D^L|\Sigma_{Y^L})$, and 
${\cal R}_{L}(\Gamma,D|\Sigma_{Y^L})$ 
using Proposition \ref{pro:MainTh1}. 

We first derive inner and outer bounds of 
${\cal R}_{L}(\DisT|\Sigma_{Y^L})$.
For $r_i\geq 0, i\in \Lambda$, let $V_{i}(r_i),$ $i\in \Lambda$ 
be $L$ independent Gaussian random variables with 
mean 0 and variance 
$\sigma_{N_i}^2/({\baseN}^{2r_i}-1)$. 
Let $\Sigma_{V^L(r^L)}$ be a covariance matrix for the random 
vector $V^L(r^L)$. Fix nonnegative vector $r^L$. For $\theta >0$ 
and for $S \subseteq  \Lambda$, define
\beqno
\Sigma_{V_S(r_{S^{\rm c}})}^{-1} 
&\defeq& 
\left. \Sigma_{V^L(r^L)}^{-1} \right|_{r_{S}={\lvc 0}} \,,
\\
\underline{\tilde{J}}_{S}(\theta, r_S|r_{\coS})
&\defeq &\frac{1}{2}\log^{+}
   \left[\ts 
   \frac{\ds |\Sigma_{Y^L}+B|\prod_{i=1}^L {\baseN}^{2r_i} }
        {\ds \theta |\Sigma_{Y^L}|\left|\Sigma_{Y^L}^{-1}
                 +\Sigma_{V_{S^{\rm c}}(r_{S^{\rm c}})}^{-1}\right|
        }
  \right],
\\
\tilde{J}_{S}\left(r_S|r_{\coS}\right)
&\defeq &\frac{1}{2}\log
   \left[\ts 
   \frac{\ds \left|\Sigma_{Y^L}^{-1}
    +\Sigma_{V^L(r^L)}^{-1}\right|
        }
        {\ds  \left|\Sigma_{Y^L}^{-1}
        +\Sigma_{V_{S^{\rm c}}(r_{S^{\rm c}})}^{-1}\right|
        }
  \right]\,.
\eeqno
Set
$$
\tilde{\cal A}_L(\DisT)
\defeq 
\left\{ r^L\geq 0:
\left[\Sigma_{Y^L}^{-1}+
\Sigma_{V^L(r^L)}^{-1}\right]^{-1}
\preceq \DisT\right\}\,.
$$
Define four regions by 
\beqno
%& &
{\cal R}_L^{({\rm out})}(\theta,r^L|\Sigma_{Y^L})
%\\
&\defeq&
\ba[t]{l}
  \left\{R^L \right. : 
  \ba[t]{l}
  \ds \sum_{i \in S} R_i 
  \geq \underline{\tilde{J}}_{S}\left(\theta,r_S|r_{\coS}\right)
  \vspace{1mm}\\
  \mbox{ for any }S \subseteq \Lambda\,. 
  \left. \right\}\,,
  \ea
\ea
\nonumber\\
{\cal R}_{L}^{({\rm out})}(\DisT|\Sigma_{Y^L})
&\defeq& 
\bigcup_{r^L \in \tilde{\cal A}_L(\DisT)}
{\cal R}_L^{({\rm out})}(|\DisT+B|,r^L|\Sigma_{Y^L})\,, 
\nonumber\\
%& &
{\cal R}_L^{({\rm in})}(r^L|\Sigma_{Y^L})
%\\
&\defeq&
\ba[t]{l}
  \left\{R^L \right.: 
  \ba[t]{l}
  \ds \sum_{i \in S} R_i 
  \geq {J}_{S}\left(r_S|r_{\coS}\right)
  \vspace{1mm}\\
  \mbox{ for any }S \subseteq \Lambda\,. 
  \left. \right\}\,,
  \ea
\ea
\nonumber\\
{\cal R}_L^{({\rm in})}(\DisT|\Sigma_{Y^L})
&\defeq&{\rm conv}
        \left\{
        \bigcup_{r^L \in {\cal A}_L(\DisT)}
        {\cal R}_L^{({\rm in})}(r^L|\Sigma_{Y^L})
        \right\}\,. 
\eeqno
The functions and sets defined above have 
properties shown in the following.   
\begin{pr}{
\label{pr:prz01zs}
$\quad$
\begin{itemize}
\item[{\rm a)}]$\:$For any positive definite 
$\DisT$,  
$\tilde{\cal G}(\DisT)=
 {\cal G}(\tilde{A}(\DisT+B){}^{\rm t}\tilde{A})$.
\vspace*{1mm}
\item[{\rm b)}]$\:$For any positive definite 
$\DisT$, we have   
$$
\hat{\cal R}_{L}^{\rm (in)}(\DisT|\Sigma_{Y^L})
=\hat{\cal R}_{L}^{\rm (in)}
(\tilde{A}(\DisT+B){}^{\rm t}\tilde{A}
|\Sigma_{X^LY^L})\,.
$$
\item[{\rm c)}]$\:$For any positive definite $\DisT$ 
and any $S\subseteq \Lambda$, we have
\beqno
& &
\underline{\tilde{J}}_S(|\DisT+B|,r_S|r_{\coS})
=\underline{J}_S(|\tilde{A}(\DisT+B){}^{\rm t}\tilde{A}|,r_S|r_{\coS}),
\\
& &\tilde{J}_S(r_S|r_{\coS})=J_S(r_S|r_{\coS}).
\eeqno
\item[{\rm d)}]$\:$For any positive definite $\DisT$, 
$
\tilde{\cal A}_L(\DisT)
={\cal A}_L(\tilde{A}(\DisT+B){}^{\rm t}\tilde{A})
\,.$
\vspace*{1mm}
\item[{\rm e)}]$\:$For any positive definite $\DisT$, we have 
$$
\ba[t]{l}
{\cal R}_{L}^{\rm (out)}(\DisT|\Sigma_{Y^L})
={\cal R}_{L}^{\rm (out)}
   (\tilde{A}(\DisT+B){}^{\rm t}\tilde{A}|\Sigma_{X^LY^L})\,,
\vspace*{1mm}\\
{\cal R}_{L}^{\rm (in)}(\DisT|\Sigma_{Y^L})
={\cal R}_{L}^{\rm (in)}
(\tilde{A}(\DisT+B){}^{\rm t}\tilde{A}|\Sigma_{X^LY^L})\,.
\ea
$$
\end{itemize}
}\end{pr}

From Theorem \ref{th:conv2}, Proposition \ref{pro:MainTh1} 
and Property \ref{pr:prz01zs}, 
we have the following.
\begin{Th}\label{tho:MainTh2a}$\:$For any positive 
definite $\DisT$, we have
\beqno
& &{\cal R}_{L}^{({\rm in})}(\DisT|\Sigma_{Y^L})
\subseteq \hat{\cal R}_{L}^{({\rm in})}(\DisT|\Sigma_{Y^L}) 
\\
&\subseteq & 
{\cal R}_{L}(\DisT|\Sigma_{Y^L}) 
\subseteq {\cal R}_{L}^{({\rm out})}(\DisT|\Sigma_{Y^L})\,. 
\eeqno
\end{Th}

Next, we derive inner and outer bounds of 
${\cal R}_{L}(\Gamma,$$D^K|\Sigma_{Y^L})$
and 
${\cal R}_{L}(\Gamma,$$D|\Sigma_{Y^L})$.
Set 
\beqno
\tilde{\cal A}_L(r^L)
&\defeq &\{\DisT:
    \DisT 
    \succeq (\Sigma_{Y^L}^{-1}
+\Sigma_{V^L(r^L)}^{-1})^{-1}\}\,,
\\
\tilde{\theta}(\Gamma,D^L,r^L)
&\defeq &  
\max_{\scs \DisT:\DisT \in \tilde{\cal A}_L(r^L), 
      \atop{\scs
      \Gamma\DisT {}^{\rm t}\Gamma \in{\cal S}_L(D^L)}
     }
\left|\DisT+B\right|\,,
\\
\tilde{\theta}(\Gamma,D,r^L)
&\defeq & 
\max_{\scs \DisT: \DisT \in \tilde{\cal A}_L({r^L}),
      \atop{\scs 
      {\rm tr}[\Gamma\DisT {}^{\rm t}\Gamma]\leq D}
      }
\left|\DisT+B\right|\,.
\eeqno
Furthermore, set 
\beqno
& &\tilde{\cal B}_L(\Gamma,D^L)
\\
&\defeq&
\left\{r^L\geq 0:
\Gamma
(\Sigma_{Y^L}^{-1}+\Sigma_{V^L(r^L)}^{-1})^{-1}
{}^{\rm t}\Gamma
\in{\cal S}_L(D^L)
\right\}\,,
\\
& &\tilde{\cal B}_L(\Gamma,D)
\\
&\defeq&
\left\{r^L\geq 0:
{\rm tr}\left[
\Gamma
(\Sigma_{Y^L}^{-1}+\Sigma_{V^L(r^L)}^{-1})^{-1}
{}^{\rm t}\Gamma
\right]
\leq D 
\right\}\,.
\eeqno
Define four regions by
\beqno
& &{\cal R}_{L}^{({\rm out})}(\Gamma,D^L|\Sigma_{Y^L})
\\
&\defeq& 
\bigcup_{r^L \in \tilde{\cal B}_L(\Gamma,D^L)}
{\cal R}_L^{({\rm out})}(\tilde{\theta}(\Gamma,D^L,r^L),r^L|\Sigma_{Y^L})\,, 
\\
& &{\cal R}_L^{({\rm in})}(\Gamma,D^L|\Sigma_{Y^L})
\\
&\defeq&
        \conv\left\{\bigcup_{r^L \in \tilde{\cal B}_L(\Gamma,D^L)}
        {\cal R}_L^{({\rm in})}(r^L|\Sigma_{Y^L})\right\}\,,
\\
& &{\cal R}_{L}^{({\rm out})}(\Gamma,D|\Sigma_{Y^L})
\\
&\defeq& 
\bigcup_{r^L \in \tilde{\cal B}_L(\Gamma,D)}
{\cal R}_L^{({\rm out})}(\tilde{\theta}(\Gamma,D,r^L),r^L|\Sigma_{Y^L})\,, 
\\
& &{\cal R}_L^{({\rm in})}(\Gamma,D|\Sigma_{Y^L})
\\
&\defeq&\conv\left\{
        \bigcup_{r^L \in \tilde{\cal B}_L(\Gamma,D)}
        {\cal R}_L^{({\rm in})}(r^L|\Sigma_{Y^L})\right\}\,.
\eeqno
It can easily be verified that the functions and sets defined 
above have the properties shown in the following.
\begin{pr}{
\label{pr:PropZ}
$\quad$
\begin{itemize}
\item[{\rm a)}]$\:$For any invertible $\Gamma$ and any $D^L>0$, 
we have  
\beqno
& & \hat{\cal R}_{L}^{\rm (in)}(\Gamma,D^L|\Sigma_{Y^L})
\\
&=&\hat{\cal R}_{L}^{\rm (in)}(\Gamma\tilde{A}^{-1},
    D^L+\tilde{b}^L|\Sigma_{X^LY^L})\,.
\eeqno
For any invertible $\Gamma$ and any $D>0$, we have    
\beqno
& & \hat{\cal R}_{L}^{\rm (in)}(\Gamma,D|\Sigma_{Y^L})
\\
&=&\hat{\cal R}_{L}^{\rm (in)}(\Gamma\tilde{A}^{-1},
    D+{\rm tr}[\tilde{B}]|\Sigma_{X^LY^L})\,.
\eeqno
\item[{\rm b)}] For any $r^L\geq 0$, we have   
\beqno
& &\DisT\in\tilde{\cal A}(r^L)
\Leftrightarrow \tilde{A}(\DisT+B){}^{\rm t}\tilde{A}
\in{\cal A}(r^L)\,,
\\
& &\tilde{\theta}(\Gamma,D^L,r^L)
=\left|\tilde{A}\right|^{-2}\theta(\Gamma\tilde{A}^{-1},D^L,r^L)\,, 
\\
& &\tilde{\theta}(\Gamma,D,r^L)
=\left|\tilde{A}\right|^{-2}\theta(\Gamma\tilde{A}^{-1},D,r^L)\,.
\eeqno

\item[{\rm c)}]$\:$For any invertible $\Gamma$ and 
any $D^L>0$, we have    
\beqno
& &{\cal R}_{L}^{\rm (out)}(\Gamma,D^L|\Sigma_{Y^L})
\\
&=&{\cal R}_{L}^{\rm (out)}(\Gamma\tilde{A}^{-1},
    D^L+\tilde{b}^L|\Sigma_{X^LY^L})\,,
\\
& &{\cal R}_{L}^{\rm (in)}(\Gamma,D^L|\Sigma_{Y^L})
\\
&=&{\cal R}_{L}^{\rm (in)}(\Gamma\tilde{A}^{-1},
    D^L+\tilde{b}^L|\Sigma_{X^LY^L})\,.
\eeqno
For any invertible $\Gamma$ and any $D>0$, we have   
\beqno
& &{\cal R}_{L}^{\rm (out)}(\Gamma,D|\Sigma_{Y^L})
\\
&=&{\cal R}_{L}^{\rm (out)}(\Gamma\tilde{A}^{-1},
    D+{\rm tr}[\tilde{B}]|\Sigma_{X^LY^L})\,,
\\
& &{\cal R}_{L}^{\rm (in)}(\Gamma,D|\Sigma_{Y^L})
\\
&=&{\cal R}_{L}^{\rm (in)}(\Gamma\tilde{A}^{-1},
    D+{\rm tr}[\tilde{B}]|\Sigma_{X^LY^L})\,.
\eeqno
\end{itemize}
}\end{pr}

From Corollary \ref{co:conv2z}, Proposition \ref{pro:MainTh1} 
and Property \ref{pr:PropZ}, we have the following theorem. 
\begin{Th}\label{th:conv4vv}$\:$For any invertible 
$\Gamma$ and any $D>0$, we have       
\beqno
&         &    {\cal R}_{L}^{({\rm in})}(\Gamma,D^L|\Sigma_{Y^L})
\subseteq  \hat{\cal R}_{L}^{({\rm in})}(\Gamma,D^L|\Sigma_{Y^L}) 
\\
&\subseteq&    {\cal R}_{L}(\Gamma,D^L|\Sigma_{Y^L}) 
\subseteq      {\cal R}_{L}^{({\rm out})}(\Gamma,D^L|\Sigma_{Y^L})\,. 
\eeqno
For any invertible $\Gamma$ and any $D>0$, we have    
\beqno
 &        &    {\cal R}_{L}^{({\rm in})}(\Gamma,D|\Sigma_{Y^L})
 \subseteq \hat{\cal R}_{L}^{({\rm in})}(\Gamma,D|\Sigma_{Y^L}) 
\\
&\subseteq&    {\cal R}_{L}(\Gamma,D|\Sigma_{Y^L}) 
\subseteq      {\cal R}_{L}^{({\rm out})}(\Gamma,D|\Sigma_{Y^L})\,. 
\eeqno
\end{Th}

Next, we derive a matching condition for 
${\cal R}_{L}^{({\rm out})}(\Gamma,D|\Sigma_{Y^L})$ to coincide
with ${\cal R}_{L}^{\rm (in)}(\Gamma,D|\Sigma_{Y^L})$. 
By Theorems \ref{th:matchTh} and \ref{th:conv4vv}, 
Proposition \ref{pro:MainTh1} and Property \ref{pr:PropZ},
we establish the following. 
\begin{Th}
\label{th:matchThZ}
Let $\mu_{\min}^{\ast}$ be the minimum eigen value of
$$
\tilde{B}
=\Gamma\left(
\Sigma_{N^L}+\Sigma_{N^L}\Sigma_{X^L}^{-1}\Sigma_{N^L}
\right){}^{\rm t}\Gamma\,.
$$
If we have 
\beqno
0<D&\leq (L+1)\mu_{\min}^{\ast}-
{\rm tr}\left[
\Gamma(\Sigma_{N^L}+\Sigma_{N^L}\Sigma_{X^L}^{-1}\Sigma_{N^L})
{}^{\rm t}\Gamma
\right]\,,
\eeqno
then 
\beqno
& &{\cal R}_L^{({\rm in})}(\Gamma,D|\Sigma_{Y^L})
=\hat{\cal R}_L^{({\rm in})}(\Gamma,D|\Sigma_{Y^L})
\\
&=&{\cal R}_L(\Gamma,D|\Sigma_{Y^L})
 ={\cal R}_L^{({\rm out})}(\Gamma,D|\Sigma_{Y^L})\,.
\eeqno
\end{Th} 

We are particularly interested in the case 
where $\Gamma$ is the following diagonal matrix:
\beq
\Gamma=
\left(
\begin{array}{cccc}
\gamma_1 &           &        & \mbox{\huge 0}\\
          & \gamma_2 &        &          \\
          &           & \ddots &          \\
\mbox{\huge 0} &      &        & \gamma_{L}\\
\end{array}
\right)
\,,\quad \sum_{i=1}^L\gamma_i^{-2}=1\,.
\label{eqn:diagZ}
\eeq
Let $\delta>0$ be an arbitrary positive constant 
specified later. 
We choose $\Sigma_{N^L}$ so that 
$\Sigma_{N^L}=\delta^2\Gamma^2$. Set
$\tilde{\Sigma}_{X^L}\defeq$ 
$\Gamma{\Sigma}_{X^L}\Gamma$. 
Then, we have
$
\tilde{B}=\delta I_L + \delta^2 \tilde{\Sigma}^{-1}_{X^L}\,.
$
Hence we have
\beq
\mu_{\min}^*\geq \delta\,.
\label{eqn:ssa99}
\eeq
Let $\lambda_{\min}$ 
be the minimum eigen value of ${\Sigma}_{X^L}$. Since 
${\Sigma}_{X^L} \succ \lambda_{\min}I_L$, we have  
$
\tilde{\Sigma}_{X^L}^{-1} \prec \lambda_{\min}^{-1}\Gamma^{-2}.  
$
Hence we have
\beq
{\rm tr}\left[\tilde{\Sigma}_{X^L}^{-1} \right]
\leq \lambda_{\min}^{-1}{\rm tr}
\left[\Gamma^{-2}\right]=\lambda_{\min}^{-1},  
\label{eqn:ssa100}
\eeq
where the last equality %of (\ref{eqn:ssa}) 
follows from the choice of $\Gamma$ specified 
with (\ref{eqn:diagZ}). From (\ref{eqn:ssa99}) 
and (\ref{eqn:ssa100}), we have
\beqa
(L+1)\mu_{\min}^{*}-{\rm tr}[\tilde{B}]
&\geq & (L+1)\delta-{\rm tr}\left[
\delta I_L +\delta^2 \tilde{\Sigma}_{X^L}^{-1}
\right]
\nonumber\\
&\geq& (L+1)\delta- L\delta-\delta^2\lambda_{\min}^{-1}
\nonumber\\
&=&\delta-\delta^2\lambda_{\min}^{-1}\,.
\eeqa
Hence if 
\beq
0<D\leq \delta-\delta^2\lambda_{\min}^{-1}\,,
\label{eqn:ssa101} 
\eeq
then the matching condition holds. The right member 
of (\ref{eqn:ssa101}) takes the maximum value 
$\frac{1}{4}\lambda_{\min}$ for $\delta=\frac{1}{2}\lambda_{\min}$.   
Summarizing the above argument, we establish 
the following corollary from Theorem \ref{th:matchThZ}.
\begin{co}
If the minimum eigen value $\lambda_{\min}$ of $\Sigma_{X^L}$ satisfies
$$
0<D\leq \frac{1}{4}\lambda_{\min},
$$
then for any diagonal matrix $\Gamma$ 
specified with (\ref{eqn:diagZ}) we have
\beqno
& &{\cal R}_L^{({\rm in})}(\Gamma,D|\Sigma_{Y^L})
=\hat{\cal R}_L^{({\rm in})}(\Gamma,D|\Sigma_{Y^L})
\\
&=&{\cal R}_L(\Gamma,D|\Sigma_{Y^L})
 ={\cal R}_L^{({\rm out})}(\Gamma,D|\Sigma_{Y^L})\,.
\eeqno

\end{co}
 
\subsection{Sum Rate Characterization 
for the Cyclic Shift Invariant Source}

In this subsection we further examine an explicit characterization 
of $R_{{\rm sum},L}($ $D|\Sigma_{Y^l})$ when the source has 
a certain symmetrical property. Let %$\tau$ be 
\beqno
\tau&=&
\left(
\ba{cccccc}
    1 &2&\cdots&      i &\cdots&    L\\ 
\tau(1)&\tau(2)&\cdots&\tau(i)&\cdots&\tau(L)
\ea
\right)
\eeqno
be a cyclic shift on 
$\Lambda$, that is,
$$
\tau(1)=2,\tau(2)=3,\cdots,\tau(L-1)=L,\tau(L)=1\,.
$$
Let
$
p_{X_{\Lambda}}(x_{\Lambda})
=p_{X_1X_2\cdots X_L}(x_1,x_2,\cdots,x_L)
$     
be a probability density function of $X^L$. 
The source ${X^L}$ is said to be cyclic 
shift invariant if we have
\beqno
p_{X_{\Lambda}}(x_{\tau(\Lambda)})
&=&p_{X_1X_2\cdots X_L}(x_2,x_3,\cdots,x_L,x_1)
\\
&=&p_{X_1X_2\cdots X_L}(x_1,x_2,\cdots,x_{L-1},x_{L})
\eeqno 
for any $(x_1,x_2,$ $\cdots,x_L)$ $\in {\cal X}^L$. 
In the following argument we assume that $X^L$ satisfies 
the cyclic shift invariant property. 
We further assume that $N_i, i\in \Lambda$ are 
%independent identically distributed 
i.i.d. Gaussian 
random variables with mean 0 and variance $\epsilon$.
Then, the observation $Y^L=X^L+N^L$ also satisfies 
the cyclic shift invariant property. 
We assume that the covariance matrix $\Sigma_{N^L}$ of $N^L$ is 
given by $\epsilon I_L$. Then $\tilde{A}$ and $B$ are given by
\beqno  
\tilde{A}&=&%\frac{1}{\epsilon}
\left(\epsilon\Sigma_{X^L}^{-1}+I_L\right)^{-1},\:
B=\epsilon\left(I_L+\epsilon\Sigma_{X^L}^{-1}\right)\,.
\eeqno
Fix $r>0$, let $N_{i}(r),$ $i\in \Lambda$ be $L$ 
i.i.d. Gaussian random variables with mean 0 and 
variance $\epsilon/(1-{\baseN}^{-2r})$. The covariance 
matrix $\Sigma_{N^L(r)}$ for the random 
vector $N^L(r)$ is given by 
$$
\Sigma_{N^L(r)}=\frac{1-{\baseN}^{-2r}}{\epsilon}I_L\,. 
$$
Let $\lambda_i, i\in\Lambda$ be 
$L$ eigen values of the matrix $\Sigma_{X^L}$ and let 
$\beta_i=\beta_i(r), i\in\Lambda$ 
be $L$ eigen values of the matrix 
$$
{}^{\rm t}\tilde{A}\left(\Sigma_{X^L}^{-1}+
\frac{1-{\baseN}^{-2r}}{\epsilon}I_L 
\right)\tilde{A}.
$$
Using the eigen values of $\Sigma_{X^L}$, 
$\beta_i(r), i\in\Lambda$ 
can be written as
$$
\beta_i(r)=\frac{1}{\epsilon}
\left[
\frac{\lambda_i}{\lambda_i+\epsilon}-
\left(\frac{\lambda_i}{\lambda_i+\epsilon}\right)^2{\rm e}^{-2r}
\right]\,.
$$
Let $\xi$ be a nonnegative number that satisfies 
$$
\sum_{i=1}^L\{[\xi-\beta_i^{-1}]^{+}+\beta_i^{-1}\}
=D+{\rm tr}[B]\,.
$$ 
Define  
\beqno
\tilde{\omega}(D,r)
&\defeq&
\prod_{i=1}^L\left\{[\xi-\beta_i^{-1} ]^{+}
+\beta_i^{-1}\right\}.
\eeqno
%$$
%\tilde{\omega}(D,r)=
%|\tilde{A}^{-1}|^2\omega(\tilde{A}^{-1},
%D+{\rm tr}[B],\underbrace{r\cdots r}_{L})\,. 
%$$
The function $\tilde{\omega}(D,r)$ has 
an expression of the so-called water filling 
solution to the following optimization problem:
\beqa
\tilde{\omega}(D,r)
=%|\Gamma|^{-2}
  \max_{\scs \xi_i\beta_{i}\geq 1,i\in\Lambda\,, 
  \atop{\scs
          \sum_{i=1}^L\xi_i\leq D+{\rm tr}[B]
       }   
      }\prod_{i=1}^L\xi_{i}\,. 
\eeqa
%Relation between $\tilde{\omega}(D,r)$ and {\omega}(D,r^L)
%re as follows.
Set
\beqno
\underline{\tilde{J}}(D,r)
&\defeq &\frac{1}{2}\log
   \left[\ts 
   \frac{\ds {\baseN}^{2Lr}\left|\Sigma_{Y^L}+B\right|}
        {\ds \tilde{\omega}(D,r)}
  \right]\,,
\\
\zeta(r)&\defeq&{\rm tr}
\left[\tilde{A}^{-1}\left(\Sigma_{X^L}^{-1}
+\frac{1-{\rm e}^{-2r}}{\epsilon}I_L
\right)^{-1}{}^{\rm t}\tilde{A}^{-1}\right]\,.
\eeqno
By definition we have
\beq
\zeta(r)  
=\sum_{i=1}^{L}\frac{1}{\beta_i(r)}\,.
\label{eqn:zeta1}
\eeq
Since $\zeta(r)$ is a monotone decreasing function 
of $r$, there exists a unique $r$ such that 
$\zeta(r)=D+{\rm tr}[B]$, we denote it by 
$r^{\ast}(D+$ ${\rm tr}[B])$.  
Note that
\beqno
& &(\underbrace{r,r,\cdots,r}_L) 
\in {\cal B}_L(\tilde{A}^{-1},D+{\rm tr}[B])
\\
&\Leftrightarrow &\zeta(r)\leq D+{\rm tr}[B]
 \Leftrightarrow r\geq r^{\ast}(D+{\rm tr}[B])\,,
\\
& &
\tilde{\omega}(D,r^\ast)
=|\tilde{A}|^{-2}
\left|\Sigma_{X^L}^{-1}
 +\frac{1-{\rm e}^{-2r^{\ast}}}{\epsilon}I_L\right|^{-1}\,.
\eeqno
Set 
$$
R_{{\rm sum},L}^{(\rm l)}(D|\Sigma_{Y^L})\defeq 
\min_{r\geq r^*(D+{\rm tr}[B])}\underline{\tilde{J}}(D,r)\,.
$$
Then, we have the following. 
\begin{Th}\label{Th:sr0} Assume that the source 
$X^L$ and its noisy version $Y^L=X^L+N^L$ are 
cyclic shift invariant. Then, we have    
$$
R_{{\rm sum},L}(D|\Sigma_{Y^L})
\geq R_{{\rm sum},L}^{(\rm l)}(D|\Sigma_{Y^L})\,.
$$
\end{Th}

Proof of this theorem will be stated in Section V.

Next, we examine a sufficient condition for 
$R_{{\rm sum},L}^{(\rm l)}(D$ $|\Sigma_{Y^L})$ 
to coincide with $R_{{\rm sum},L}($ $D|\Sigma_{Y^L})$. 
It is obvious from 
the definition of $\underline{\tilde{J}}(D,r)$ 
that when ${\rm e}^{-2Lr}\tilde{\omega}(D,r)$ 
is a monotone decreasing function of 
$r\in [r^*($ $D+{\rm tr}[B]),+\infty)$, 
we have 
$R_{{\rm sum},L}^{(\rm l)}(D|\Sigma_{Y^L})$ 
$=R_{{\rm sum},L}(D|\Sigma_{Y^L})$. 
Let $\lambda$ be the maximum eigen value of $\Sigma_{X^L}$. 
Set
\beqno
\beta_{i_0}(r)\defeq \min_{1\leq i\leq L}\beta_{i}(r)\,,
\beta_{i_1}(r)\defeq \max_{1\leq i\leq L}\beta_{i}(r)\,.
\eeqno
Then we have the following two lemmas.
\begin{lm}\label{lm:pro3b}
If  
$$
{\beta_{i_1}(r)}-{\beta_{i_0}(r)}
\leq \epsilon{\rm e}^{2r}\cdot\frac{L}{L-1}
\left(\frac{\lambda_{\max}+\epsilon}{\lambda_{\max}}\right)^2
(\beta_{i_0}(r))^2
$$
or equivalent to
\beqa
& &
  \left(\frac{\lambda_{i_1}}{\lambda_{i_1}+\epsilon}
 -\frac{\lambda_{i_0}}{\lambda_{i_0}+\epsilon}\right)
%\\
%& &\times
\left[{\rm e}^{2r}-\left(
\frac{\lambda_{i_0}}{\lambda_{i_0}+\epsilon}+
\frac{\lambda_{i_1}}{\lambda_{i_1}+\epsilon}
\right)
\right]
\nonumber\\
&\leq& 
\frac{L}{L-1}
\left(\frac{\lambda_{\max}+ \epsilon}{\lambda_{\max}}\right)^2
\left(\frac{\lambda_{i_0}}{\lambda_{i_0}+\epsilon}\right)^2
\left(
{\rm e}^{2r}-\frac{\lambda_{i_0}}{\lambda_{i_0}+\epsilon}
\right)^2
\label{eqn:lemCond}
\eeqa
holds for $r\geq r^*(D+{\rm tr}[B])$, then 
${\rm e}^{-2Lr}\tilde{\omega}(D,r)$ 
is a monotone decreasing function of 
$r\in [r^*(D+{\rm tr}[B]),\infty)$.
\end{lm}

\begin{lm}\label{lm:pro3c} If we have
\beqa
& &
  \frac{\lambda_{i_1}}{\lambda_{i_1}+\epsilon}
 -\frac{\lambda_{i_0}}{\lambda_{i_0}+\epsilon}
\nonumber\\
&\leq& 
\frac{4L}{L-1}
\left(\frac{\lambda_{\max}+\epsilon}{\lambda_{\max}}\right)^2
\left(\frac{\lambda_{i_0}}{\lambda_{i_0}+\epsilon}\right)^2
\left(\frac{\lambda_{i_1}}{\lambda_{i_1}+\epsilon}\right),
\label{eqn:aasss00}
\eeqa
then the sufficient condition (\ref{eqn:lemCond}) 
in Lemma \ref{lm:pro3b} holds for any nonnegative $r$.
\end{lm}

%From Lemma \ref{lm:pro3b} and an elementary 
%computation we obtain the following. 
Proofs of Lemmas \ref{lm:pro3b} and \ref{lm:pro3c} 
will be given in Section V. 
If we take $\epsilon$ sufficiently small in 
(\ref{eqn:aasss00}) in Lemma \ref{lm:pro3c}, 
then the left hand side of this inequality %(\ref{eqn:aasss00}) 
becomes close to zero. On the other hand,
the right hand side of (\ref{eqn:aasss00}) 
becomes close to $\frac{4L}{L-1}$. Hence if 
we choose $\epsilon$ sufficiently small, then     
the inequality (\ref{eqn:lemCond}) in Lemma \ref{lm:pro3b} 
always holds. 
%
%This implies that for the Gaussian 
%source with the cyclic shift invariant property 
%the matching condition on the sum rate part 
%always holds. 
%
Next we suppose that the Gaussian source $Y^L$ satisfies the cyclic shift
invariant property. It is obvious that for arbitrarily prescribed
small positive $\epsilon$, we can always choose a Gaussian random vector
$N^L$ so that $\Sigma_{N^L}=\epsilon I_L$ and $Y^L=X^L+N^L$. For the
above choice of $N^L$, the Gaussian remote source $X^L$ also satisfies
the cyclic shift invariant property.  
%Hence, we have the following
%corollary form Theorem\ref{th:matchTh2}.
Summarizing those arguments we obtain 
the following theorem.  

\begin{Th}\label{th:matchTh2}
%Assume that $X^L$ and $Y^L=X^L+N^L$ are cyclic shift invariant. 
%The quantities $a$,$c$,$\lambda_{\min}$, and $\lambda_{\max}$    
%are the same as those of previous definitions.
%If noise variance $\epsilon$ is sufficiently small, 
%then  
If $Y^L$ is cyclic shift invariant, then  
${R}_{{\rm sum},L}^{\rm (l)}(D|$
$\Sigma_{Y^L})={R}_{{\rm sum},L}(D|\Sigma_{Y^L}).
$ 
Furthermore, the curve $R=R_{{\rm sum},L}(D|\Sigma_{Y^L})$ has 
the following parametric form:
$$
\left.
\ba{rcl}
R&=&\ds\frac{1}{2}
\log\left[{|\Sigma_{Y^L}+B|}{\rm e}^{2Lr}
\prod_{i=1}^L\beta_{i}(r)\right]\,,
\vspace*{2mm}\\
D&=&\ds\sum_{i=1}^L\frac{1}{\beta_i(r)}
-{\rm tr}[B]\,.
\ea
\right\}
$$
\end{Th}
% if the identical varaince $\sigma^2$ 
%is relatively high or correlation coefficient $\rho$ 
%is relatively small. 
%Proofs of Lemma \ref{lm:pro3b} and Theorem \ref{th:matchTh2} 
%will be stated in Section V. 
%Note that the condition 
%(\ref{eqn:zaa0}) depends only on the correlation 
%property of $X^L$ and $N^L$. 
%\input{SI.tex}

%{\it Proof}
%\begin{co}\label{co:MainCo}

\section{Proofs of the Results}

\renewcommand{\irb}[1]{{\color[named]{Black}#1\normalcolor}}
\renewcommand{\irg}[1]{{\color[named]{Black}#1\normalcolor}}
\renewcommand{\irBr}[1]{{\color[named]{Black}#1\normalcolor}}
\renewcommand{\irBw}[1]{{\color[named]{Black}#1\normalcolor}}

%In this section we state the proofs of the theorems 
%stated in the previous section.

\subsection{Derivation of the Outer Bounds
}

In this subsection we prove the results on outer bounds of 
the rate distortion region.    
We first state two important lemmas which are mathematical cores of 
the converse coding theorem. For $i=1,2,\cdots, L$, set 
\beq
\irb{W_i}=\varphi_i(\irBw{\vc Y}_i), 
\irb{r_i^{(n)}}
=\frac{1}{n}I(\irBw{\vc Y}_i;\irb{W_i}|{\vc X}^K)\,. 
\eeq
For $\irBr{S}\subseteq \Lambda$, let ${Q}_\irBr{S}$ 
be a unitary matrix which transforms $X^K$ 
into $\irg{Z}^K= QX^K$. For 
$$
{\vc X}^K=(X^K(1),X^K(2),\cdots, X^K(n))
$$ 
we set 
$$
 {\vc Z}^K=Q{\vc X}^K
=(QX^K(1), QX^K(2), \cdots,QX^K(n))\,.
$$
Furthermore, for 
$\hat{\vc X}^K=(\hat{X}^K(1),$ 
           $\hat{X}^K(2),$ 
   $\cdots, \hat{X}^K(n))$,
we set
$$
 \hat{\vc Z}^K=Q\hat{\vc X}^K
=(Q\hat{X}^K(1), 
  Q\hat{X}^K(2), \cdots, 
  Q\hat{X}^K(n))\,.
$$
We have the following two lemmas. 
\begin{lm}\label{lm:lm1}
For any $i=1,2,\cdots, K$, we have 
\beqno
& &\left.h(\irg{\vc Z}_i \right| \irg{\vc Z}_{[i]}^K \irb{W}^L)
   \leq  h(\irg{{\vc Z}}_i -\hat{\irg{{\vc Z}}}_{i}
   \left.\right| \irg{{\vc Z}}_{[i]}^K -\hat{\irg{{\vc Z}}}_{[i]}^K)
\\
&\leq& 
\frac{n}{2}\log 
\left\{
(2\pi {\rm e})
\left[
Q\left({\ts \frac{1}{n}}
\Sigma_{{\lvc X}^K-\hat{\lvc X}^K}^{-1}\right){}^{\rm t}Q
\right]_{ii}^{-1}
\right\}\,,
\eeqno
where $h(\cdot)$ stands for the differential entropy. 
\end{lm}
%
%and 
%$[C]_{ij}$ stands for the $(i,j)$ entry of the matrix $C$.

\begin{lm}\label{lm:lm2} 
For any $i=1,2,\cdots, K$, we have
\beqno
& &
h(\irg{\vc Z}_i|\irg{\vc Z}_{[i]}^K\irb{W}^L)
\\
&\geq& 
\frac{n}{2}\log 
\left\{
(2\pi {\rm e})\left[
Q\left(\Sigma_{X^K}^{-1}
 +{}^{\rm t}A
\Sigma_{N_{\Lambda}(\irb{r_{\irBr{\Lambda}}^{(n)}})}^{-1}A\right){}^{\rm t}Q
\right]_{ii}^{-1}
\right\}\,.
\eeqno
\end{lm}

%The above lemma is mathematical core of the proof of the converse 
%coding theorem. 
%This lemma yields the sharper result on an 
%outer bound of the rate distortion region. 
%Lemma 2 is new! 
%This lemma play 
%A coupling of those two lemmas is necessary 
%to establish converse coding theorem. 
Proofs of Lemmas \ref{lm:lm1} and \ref{lm:lm2} will 
be stated in Appendixes A and B, respectively. 
The following corollary immediately follows 
from Lemmas \ref{lm:lm1} and \ref{lm:lm2}. 
\begin{co}\label{co:co0} For any $\Sigma_{X^KY^L}$ and for 
any 
$(\varphi_1^{(n)},$ 
$\varphi_2^{(n)}, \cdots,$ 
$\varphi_L^{(n)},$ $\psi^{(n)})$, we have 
$$
{\ts \frac{1}{n}}\Sigma_{{\lvc X}^K-\hat{\lvc X}^K}^{-1}
\preceq
\Sigma_{X^K}^{-1}
 +{}^{\rm t}A\Sigma_{N_{\Lambda}(\irb{r_{\Lambda}^{(n)}})}^{-1}A\,.
$$
\end{co}

%{\it Proof:} 
From Lemma \ref{lm:lm2}, we obtain the following corollary.
\begin{co}\label{co:co1} For any $S\subseteq \Lambda $, we have  
\beq
%\frac{1}{n}
I({\vc X}^K;W_S) \leq 
\frac{n}{2}
\log 
\left|I+\Sigma_{X^K}{}^{\rm t}A\Sigma_{N_{S}(r_S^{(n)})}^{-1}A\right|\,.
\eeq
\end{co}

{\it Proof:} For each $i\in \Lambda-S$, we choose 
$W_i$ so that it takes a constant value. In this case
we have $r_i^{(n)}=0$ for $i\in \Lambda-S$. 
Then by Lemma \ref{lm:lm2}, 
for any $i=1,2,\cdots, K$, we have
\beqa
\hspace*{-6mm}& &
h(\irg{\vc Z}_i|\irg{\vc Z}_{[i]}^K\irb{W}_{\irBr{S}})
\nonumber\\
\hspace*{-6mm}&\geq& 
\frac{n}{2}\log 
\left\{
(2\pi {\rm e})\left[
Q\left(\Sigma_{X^K}^{-1}
 +{}^{\rm t}A
\Sigma_{N_{S}(\irb{r_{\irBr{S}}^{(n)}})}^{-1}A\right){}^{\rm t}Q
\right]_{ii}^{-1}
\right\}.
\label{eqn:conv999}
\eeqa
We choose a unitary matrix $Q$ so that 
$$
Q\left(\Sigma_{X^K}^{-1}
+{}^{\rm t}A\Sigma_{N_S(r_S^{(n)})}^{-1}A\right){}^{\rm t}Q
$$ 
becomes the following diagonal matrix: 
\beq
Q\left(\Sigma_{X^K}^{-1}
+{}^{\rm t}A\Sigma_{N_S(r_S^{(n)})}^{-1}A\right){}^{\rm t}Q
=\left(
\begin{array}{cccc}
\lambda_1 &           &        & \mbox{\huge 0}\\
          & \lambda_2 &        &          \\
          &           & \ddots &          \\
\mbox{\huge 0} &      &        & \lambda_{K}\\
\end{array}
\right)\,.
\label{eqn:diag}
\eeq
Then we have the following chain of inequalities:  
\beqa
 &    &   
I({\vc X}^K;W_S)
=h({\vc X}^K)-h({\vc X}^K|W_S )
\nonumber\\ 
&\MEq{a}& h({\vc X}^K) 
         -h({\vc Z}^K|W_S )
\nonumber\\
&\leq  & h\left({\vc X}^K \right) 
         -\sum_{i=1}^{K} 
         h\left({\vc Z}_i|{\vc Z}_{[i]}^KW_S\right)
\nonumber\\
&\MLeq{b} &\frac{n}{2} 
         \log \left[ (2\pi {\rm e})^{K} 
\left|\Sigma_{X^K}\right| \right]
\nonumber\\
&     &\hspace*{-3mm}
             +\sum_{i=1}^{K} \frac{n}{2} \log 
             \left\{
             \frac{1}{2\pi {\rm e}}\left[
             Q\hspace*{-1mm}\left(\Sigma_{X^K}^{-1}
             +{}^{\rm t}A\Sigma_{N_S(r_S^{(n)})}^{-1}A\right){}^{\rm t}Q
             \right]_{ii}
             \right\}
\nonumber\\
&\MEq{c}& \frac{n}{2} \log \left| \Sigma_{X^K} \right| 
         +\sum_{i=1}^{K}\frac{n}{2}\log \lambda_i 
\nonumber\\
&=&\frac{n}{2} \log \left|\Sigma_{X^K} \right| 
   +\frac{n}{2}\log \left|\Sigma_{X^K}^{-1}
   +{}^{\rm t}A\Sigma_{N_S(r_S^{(n)})}^{-1}A\right| 
\nonumber\\
&=& \frac{n}{2}\log \left|I+\Sigma_{X^K}
{}^{\rm t}A\Sigma_{N_{S}(r_S^{(n)})}^{-1}A\right|\,. 
\nonumber
\eeqa
Step (a) follows from the rotation invariant property 
of the (conditional) differential entropy. 
Step (b) follows from (\ref{eqn:conv999}).
Step (c) follows from (\ref{eqn:diag}).
\hfill \IEEEQED 

%\end{document}
We first prove the inclusion 
$
{\cal R}_{L}(\DisT|$ $\Sigma_{X^KY^L})
\subseteq 
{\cal R}_{L}^{({\rm out})}(\DisT$ $|\Sigma_{X^KY^L})
$
stated in Theorem \ref{th:conv2}. 
Using Lemmas \ref{lm:lm1}, \ref{lm:lm2}, Corollary \ref{co:co1} 
and a standard argument on the proof of converse coding theorems, 
we can prove the above inclusion. 
%$
%{\cal R}_L(\DisT|\Sigma_{X^KY^L})
%\subseteq {\cal R}_L^{ ({\rm out})}(\DisT|\Sigma_{X^KY^L})\,.
%$

{\it Proof of 
${\cal R}_L(\DisT|\Sigma_{X^KY^L})
\subseteq {\cal R}_L^{ ({\rm out})}(\DisT|\Sigma_{X^KY^L})$:}
We first observe that %by virtue of the CI condition, 
\beq
 W_S\to {\vc Y}_S\to {\vc X}^K \to {\vc Y}_{\coS}\to W_{\coS}
\label{eqn:MkPr1}
\eeq
hold for any subset $S$ of $\Lambda$. 
Assume $(R_1, R_2,$ $\!\cdots, R_L) 
\in {\cal R}_{L}(\DisT|\Sigma_{X^KY^L})$. 
Then, there exists a sequence
$\{(\varphi_1^{(n)},\varphi_2^{(n)},$ 
$\cdots,\varphi_L^{(n)},\psi^{(n)}\}_{n=1}^{\infty}$
such that 
\beq
\left.
\ba{l}
\ds\limsup_{n\to\infty}R_i^{(n)}\leq R_i, i\in \Lambda
\vspace{1mm}\\
\ds \limsup_{n\to\infty}
\frac{1}{n}\Sigma_{{\lvc X}^K-\hat{{\lvc X}}^K}
\preceq \DisT
\ea
\right\}
\label{eqn:goddz1}
\eeq
We set
\beq
r_i\defeq \limsup_{n\to\infty}r_i^{(n)}
=\limsup_{n\to\infty}
\frac{1}{n}I({\vc Y}_i;W_S|{\vc X}^K)\,.
\eeq
For any subset $S \subseteq \Lambda$, we have 
the following chain of inequalities:
\beqa
& &
\sum_{i\in S}nR_i^{(n)}
%\nonumber\\
\geq  
\sum_{i\in S}\log M_i
\nonumber\\
&\geq & \sum_{i\in S}H(W_i)
%\nonumber\\
\geq H(W_S|W_{\coS})
\nonumber\\
&=&I({\vc X}^K;W_S|W_{\coS}) + H(W_S|W_{\coS}{\vc X}^K)
\nonumber\\
&{\stackrel{({\rm a})}{=}}&I({\vc X}^K;W_S|W_{\coS}) 
+\sum_{i\in S}H(W_i|{\vc X}^K) 
\nonumber\\
&{\stackrel{({\rm b})}{=}}&I({\vc X}^K;W_S|W_{\coS}) 
+\sum_{i\in S}H(W_i|{\vc X}^K) 
\nonumber\\
&{\stackrel{({\rm c})}{=}}&I({\vc X}^K;W_S|W_{\coS})
+n\sum_{i\in S}r_i^{(n)}, 
\label{eqn:conv1zf}
\eeqa
where steps (a),(b) and (c) follow from (\ref{eqn:MkPr1}). 
We estimate a lower bound of $I({\vc X}^K;W_S|W_{\coS})$. 
Observe that
\beqa
I({\vc X}^K;W_S|W_{S^{\rm c}})
& =& I({\vc X}^K;W^L)-I({\vc X}^K;W_{\coS}) 
\label{eqn:prthc0}
\eeqa
Since an upper bound of $I({\vc X}_{\coS};W_{\coS})$ 
is derived by Corollary \ref{co:co1}, it suffices 
to estimate a lower bound of $I({\vc X}^K;$ $W^L)$.
We have the following chain of inequalities:
\beqa
& &
I({\vc X}^K;W^L)
=h({\vc X}^K)-h({\vc X}^K|W^L)
\nonumber\\
&\geq &h({\vc X}^K)-h({\vc X}^K|\hat{{\vc X}}^K)
\nonumber\\
&\geq &h({\vc X}^K)-h({\vc X}^K-\hat{{\vc X}}^K)
\nonumber\\
&\geq &\frac{n}{2}\log\left[(2\pi {\rm e})^K
 \left|\Sigma_{{X}^K}\right|\right] 
%\nonumber\\
%& &\qquad 
-\frac{n}{2}\log \left[(2\pi {\rm e})^K
\left| {\ts\frac{1}{n}}\Sigma_{{\lvc X}^K-\hat{{\lvc X}}^K}
\right|\right]
\nonumber\\
&=&\frac{n}{2}\log\left[\frac{\left|\Sigma_{{X}^K}\right|}
{\left| {\ts\frac{1}{n}}\Sigma_{{\lvc X}^K-\hat{{\lvc X}}^K}\right|}\right]\,.
\label{eqn:prthz120}
\eeqa
Combining (\ref{eqn:prthc0}), (\ref{eqn:prthz120}), and  
Corollary \ref{co:co1}, we have 
\beqno
& & 
I({\vc X}^K;W_S|W_{S^{\rm c}}) +n\sum_{i\in S}r_i^{(n)}
\nonumber\\
&\geq & \frac{n}{2}\log 
\left[
\frac{\prod_{i\in S}{\baseN}^{2r_i^{(n)}} \left|\Sigma_{X^K}\right|}
{
\left|I+ \Sigma_{X^K}{}^{\rm t}A
\Sigma_{N_{S^{\rm c}}(r_{S^{\rm c}}^{(n)})}^{-1}A
\right|
\left|{\ts \frac{1}{n}}\Sigma_{{\lvc X}^K-\hat{{\lvc X}}^K}\right|
}
\right]
\\
&=& \frac{n}{2}\log 
\left[
\frac{\prod_{i\in S}{\baseN}^{2r_i^{(n)}} }
{\left|\Sigma_{X^K}^{-1} 
+ {}^{\rm t}A\Sigma_{N_{S^{\rm c}}(r_{S^{\rm c}}^{(n)}) }^{-1}A\right|
\left|{\ts \frac{1}{n}}\Sigma_{{\lvc X}^K-\hat{{\lvc X}}^K}\right|}
\right]\,.
\eeqno
Note here that
$
I({\vc X}^K;W_S|W_{\coS})+n\sum_{i\in S}r_i^{(n)}
$
is nonnegative. Hence, we have 
\beqa
& &
I({\vc X}^K;W_S|W_{\coS})+n\sum_{i\in S}r_i^{(n)}
\nonumber\\
&\geq & n\underline{J}_{S}\left(
\left.
\left|
{\ts\frac{1}{n}}\Sigma_{{\lvc X}^K-\hat{{\lvc X}}^K}
\right|,r_S^{(n)}\right|r_{\coS}^{(n)}\right)\,.
\label{eqn:prthdzq}
\eeqa
Combining (\ref{eqn:conv1zf}) and (\ref{eqn:prthdzq}), we obtain
\beqa
\sum_{i\in S}R_i^{(n)} 
&\geq & \underline{J}_{S}\left(
\left.
\left|
{\ts\frac{1}{n}}\Sigma_{{\lvc X}^K-\hat{{\lvc X}}^K}
\right|,r_S^{(n)}\right|r_{\coS}^{(n)}\right)
\label{eqn:prth99}
\eeqa
for $S\subseteq \Lambda$. On the other hand, 
by Corollary \ref{co:co0}, we have 
\beqa
\Sigma_{X^K}^{-1}
+{}^{\rm t}A\irBr{\Sigma_{N_\Lambda(r_{\Lambda}^{(n)})}^{-1}}A
&\succeq&
{\ts \frac{1}{n}}\Sigma_{{\lvc X}^K-\hat{\lvc X}^K}^{-1}
\label{eqn:prth101}
\eeqa
By letting $n\to\infty$ 
in (\ref{eqn:prth99}) and (\ref{eqn:prth101}) 
and taking (\ref{eqn:goddz1}) into account, 
we have for any $S\subseteq \Lambda$
\beqa
\sum_{i\in S}R_i 
&\geq &\underline{J}_{S}(\left|\DisT\right|,r_S|r_{\coS})\,,
\label{eqn:prth104}
\eeqa
and 
\beq
\Sigma_{X^K}^{-1}+{}^{\rm t}A\irBr{\Sigma_{N^L(r^L)}^{-1}}A
\succeq \DisT^{-1}\,.
\label{eqn:prth103} 
\eeq
From (\ref{eqn:prth104}) and (\ref{eqn:prth103}),
${\cal R}_{\Iset}(\DisT|\Sigma_{X^KY^L})
\subseteq {\cal R}_{\Iset}^{({\rm out})}(\DisT|$$\Sigma_{X^KY^L})$
is concluded. \hfill \IEEEQED

{\it Proof of Theorem \ref{th:conv2a}:} 
We choose a unitary matrix $Q$ so that 
\beqno
& &Q\Gamma^{-1}\left(\Sigma_{X^K}^{-1}
+{}^{\rm t}A\Sigma_{N^L(r^L)}^{-1}A\right){}^{\rm t}\Gamma^{-1} {}^{\rm t}Q
\\
&=&\left[
\begin{array}{cccc}
\alpha_1 &           &        & \mbox{\huge 0}\\
          & \alpha_2 &        &               \\
          &           & \ddots &              \\
\mbox{\huge 0} &      &        & \alpha_{K}   \\
\end{array}
\right]\,.
\eeqno
Then we have
\beqa
&&Q\Gamma\left(\Sigma_{X^K}^{-1}
+{}^{\rm t}A\Sigma_{N^L(r^L)}^{-1}A\right)^{-1}{}^{\rm t}\Gamma {}^{\rm t}Q
\nonumber\\
&=&\left[
\begin{array}{cccc}
\alpha_1^{-1} &           &        & \mbox{\huge 0}\\
          & \alpha_2^{-1} &        &           \\
          &           & \ddots &               \\
\mbox{\huge 0} &      &        & \alpha^{-1}_{K}\\
\end{array}
\right]\,.
\label{eqn:diagaa}
\eeqa
For $\DisT \in {\cal A}(r^L)$, set 
$$
\tilde{\Sigma}_d 
\defeq Q\Gamma\DisT {}^{\rm t}\Gamma {}^{\rm t}Q\,,\quad 
\xi_i\defeq \left[\tilde{\Sigma}_d\right]_{ii}\,.
$$ 
Since 
$$
\Gamma \DisT {}^{\rm t}\Gamma \succeq \Gamma(\Sigma_{X^L}^{-1}
+{}^{\rm t}A\Sigma_{N^L(r^L)}^{-1}A)^{-1}{}^{\rm t}\Gamma\,,
$$
$(\ref{eqn:diagaa})$, 
and ${\rm tr}[\Gamma\Sigma_d{}^{\rm t}\Gamma]\leq D$, 
we have
\beq
\left.
\ba{l}
\xi_i\geq \alpha_i^{-1},\mbox{ for }i=1,2,\cdots,K\,,
\vspace{1mm}\\
\ds \sum_{i=1}^K\xi_i={\rm tr}\left[\tilde{\Sigma}_d\right]
={\rm tr}\left[\Gamma\Sigma_d{}^{\rm t}\Gamma\right]\leq D\,.
\ea
\right\}
\label{eqn:aa2}
\eeq
Furthermore, by Hadamard's inequality we have 
\beqa
|\Sigma_d|=|\Gamma|^{-2}|\tilde{\Sigma}_d| \leq 
|\Gamma|^{-2}\prod_{i=1}^K[\tilde{\Sigma}_d]_{ii}
=|\Gamma|^{-2}\prod_{i=1}^K\xi_{i}\,. 
\label{eqn:aa3}
\eeqa
Combining (\ref{eqn:aa2}) and (\ref{eqn:aa3}), we obtain 
\beqno
& &\theta(\Gamma,D,r^L)
\\
&=&\max_{\scs \DisT: \DisT \in {\cal A}_L({r^L}),
      \atop{\scs 
       {\rm tr}[\Gamma\DisT {}^{\rm t}\Gamma]\leq D}
      }
    \left|\DisT \right|
\\
&\leq& |\Gamma|^{-2}\max_{\scs \xi_i\alpha_{i}\geq 1,i=1,2,\cdots,K\,, 
          \atop{\scs
          \sum_{i=1}^K\xi_i\leq D
          }   
      }\prod_{i=1}^K\xi_{i} 
={\omega}(\Gamma,D,r^L)\,.
\eeqno
The equality holds when 
$\tilde{\Sigma}_d$ is a diagonal matrix.
\hfill\IEEEQED

{\it Proof of Theorem \ref{Th:sr0}:}
Assume that
$(R_1, R_2,$ $\!\cdots, R_L) \in {\cal R}_{L}(D|\Sigma_{Y^L})$. 
Then, there exists a sequence 
$\{(\varphi_1^{(n)},\varphi_2^{(n)},$ 
$\cdots,\varphi_L^{(n)},\phi^{(n)}\}_{n=1}^{\infty}$
such that 
\beq
\left.
\ba{l}
\ds\limsup_{n\to\infty}R_i^{(n)}\leq R_i, i\in \Lambda
\vspace{1mm}\\
\ds \limsup_{n\to\infty}
\frac{1}{n}\Sigma_{{\lvc Y}_{\Lambda}-\hat{{\lvc Y}}_{\Lambda}}
\preceq \DisT,\:
%\vspace{1mm}\\
{\rm tr }[\DisT]\leq D
\vspace{1mm}\\
\mbox{ for some }\DisT.
\ea
\right\}
\label{eqn:godd}
\eeq
%We set
%\beq
%r_i\defeq \limsup_{n\to\infty}r_i^{(n)}
%=\limsup_{n\to\infty}
%\frac{1}{n}I({\vc Y}_i;W_i|{\vc X}^L)\,.
%\eeq
For each $l=0,1,\cdots,L-1$, we use 
$(\varphi_{\tau^l{(1)}}^{(n)}, \varphi_{\tau^l(2)}^{(n)}, \cdots,$ 
 $\varphi_{\tau^l(L)}^{(n)})$ 
for the encoding of $({\vc Y}_1, {\vc Y}_2, \cdots, {\vc Y}_L)$. 
For $i\in {\Lambda}$ and for $l=0,1,\cdots,L-1$, set 
\beqno
W_{l,i}&\defeq &\varphi_{\tau^l(i)}({\vc Y}_i),
\quad
\hat{\vc Y}_{l,i}
\defeq \phi_{\tau^l(i)}(\varphi_{\tau^l(i)}({\vc Y}_i)),
\\
r_{l,i}^{(n)} &\defeq& 
\frac{1}{n}I({\vc Y}_i;W_{l,i}|{\vc X}^L).
\eeqno
In particular,
$$
r_{0,i}^{(n)}=r_i^{(n)}=\frac{1}{n}I({\vc Y}_i;W_{i}|{\vc X}_i), 
\quad\mbox{for }i\in\Lambda. 
$$
Furthermore, set 
\beqno
r_{\tau^l(\Lambda)}^{(n)}
&\defeq&(r^{(n)}_{l,1},r_{l,2}^{(n)},\cdots,r_{l,L}^{(n)})\,,
\mbox{ for }l=0,1,\cdots,L-1\,, 
\\
r^{(n)}&\defeq& \frac{1}{L}\sum_{i=1}^{L}r_i^{(n)}\,.
\eeqno 
By the cyclic shift invariant property of 
${\vc X}_{\Lambda}$ and ${\vc Y}_{\Lambda}$, we have
for $l=0,1,\cdots,L-1$,
\beq
 \frac{1}{L}\sum_{i=1}^Lr^{(n)}_{l,i}
=\frac{1}{L}\sum_{i=1}^Lr^{(n)}_{0,i}=r^{(n)}\,.
\label{eqn:godd0}
\eeq
 For $\DisT=[d_{ij}]$, set 
\beqno
& &\tau^l(\DisT)\defeq [d_{\tau^l(i)\tau^l(j)}],\:
\overline{\DisT}\defeq
\frac{1}{L}\sum_{l=0}^{L-1}\tau^l(\DisT)\,.
\eeqno
Then, we have 
\beqa
& &\limsup_{n\to\infty}
\frac{1}{L}\sum_{l=0}^{L-1}
{\ts \frac{1}{n}}\Sigma_{{\lvc Y}_{\Lambda}-\hat{{\lvc Y}}_{\tau^l(\Lambda)}}
\nonumber\\
&\MEq{a}&\limsup_{n\to\infty}
\frac{1}{L}\sum_{l=0}^{L-1}
{\ts \frac{1}{n}}\Sigma_{{\lvc Y}_{\tau^l(\Lambda)}-\hat{{\lvc Y}}_{\tau^l(\Lambda)}}
\nonumber\\
&\MPreq{b}&
\frac{1}{L}\sum_{l=0}^{L-1}\tau^l(\DisT)\MEq{c}\overline{\DisT}\,.
\label{eqn:godd2}
\eeqa
Step (a) follows from the cyclic shift invariant property of
${\vc Y}_{\Lambda}$.
Step (b) follows from (\ref{eqn:godd}). 
Step (c) follows from the definition of $\overline{\DisT}$.
From ${\vc Y}_{\Lambda}$, we construct an 
estimation $\hat{\vc X}_{\Lambda}$ of 
${\vc X}_{\Lambda}$ by 
$\hat{\vc X}_{\Lambda}=\tilde{A}\hat{\vc Y}_{\Lambda}\,.$
Then for $l=0,1,\cdots,L-1$, we have the following.
\beqa
& &\Sigma_{X_{\Lambda}}^{-1}
    +\Sigma_{N_{\tau^l(\Lambda)}(r_{\tau^l(\Lambda)}^{(n)})}^{-1}
    \MEq{a}\Sigma_{X_{\tau^l(\Lambda)}}^{-1}
    +\Sigma_{N_{\tau^l(\Lambda)}(r_{\tau^l(\Lambda)}^{(n)})}^{-1}
\nonumber\\
&\MSueq{b}&
{\ts \frac{1}{n}}
\Sigma_{{\lvc X}_{\tau^l(\Lambda)}-\hat{\lvc X}_{\tau^l(\Lambda)}}^{-1}
 \MEq{c}{\ts \frac{1}{n}}
\Sigma_{{\lvc X}_{\Lambda}-\hat{\lvc X}_{\tau^l(\Lambda)}}^{-1}
\nonumber\\
&=&\left[\tilde{A}
\left(
{\ts \frac{1}{n}}\Sigma_{{\lvc Y}_{\Lambda}-\hat{\lvc Y}_{\tau^l(\Lambda)}}
\right)
{}^{\rm t}\tilde{A}+\Sigma_{X_{\Lambda}|Y_{\Lambda}}
\right]^{-1}\,.
\label{eqn:zaaa0}
\eeqa
Steps (a) and (c) follow from the cyclic shift invariant 
property of ${X}_{\Lambda}$ and ${\vc X}_{\Lambda}$, 
respectively. 
Step (b) follows from 
Corollary \ref{co:co0}. From (\ref{eqn:zaaa0}), we have
\beqa
& &\frac{1}{L}\sum_{l=0}^{L-1}
\left[\Sigma_{X_{\Lambda}}^{-1}
     +\Sigma_{N_{\tau^l(\Lambda)}(r_{\tau^l(\Lambda)}^{(n)})}^{-1}\right]
\nonumber\\
&\succeq&
\frac{1}{L}\sum_{l=0}^{L-1}
\left[\tilde{A}
\left(
{\ts \frac{1}{n}}\Sigma_{{\lvc Y}_{\Lambda}-\hat{\lvc Y}_{\tau^l(\Lambda)}}
\right)
{}^{\rm t}\tilde{A}+\Sigma_{X_{\Lambda}|Y_{\Lambda}}
\right]^{-1}
\nonumber\\
&\MSueq{a}&
\left[\tilde{A}
\left(
\frac{1}{L}\sum_{l=0}^{L-1}
{\ts \frac{1}{n}}\Sigma_{{\lvc Y}_{\Lambda}-\hat{\lvc Y}_{\tau^l(\Lambda)}}
\right)
{}^{\rm t}\tilde{A}+\Sigma_{X_{\Lambda}|Y_{\Lambda}}
\right]^{-1}
\nonumber\\
&=&
\left[\tilde{A}
\left(
\frac{1}{L}\sum_{l=0}^{L-1}
{\ts \frac{1}{n}}\Sigma_{{\lvc Y}_{\Lambda}-\hat{\lvc Y}_{\tau^l(\Lambda)}}
+B\right)
{}^{\rm t}\tilde{A}
\right]^{-1}\,.
\label{eqn:zaaa1}
\eeqa
Step (a) follows form that
$
(\tilde{A}
\Sigma
{}^{\rm t}\tilde{A}+\Sigma_{X_{\Lambda}|Y_{\Lambda}}
)^{-1}
$
is convex with respect to $\Sigma$. On the 
other hand, we have 
\beqa
& &\frac{1}{L}\sum_{l=0}^{L-1}
   \left[\Sigma_{X_{\Lambda}}^{-1}
         +\Sigma_{N_{\tau^l(\Lambda)} (r_{\tau^l(\Lambda)}^{(n)}) }^{-1}
   \right]
\nonumber\\
&=&\Sigma_{X_{\Lambda}}^{-1}
    +\left(\frac{1}{L}\sum_{i=1}^{L}
    \frac{1-{\rm e}^{-2r_{i}^{(n)}}}{\epsilon}\right)I_L
\nonumber\\   
&\MPreq{a}&
 \Sigma_{X_{\Lambda}}^{-1}
    +\left(
    \frac{1-{\rm e}^{-2\frac{1}{L}\sum_{i=1}^{L}r_{i}^{(n)}}}
    {\epsilon}
    \right)I_L
\nonumber\\
&=&\Sigma_{X_{\Lambda}}^{-1}
     +\left(
     \frac{1-{\rm e}^{-2r^{(n)}}}
    {\epsilon}
    \right)I_L\,.
\label{eqn:zaaa2}
\eeqa
Step (a) follows from that $1-{\rm e}^{-2a}$ is a concave
function of $a$. Combining (\ref{eqn:zaaa1}) 
and (\ref{eqn:zaaa2}), we obtain
\beqno
& &\Sigma_{X_{\Lambda}}^{-1}
    +\left(
    \frac{1-{\rm e}^{-2r^{(n)}}}{\epsilon}
    \right)I_L
\\
&\succeq &
\left[
\tilde{A}
\left(
\frac{1}{L}\sum_{l=0}^{L-1}
{\ts \frac{1}{n}}\Sigma_{{\lvc Y}_{\Lambda}-\hat{\lvc Y}_{\tau^l(\Lambda)}}
+B\right)
{}^{\rm t}\tilde{A}
\right]^{-1}\,,
\eeqno
from which we obtain
\beqa
& &\frac{1}{L}\sum_{l=0}^{L-1}
{\ts \frac{1}{n}}\Sigma_{{\lvc Y}_{\Lambda}-\hat{\lvc Y}_{\tau^l(\Lambda)}}+B
\nonumber\\
&\succeq &
\left[
{}^{\rm t}\tilde{A}\left\{
\Sigma_{X_{\Lambda}}^{-1}
    +\left(
    \frac{1-{\rm e}^{-2r^{(n)}}}{\epsilon}
     \right)I_L
\right\}\tilde{A}
\right]^{-1}\,.
\label{eqn:coverse1001}
\eeqa
Next we derive a lower bound of the sum rate part. 
For each $l=0,1,\cdots,L-1$, we have the following 
chain of inequalities.
\beqa
& &
\sum_{i\in \Lambda}nR_i^{(n)}\geq \sum_{i\in \Lambda}\log M_i 
\geq \sum_{i\in \Lambda}H(W_{l,i})\geq H(W_{\tau^l(\Lambda)})
\nonumber\\
&=&I({\vc X}_\Lambda;W_{\tau^l(\Lambda)}) 
   +H(W_{\tau^l(\Lambda)}|{\vc X}_{\Lambda}) 
\nonumber\\
&{\stackrel{({\rm a})}{=}}&I({\vc X}_\Lambda;W_{\tau^l(\Lambda)}) 
+\sum_{i\in \Lambda}H(W_{l,i}|{\vc X}_{\Lambda}) 
\nonumber\\
&=&I({\vc X}_\Lambda;W_{\tau^l(\Lambda)}) 
+\sum_{i\in \Lambda}I({\vc Y}_{\Lambda};W_{l,i}|{\vc X}_{\Lambda}) 
\nonumber\\
&\MEq{b}& I({\vc X}_\Lambda;W_{\tau^l(\Lambda)})+nLr^{(n)}
\nonumber\\ 
&\MGeq{c}&
\frac{n}{2}\log\left[\frac{\left|\Sigma_{{X}_{\Lambda}}\right|}
{\left|{\ts\frac{1}{n}}
\Sigma_{{\lvc X}_{\Lambda}-\hat{{\lvc X}}_{\tau^l(\Lambda)}}\right|}\right]
+nLr^{(n)}  
\nonumber\\
&=&\frac{n}{2}\log
\left[
     \frac{\left|\tilde{A}\Sigma_{{Y}_{\Lambda}}{}^{\rm t}\tilde{A}
      +\Sigma_{X_{\Lambda}|Y_{\Lambda}}\right|}
      {\left|
      \tilde{A}
      \left(
      {\ts\frac{1}{n}}\Sigma_{{\lvc Y}_{\Lambda}-\hat{{\lvc Y}}_{\tau^l(\Lambda)}}
      \right)
      {}^{\rm t}\tilde{A}+\Sigma_{X_{\Lambda}|Y_{\Lambda}}
      \right|}
      \right]+nLr^{(n)} 
\nonumber\\
&=&\frac{n}{2}\log
\left[
     \frac{\left|\Sigma_{{Y}_{\Lambda}}+B\right|}
      {\left|
      {\ts\frac{1}{n}}\Sigma_{{\lvc Y}_{\Lambda}-\hat{{\lvc Y}}_{\tau^l(\Lambda)}}+B
      \right|}
      \right]+nLr^{(n)}\,. 
\label{eqn:prthz0zz}
\eeqa
Step (a) follows from (\ref{eqn:MkPr1}). 
Step (b) follows from (\ref{eqn:godd2}). 
Step (c) follows from (\ref{eqn:prthz120}). 
From (\ref{eqn:prthz0zz}), we have
\beqa
& &\sum_{i\in \Lambda}R_i^{(n)}
=\frac{1}{L}\sum_{l=1}^{L-1}\sum_{i\in \Lambda}R_i^{(n)}
\nonumber\\
&\geq&
{\ds \frac{1}{L}\sum_{l=1}^{L-1}}
\frac{1}{2}
      \log \left[
      \frac{\left|\Sigma_{{Y}_{\Lambda}}+B\right|}
      {\left|
      {\ts\frac{1}{n}}\Sigma_{{\lvc Y}_{\Lambda}-\hat{{\lvc Y}}_{\tau^l(\Lambda)}}+B
      \right|}
      \right]+Lr^{(n)}
\nonumber\\
&\MGeq{a}&\frac{1}{2}
\log \left[
      \frac{\left|\Sigma_{{Y}_{\Lambda}}+B\right|}
      {\left|{\ds \frac{1}{L}\sum_{l=1}^{L-1}}
      {\ts\frac{1}{n}}\Sigma_{{\lvc Y}_{\Lambda}-\hat{{\lvc Y}}_{\tau^l(\Lambda)}}+B
      \right|}
      \right]+Lr^{(n)}\,. 
\label{eqn:prthz20}
\eeqa
Step (a) follows from that $-\log|\Sigma+B|$ is convex 
with respect to $\Sigma$. 
Letting $n\to\infty$ in (\ref{eqn:coverse1001}) 
and (\ref{eqn:prthz20}) and taking (\ref{eqn:godd2}) 
into account, we have
\beqa
\ds\sum_{i\in \Lambda}R_i
&\geq&
\ds \frac{1}{2}
      \log \left[
      \frac{\left|\Sigma_{{Y}_{\Lambda}}+B\right|}
      {\left|
      \overline{\DisT}+B
      \right|}
      \right]+Lr\,,
\label{eqn:prthz17}
\\
\overline{\DisT}+B
&\succeq&
\ds \left[
  {}^{\rm t}\tilde{A}\left\{
  \Sigma_{X_{\Lambda}}^{-1}
    +\left(
    \frac{1-{\rm e}^{-2r}}{\epsilon}
     \right)I_L\right\}\tilde{A}\right]^{-1},
\label{eqn:prthz18}
\\
{\rm tr}[\overline{\DisT}+B]&=&{\rm tr}[\DisT]+{\rm tr}[B]
\leq D+{\rm tr}[B]\,.
\label{eqn:prthz21}
\eeqa
Now we choose a unitary matrix $Q$ so that 
\beqno
%& &
Q
{}^{\rm t}\tilde{A}\left\{\Sigma_{X_{\Lambda}}^{-1}
+\left(\frac{1-{\rm e}^{-2r}}{\epsilon}\right)I_L\right\}\tilde{A}{}^{\rm t}Q
%\\
&=&\left[
\begin{array}{cccc}
\beta_1 &           &        & \mbox{\huge 0}\\
          & \beta_2 &        &               \\
          &           & \ddots &             \\
\mbox{\huge 0} &      &        & \beta_{L}   \\
\end{array}
\right]\,.
\eeqno
%Then we have
Set 
\beqno
\hat{\Sigma}_d &\defeq&Q\DisT {}^{\rm t}Q\,,
\hat{B}_d \defeq QB{}^{\rm t}Q\,,
\xi_i\defeq \left[\hat{\Sigma}_d+\hat{B}\right]_{ii}\,.
\eeqno 
From (\ref{eqn:prthz18}) and (\ref{eqn:prthz21}) 
we have
\beq
\left.
\ba{l}
\xi_i\geq \beta_i^{-1}(r), i\in\Lambda\,,
\vspace{1mm}\\
\ds \sum_{i=1}^L\xi_i
={\rm tr}\left[\hat{\Sigma}_d+\hat{B}\right]
={\rm tr}\left[\Sigma_d+B\right]\leq D+{\rm tr}[B]\,.
\ea
\right\}
\label{eqn:aa2z}
\eeq
From (\ref{eqn:aa2z}), we have
\beqa
\sum_{i=1}^L\frac{1}{\beta_i(r)}
    &\leq&\sum_{i=1}^L\xi_i
    ={\rm tr}[\hat{\Sigma_d}+\hat{B}]
\leq D+{\rm tr}[B]
\nonumber\\
&\Leftrightarrow& r\geq r^*(D+{\rm tr}[B])\,.
\label{eqn:aa03zz}
\eeqa
Furthermore, by Hadamard's inequality we have 
\beqa
|\Sigma_d+B|=|\hat{\Sigma}_d +\hat{B}|
\leq \prod_{i=1}^L[\hat{\Sigma}_d +\hat{B}]_{ii}
=\prod_{i=1}^L\xi_{i}\,. 
\label{eqn:aa3z}
\eeqa
Combining (\ref{eqn:aa2z}) and (\ref{eqn:aa3z}), we obtain 
\beq
|\Sigma_d+B|
\leq\max_{\scs \xi_i\beta_{i}\geq 1,i\in{\Lambda}\,, 
          \atop{\scs
          \sum_{i=1}^L\xi_i\leq D+{\rm tr}[B]
          }   
      }\prod_{i=1}^L\xi_{i} 
=\tilde{\omega}(D,r)\,.
\label{eqn:prthz22}
\eeq
%The equality holds when $\hat{\Sigma}_d+\hat{B}$ 
%is a diagonal matrix.
Hence, from (\ref{eqn:prthz17}), (\ref{eqn:aa03zz}), 
and (\ref{eqn:prthz22}) we have 
\beqno
\sum_{i=1}^L R_i
&\geq& 
\min_{r\geq r^*(D+{\rm tr}[B])}
\frac{1}{2}
\log
\left[
\frac{{\rm e}^{Lr}|\Sigma_Y+B|}{\tilde{\omega}(D,r)}
\right]
\\
&=&\min_{r\geq r^*(D+{\rm tr}[B])}\underline{\tilde{J}}(D,r)
=R_{{\rm sum},L}(D|\Sigma_{Y^L})\,,
\eeqno
completing the proof. 
\hfill \IEEEQED

\subsection{Derivation of the Inner Bounds}

%Proof of Theorem \ref{th:conv2}

In this subsection we prove 
${\cal R}_L^{({\rm in})}(\DisT$ $|\Sigma_{X^KY^L})$
$\subseteq$ ${\cal R}_L(\DisT$ $|\Sigma_{X^KY^L})$ 
stated in Theorem \ref{th:conv2}. 

{\it Proof of ${\cal R}_L^{({\rm in})}(\DisT|\Sigma_{X^KY^L})$
$\subseteq$ ${\cal R}_L(\DisT|\Sigma_{X^KY^L})$:} 
Since $\hat{\cal R}_L^{({\rm in})}($ $\DisT|\Sigma_{X^KY^L})$ $\subseteq$ 
${\cal R}_L(\DisT|\Sigma_{X^KY^L})$ is proved by Theorem \ref{th:direct}, 
it suffices to show 
${\cal R}_L^{({\rm in})}(\DisT|\Sigma_{X^KY^L})$
$\subseteq$ 
$\hat{\cal R}_L^{({\rm in})}(\DisT|\Sigma_{X^KY^L})$
to prove ${\cal R}_L^{({\rm in})}(\DisT|\Sigma_{X^KY^L})$
$\subseteq$ ${\cal R}_L(\DisT|\Sigma_{X^KY^L})$. 
We assume that 
$R^L\in {\cal R}_L^{({\rm in})}$ $(\DisT|\Sigma_{X^KY^L})$. 
Then, there exists nonnegative vector $r^L$ such that
\beqno
& &\left(\Sigma_{X^K}^{-1}
  +{}^{\rm t}{A}\Sigma_{N^L(r^L)}^{-1}A\right)^{-1}\preceq \DisT
\eeqno
and 
\beq
\sum_{i\in S} R_i\geq K(r_S|r_{\coS})
\mbox{ for any }S \subseteq \Lambda\,. 
\label{eqn:zsa1}
\eeq
Let $V_i, i\in \Lambda$ be $L$ independent zero mean Gaussian 
random variables with variance $\sigma_{V_i}^2$.
Define Gaussian random variables $U_i, i\in \Lambda$ by 
$
U_i=X_i+N_i+V_i. %,\:\: i\in \Lambda.
$
By definition it is obvious that
\beq
\left.
\ba{l}
U^L\to Y^L \to X^K \\
U_S\to Y_S \to X^K \to Y_{\coS}\to U_{\coS}\\
\mbox{ for any } S\subseteq \Lambda\,.  
\ea
\right\}
\label{eqn:gau00sz} 
\eeq
For given $r_i \geq 0, i\in \Lambda$, choose 
$\sigma_{V_i}^2$ so that 
$\sigma_{V_i}^2=\sigma_{N_i}^2/({\baseN}^{2r_i}-1)$
when $r_i>0$. When $r_i=0$ we choose $U_i$ so that $U_i$
takes constant value zero. In the above choice, the covariance 
matrix of $N^L+V^L$ becomes 
$\Sigma_{N^L(r^L)}$. Define 
the linear function ${\psi}$ of $U^L$ by 
$$
{\psi}\left(U^L\right) = 
(\Sigma_{X^K}^{-1} +{}^{\rm t}A\Sigma_{N^L(r^L)}^{-1}A)^{-1}
{}^{\rm t}A\Sigma_{N^L(r^L)}^{-1}U^L
\,.
$$
Set $\hat{X}^L={\psi}\left(U^L\right)$ and 
\beqa 
d_{ii}%({X}_i,\hat{X}_i)
& \defeq & {\rm E}\left[||{X}_i-\hat{X}_i||^2\right]\,,
\nonumber\\
d_{ij}%({X}_i,\hat{X}_j) 
& \defeq &
{\rm E}\left[
       \left({X}_i-\hat{X}_i\right)
       \left({X}_j-\hat{X}_j\right)
       \right]\,,
%\nonumber\\
%& &
1 \leq i\ne j \leq K.
\nonumber
\eeqa
Let $\Sigma_{{X}^K-\hat{X}^K}$ be a covariance matrix 
with $d_{ij}$ in its $(i,j)$ entry. 
By simple computations we can show that
\beq
\Sigma_{X^K-\hat{X}^K}
=(\Sigma_{X^K}^{-1} +{}^{\rm t}A\Sigma_{N^L(r^L)}^{-1}A)^{-1}\preceq \DisT
\label{eqn:gau2zz} 
\eeq
and that for any $S\subseteq \Lambda$, 
\beqa
%\hspace*{-7mm}
%\underline{J}_S(\DisT^{-1},r_S|r_{\coS}) 
& &J_S(r_S|r_{\coS})=I(Y_S;U_S|U_{\coS})\,.
\label{eqn:gau1z} 
\eeqa
From (\ref{eqn:gau00sz}) and (\ref{eqn:gau2zz}), we have 
$U^L\in {\cal G}(\DisT)$. Thus, from (\ref{eqn:gau1z}) 
$
{\cal R}_L^{({\rm in})}
(\DisT|\Sigma_{X^KY^L})\subseteq 
\hat{\cal R}_L^{({\rm in})}(\DisT|\Sigma_{X^KY^L})
$
is concluded. \hfill \IEEEQED 

\subsection{Proofs of the Results on Matching Conditions}

%Proof of Lemma \ref{lm:lem1}
%In this subsection we prove Lemma \ref{lm:lem1}.
%We first present a preliminary observation on 
%${\cal R}_L^{(\rm out)}(D)$. For $r^L\in {\cal B}_L(D)$, 
%we examine a form of the region 

We first observe that the condition
$$
{\rm tr}\left[
\Gamma
\left(\Sigma_{X^K}^{-1}+{}^{\rm t}
A\Sigma_{N^L(r^L)}^{-1}A
\right)^{-1}{}^{\rm t}\Gamma
\right]\leq D
$$
is equivalent to 
\beq
\sum_{j=1}^K\frac{1}{\alpha_{j}(r^L)}\leq D\,.
\label{eqn:aa00z}
\eeq
 
{\it Proof of Lemma \ref{lm:pro3}:}
Let $\tilde{\Lambda}=\{1,2,\cdots,K\}$ and let $S \subseteq \tilde{\Lambda}$ 
be a set of integers that satisfies 
$\alpha_i^{-1}\geq \xi$ in the definition of 
$\theta(\Gamma,D,u^L)$. Then, $\theta(\Gamma,D,u^L)$ 
is computed as 
\beqno
& &\theta(\Gamma,D,u^L)
\\
&=&{\ts \frac{1}{(K-|S|)^{K-|S|}}}
   \left(\prod_{i\in S}\frac{1}{\alpha_i}\right)
   \left(D-\sum_{i\in S} \frac{1}{\alpha_i}\right)^{K-|S|}\,.
\eeqno
Fix $i\in\Lambda$ arbitrary. For simplicity of notation 
we set 
$$
\chi_i \defeq ||\hat{\vc a}_i||^2\frac{1}{\sigma^2_{N_i}}
+\eta_i(u_{[i]}^L) 
$$
and set 
$$
\Psi\defeq \log 
\frac{\frac{1}{\sigma^2_{N_i}}}{\frac{1}{\sigma^2_{N_i}}-u_i}
-\log \theta(\Gamma,D,u^L)\,.
$$
Computing the partial derivative of $\Psi$ by $u_i$, we obtain
\beqa
%& &
\frac{\partial \Psi}{\partial u_i}
%\nonumber\\
&=&\sum_{j\in S}
\left(
\frac{\partial \alpha_j}{\partial u_i}
\right)
\left[\frac{1}{\alpha_j}
-\frac{K-|S|}
{D-{\ds \sum_{j\in S}}\frac{1}{\alpha_j}}\frac{1}{\alpha_j^2}
\right]
+\frac{1}{\frac{1}{\sigma^2_{N_i}}-u_i}\,.
\nonumber\\
&=&\sum_{j\in S}
\left(
\frac{\partial \alpha_j}{\partial u_i}
\right)
\left[\frac{1}{\alpha_j}
-\frac{K-|S|}
{D-{\ds \sum_{j\in S}}\frac{1}{\alpha_j}}\frac{1}{\alpha_j^2}
\right]
\nonumber\\
& &
+\frac{||\hat{\vc a}_i||^2}{
 \frac{1}{\sigma^2_{N_i}}||\hat{\vc a}_i||^2
+\eta_i -(||\hat{\vc a}_i||^2u_i+\eta_i)}
\nonumber\\
&=&
\sum_{j\in S}
\left(
\frac{\partial \alpha_j}{\partial u_i}
\right)
\left[\frac{1}{\alpha_j}
-\frac{K-|S|}
{D-{\ds \sum_{j\in S}}\frac{1}{\alpha_j}}\frac{1}{\alpha_j^2}
\right]
\nonumber\\
& &\quad +\frac{||\hat{\vc a}_i||^2}{
 \chi_i -(||\hat{\vc a}_i||^2u_i+\eta_i)}\,.
\label{zsz0}
\eeqa
From Lemma \ref{lm:Egn1} and (\ref{zsz0}), we obtain
$$
%& &
\hspace*{-2mm}\frac{\partial \Psi}{\partial u_i}
%\nonumber\\
\geq \sum_{j\in S}
\left(
\frac{\partial \alpha_j}{\partial u_i}
\right)
\left[\frac{1}{\alpha_j}
-\frac{K-|S|}{D-{\ds \sum_{k\in S}} \frac{1}{\alpha_j}} 
\frac{1}{\alpha_j^2}
+\frac{1}{\chi_i-\alpha_{\min}}
\right]\,.
$$
To examine signs of contents of the above summation we 
set  
\beqno
\Phi_j 
&\defeq& 
\left\{
D-\sum_{j\in S}{\frac{1}{\alpha_j}-\frac{K-|S|}{\alpha_j}}
\right\}(\chi_i-\alpha_{\min})
\\
& &+\alpha_{j}\left(D-\sum_{j\in S}{\frac{1}{\alpha_j}}\right).
\eeqno
If $|S|=K$, $\Phi_j \geq 0, j \in \Lambda$ is obvious. 
We hereafter assume $|S|\leq K-1$. Computing $\Phi_j$, we obtain
\beqa
\Phi_j&=&
  \chi_i\left(D-\sum_{j\in S} {\frac{1}{\alpha_j}}\right)
  -{\frac{K-|S|}{\alpha_j}}\cdot(\chi_i-\alpha_{\min})
\nonumber\\
& &+(\alpha_{j}-\alpha_{\min})
\left(D-\sum_{j\in S}{\frac{1}{\alpha_j}}\right)
\nonumber\\
&\geq & \chi_i\left(D-\sum_{j\in S}{\frac{1}{\alpha_j}}\right)
         -{\frac{K-|S|}{\alpha_j}}\cdot(\chi_i-\alpha_{\min})
\nonumber\\
&\MGeq{a}& \chi_i\sum_{j\in \tilde{\Lambda}-S} \frac{1}{\alpha_{j}}
-{\frac{K-|S|}{\alpha_{j}}}\cdot(\chi_i-\alpha_{\min})
\nonumber\\
&\geq & \chi_i\cdot\frac{K-|S|}{\alpha_{\max}}
-{\frac{K-|S|}{\alpha_{\min}}}\cdot(\chi_i-\alpha_{\min})
\nonumber\\
& = & 
{\chi_i}(K-|S|)
\left(\frac{1}{\alpha_{\max}}
-\frac{1}{\alpha_{\min}}+\frac{1}{\chi_i}\right)\,.
\label{eqn:zsd0}
\eeqa
Step (a) follows from the inequality (\ref{eqn:aa00z}). 
From (\ref{eqn:zsd0}), we can see that if 
$$
 \frac{1}{\alpha_{\min}(r^L)}
-\frac{1}{\alpha_{\max}(r^L)} \leq \frac{1}{\chi_i} 
\mbox{ for }i\in \Lambda,
$$
then, $\Phi_j\geq 0$ for $j\in S\,.$ \hfill\IEEEQED

{\it Proof of Theorem \ref{th:matchTh}: } 
%Let $\alpha_{\max}(r^L)$ be the maximum eigen value 
%of $D(\Sigma_{X^L}^{-1}+$ $\Sigma_{N^L}^{-1})$. 
By (\ref{eqn:aa00z}), we have 
\beqno
\frac{1}{\alpha_{\min}(r^L)}
&\leq &D-\frac{K-1}{\alpha_{\max}(r^L)}
\\
&=&\frac{1}{\alpha_{\max}(r^L)}+D-\frac{K}{\alpha_{\max}(r^L)}
\,.
\eeqno
Hence, if 
$$
D-\frac{K}{\alpha_{\max}(r^L)} \leq \frac{1}{\chi_{i}}\,, 
$$
or equivalent to 
\beq
\left(D-\frac{1}{\chi_i}\right) 
\alpha_{\max}(r^L) \leq K
\label{eqn:zxz}
\eeq
holds for $r^L\in {\cal B}_L(\Gamma,D)$ and 
$i\in \Lambda$, the condition on $\alpha_{\min}$ and $\alpha_{\max}$ 
in Lemma \ref{lm:pro3} holds. 
By Lemma \ref{lm:Egn1}, we have 
\beq
\alpha_{\max}(r^L)\leq \alpha_{\max}^{\ast}
\mbox{ for }r^L\in {\cal B}_L(\Gamma,D). 
\label{eqn:zsd00}
\eeq
It can be seen  from (\ref{eqn:zxz}) and (\ref{eqn:zsd00}) that 
\beq
\left(D-\frac{1}{\chi_i}\right) 
\alpha_{\max}^{\ast} \leq K
\mbox{ for }i \in \Lambda\,.
\label{eqn:zsd}
\eeq
is a sufficient condition for (\ref{eqn:zxz}) to hold.
%Since 
By Lemma \ref{lm:Egn1}, we have
\beqno
\chi_i&=&||\hat{\vc a}_i||^2\frac{1}{\sigma^2_{N_i}}+\eta_i(u_{[i]}^L) 
\leq \lim_{u_i\to \frac{1}{\sigma_{N_i}^2}}\alpha_{\max}(u^L)
\nonumber\\
&\leq&\alpha_{\max}^{\ast}
\mbox{ for }i\in \Lambda, 
%\label{eqn:zsdzz}
\eeqno
from which we have 
%(\ref{eqn:zsd}) and 
%(\ref{eqn:zsdzz}), we have 
$$ 
\left(D-\frac{1}{\chi_i}\right) 
        \alpha_{\max}^{\ast}
\leq D\alpha_{\max}^{\ast}-1\,.
$$
Thus, if we have 
$
D\alpha_{\max}^{\ast}-1 \leq K
$
or equivalent to $D\leq (K+1)/\alpha_{\max}^{\ast}$, 
we have (\ref{eqn:zsd}). 
\hfill\IEEEQED

{\it Proof of Lemma \ref{lm:pro3b}:}
We first derive expression of $\tilde{\omega}(D,r)$ using 
$\beta_i=\beta_i(r),$ $i\in \Lambda$. 
Let $S$ be a set of integers that 
satisfies $\beta_i^{-1}\geq \xi$ in the definition 
of $\tilde{\omega}(D,r)$.
Then %we have 
%$$
$\tilde{\omega}(D,r)$ is computed as 
\beqno
\tilde{\omega}(D,r)&=&{\ts \frac{1}{(L-|S|)^{L-|S|}}}
   \left(\prod_{k\in S}\frac{1}{\beta_k}\right)
   \left(D-\sum_{k\in S} \frac{1}{\beta_k}\right)^{L-|S|}\,.
\eeqno
Fix $i \in \Lambda$ arbitrary and set
$$
\Psi\defeq Lr-\log \tilde{\omega}(D,r)\,.
$$
Computing the derivative of $\Psi$ by $r$, we obtain
\beqno
& &\frac{{\rm d}\Psi}{{\rm d}r}
\\
&=&\sum_{k\in S}
\frac{{\rm e}^{-2r}}{\epsilon}
\left(\frac{\lambda_k}{\lambda_k+\epsilon}\right)^2
 \left[\frac{1}{\beta_k}
-\frac{L-|S|}{D-{\ds \sum_{k\in S}}\frac{1}{\beta_k}}\frac{1}{\beta_k^2}
\right]
+L
\nonumber\\
&=&
\sum_{k\in S}
\left\{
\frac{{\rm e}^{-2r}}{\epsilon}
\left(\frac{\lambda_k}{\lambda_k+\epsilon}\right)^2
\hspace*{-1mm}\left[\frac{1}{\beta_k}
-\frac{L-|S|}{D-{\ds \sum_{k\in S}} \frac{1}{\beta_k}} 
\frac{1}{\beta_k^2}
\right]
\hspace*{-1mm}+\frac{L}{|S|}
\right\}\,.
\eeqno
To examine signs of contents of the above summation we 
set  
\beqno
\Phi_k 
&\defeq&\frac{{\rm e}^{-2r}|S|}{\epsilon L}
\left(\frac{\lambda_k}{\lambda_k+\epsilon}\right)^2
\left\{
D-\sum_{k\in S}{\frac{1}{\beta_k}-\frac{L-|S|}{\beta_k}}
\right\}
\\
& &+\beta_{k}\left(D-\sum_{k\in S}{\frac{1}{\beta_k}}\right).
\eeqno
If $|S|=L$, $\Phi_k \geq 0, k \in \Lambda$ is obvious. 
We hereafter assume $|S|\leq L-1$. Computing $\Phi_k$, we obtain
\beqa
\Phi_k&=& 
\left\{
\frac{{\rm e}^{-2r}|S|}{\epsilon L}
\left(\frac{\lambda_k}{\lambda_k+\epsilon}\right)^2
+\beta_k\right\}
\left(D-\sum_{k\in S}\frac{1}{\beta_k}\right)
\nonumber\\
& &-\frac{{\rm e}^{-2r}|S|}{\epsilon L}
\left(\frac{\lambda_k}{\lambda_k+\epsilon}\right)^2
\cdot \frac{L-|S|}{\beta_k}
\nonumber\\
&\MGeq{a}&
\left\{
\frac{{\rm e}^{-2r}|S|}{\epsilon L}
\left(\frac{\lambda_k}{\lambda_k+\epsilon}\right)^2
+\beta_k\right\}
\left(
\sum_{k\in \Lambda-S}{\frac{1}{\beta_k}}
\right)
\nonumber\\
& &-\frac{{\rm e}^{-2r}|S|}{\epsilon L}
   \left(\frac{\lambda_k}{\lambda_k+\epsilon}\right)^2
   \cdot \frac{L-|S|}{\beta_k}
\nonumber\\
&\geq&
\left\{
\frac{{\rm e}^{-2r}|S|}{\epsilon L}
\left(\frac{\lambda_k}{\lambda_k+\epsilon}\right)^2
+\beta_{i_0}\right\}
 \frac{L-|S|}{\beta_{i_1}}
\nonumber\\
& &-\frac{{\rm e}^{-2r}|S|}{\epsilon L}
   \left(\frac{\lambda_k}{\lambda_k+\epsilon}\right)^2
   \cdot \frac{L-|S|}{\beta_{i_0}}\,.
\label{eqn:zsd0z}
\eeqa
Step (a) follows from %that is equivalent
$$
D-\sum_{k=1}^L\frac{1}{\beta_k}\geq 0
\Leftrightarrow D-\sum_{k\in S}\frac{1}{\beta_k}
\geq \sum_{k\in \Lambda-S}\frac{1}{\beta_k}\,.
$$
%(\ref{eqn:aa00z}). 
From (\ref{eqn:zsd0z}), we can see that if 
\beqa
& &\left\{
\frac{{\rm e}^{-2r}|S|}{\epsilon L}
\left(\frac{\lambda_k}{\lambda_k+\epsilon}\right)^2
+\beta_{i_0}\right\}\frac{1}{\beta_{i_1}}
\nonumber\\
& &-\frac{{\rm e}^{-2r}|S|}{\epsilon L}
   \left(\frac{\lambda_k}{\lambda_k+\epsilon}\right)^2
   \cdot \frac{1}{\beta_{i_0}}
\nonumber\\
&=&\frac{\beta_{i_0}}{\beta_{i_1}}
   -\frac{{\rm e}^{-2r}|S|}{\epsilon L}
    \left(\frac{\lambda_k}{\lambda_k+\epsilon}\right)^2
    \left(\frac{1}{\beta_{i_0}}-\frac{1}{\beta_{i_1}}\right)
\geq 0,
\label{eqn:saa}
\eeqa
then $\Phi_k\geq 0$ for $k\in\Lambda$. 
The inequality (\ref{eqn:saa}) is equivalent to
$$
\frac{1}{\beta_{i_0}}-\frac{1}{\beta_{i_1}}
\leq \left(\frac{\lambda_k+\epsilon}{\lambda_k}\right)^2
\frac{\epsilon{\rm e}^{2r}L}{|S|}\frac{\beta_{i_0}}{\beta_{i_1}}\,.
$$   
Hence 
\beq
\frac{1}{\beta_{i_0}}-\frac{1}{\beta_{i_1}}
\leq \left(\frac{\lambda_{\max}+\epsilon}{\lambda_{\max}}\right)^2
\frac{\epsilon{\rm e}^{2r}L}{L-1}\frac{\beta_{i_0}}{\beta_{i_1}}
\label{eqn:za100a}
\eeq
is a sufficient condition for $\Phi_k\geq 0,$ $k\in\Lambda$. 
The condition (\ref{eqn:za100a}) is equivalent to
$$
{\beta_{i_1}(r)}-{\beta_{i_0}(r)}
\leq \epsilon{\rm e}^{2r}\cdot\frac{L}{L-1}
\left(\frac{\lambda_{\max}+\epsilon}{\lambda_{\max}}\right)^2
(\beta_{i_0}(r))^2\,,
$$
completing the proof.
\hfill\IEEEQED

{\it Proof of Lemma \ref{lm:pro3c}:}
Set
\beqno
& &F(r)
\\
&\defeq &
\left[
{\rm e}^{2r}-\left(
\frac{\lambda_{i_0}}{\lambda_{i_0}+\epsilon}+
\frac{\lambda_{i_1}}{\lambda_{i_1}+\epsilon}
\right)
\right]^{-1}
\left(
{\rm e}^{2r}-\frac{\lambda_{i_0}}{\lambda_{i_0}+\epsilon}
\right)^2\,.
\eeqno
Then, the sufficient condition stated in
Lemma \ref{lm:pro3b} is equivalent to
\beqa
& &
  \frac{\lambda_{i_1}}{\lambda_{i_1}+\epsilon}
 -\frac{\lambda_{i_0}}{\lambda_{i_0}+\epsilon}
%\\
%& &\times
%\left[{\rm e}^{2r}-\left(
%\frac{\lambda_{i_0}}{\lambda_{i_0}+\epsilon}+
%\frac{\lambda_{i_1}}{\lambda_{i_1}+\epsilon}
%\right)
%\right]
\nonumber\\
&\leq& 
\frac{L}{L-1}
\left(\frac{\lambda_{\max}+ \epsilon}{\lambda_{\max}}\right)^2
\left(\frac{\lambda_{i_0}}{\lambda_{i_0}+\epsilon}\right)^2
\cdot F(r)\,.
\label{eqn:aasss}
\eeqa
To derive an explicit sufficient condition for
(\ref{eqn:aasss}) to hold, we estimate a lower bound 
of $F(r)$. Set 
\beqno
T(r)&\defeq & {\rm e}^{2r}-\left(
\frac{\lambda_{i_0}}{\lambda_{i_0}+\epsilon}+
\frac{\lambda_{i_1}}{\lambda_{i_1}+\epsilon}
\right),\:
P\defeq \frac{\lambda_{i_1}}{\lambda_{i_1}+\epsilon}.
\eeqno
Then 
\beqno
F(r)&=&[T(r)]^{-1}[T(r)+P]^2=T(r)+\frac{P^2}{T(r)}+2P
\nonumber\\
   &\geq&4P=\frac{4\lambda_{i_1}}{\lambda_{i_1}+\epsilon}\,.
\eeqno
Hence, 
\beqa
& &
  \frac{\lambda_{i_1}}{\lambda_{i_1}+\epsilon}
 -\frac{\lambda_{i_0}}{\lambda_{i_0}+\epsilon}
\nonumber\\
&\leq& 
\frac{4L}{L-1}
\left(\frac{\lambda_{\max}+\epsilon}{\lambda_{\max}}\right)^2
\left(\frac{\lambda_{i_0}}{\lambda_{i_0}+\epsilon}\right)^2
\left(\frac{\lambda_{i_1}}{\lambda_{i_1}+\epsilon}\right)
%\label{eqn:aasss00}
\nonumber 
\eeqa
is a sufficient condition for (\ref{eqn:aasss}) to hold.
\hfill\IEEEQED
%\input{ExtSys2.tex}
%\input{ExtSys.tex}

%\section{Conclusion}
%\noindent
%
%We have considered the distributed source coding of correlated
%Gaussian observation and given a complete solution to this problem by
%deriving an explicit form of the rate distortion region.
%
%Furthermore, we established a sufficient condition 
%under which this outer bound is tight. A complete solution is still lacking. 
%\end{document}
%\input{apdxFinal.tex}
\section*{\empty}
\appendix
%\newpage

\subsection{
Proof of Lemma \ref{lm:lm1}
}

In this appendix we prove Lemma \ref{lm:lm1}. 
To prove this lemma we need some preparations.
For $i\in\Lambda$, set
$$
F_i(\Sigma|Q)\defeq 
\sup_{\scs p_{\hat{X}^K|X^K}:
\atop{\scs \Sigma_{X^K-\hat{X}^K}\preceq \Sigma }}
h(Z_i-\hat{Z}_i|Z_{[i]}^K-\hat{Z}_{[i]}^K)\,.
$$
To compute $F_i(\Sigma|Q)$, define two random variables 
by %\tilde{X}^K 
$$
\tilde{X}^K\defeq X^K-\hat{X}^K, \tilde{Z}^K\defeq Z^K-\hat{Z}^K\,.
$$
Note that by definition we have $\tilde{Z}^K=Q\tilde{X}^K$. 
Let $p_{X^K\tilde{X}^K}$ $(x^K,\tilde{x}^K)$ be a density 
function of $(X^K,\tilde{X}^K)$. 
Let $q_{Z^K\tilde{Z}^K}$ $(z^K,\tilde{z}^K)$ be a density 
function of $(Z^K,\tilde{Z}^K)$ induced by the unitary 
matrix $Q$, that is,
$$
q_{Z^K\tilde{Z}^K}(z^K,\tilde{z}^K)
\defeq p_{{}^{\rm t}QZ^K {}^{\rm t}Q\tilde{Z}^K}({}^{\rm t}Qz^K,{}^{\rm t}Q\tilde{z}^K)\,.
$$
%Using the above Explicit function 
Expression of $F_i(\Sigma|Q)$ using the above density 
functions is the following.
\beqno
&  &F_i(\Sigma|Q)
\\
&=&
\sup_{\scs p_{\tilde{X}^K|X^K}:
\atop{\scs \Sigma_{\tilde{X}^K}\preceq \Sigma }}
h(\tilde{Z}_i|\tilde{Z}_{[i]}^K)
\nonumber\\
&=&\sup_{\scs p_{\tilde{X}^K|X^K}:
\atop{\scs \Sigma_{\tilde{X}^K}\preceq \Sigma }}-\int
q_{\tilde{Z}^K}(z^K)
\log q_{\tilde{Z}_{i}|\tilde{Z}_{[i]}^K}(z_i|z_{[i]}^K){\rm d}z^K
\nonumber\\
&=&\sup_{\scs p_{\tilde{X}^K|X^K}:
\atop{\scs \Sigma_{\tilde{X}^K}\preceq \Sigma }}-\int
q_{\tilde{Z}^K}(z^K)
\log \frac{ q_{\tilde{Z}^K}(z^K)}
          { q_{\tilde{Z}_{[i]}^K}(z_{[i]}^K)}{\rm d}z^K\,.
\eeqno
The following two properties on $F_i(\Sigma|Q)$ are 
useful for the proof of Lemma \ref{lm:lm1}.
\begin{lm}\label{lm:lm7}
$F_i(\Sigma|Q)$ is concave with respect to $\Sigma$.
\end{lm}
\begin{lm}\label{lm:lm8}
$$
F_i(\Sigma|Q)=\frac{1}{2}
\log
 \left\{
 {(2\pi{\rm e})}\left[Q\Sigma {}^{\rm t}Q\right]_{ii}^{-1}
\right\}\,.
$$ 
\end{lm}

%We postpone the proofs of Lemmas \ref{lm:lm7} and \ref{lm:lm8}.
We first prove Lemma \ref{lm:lm1} using those two lemmas and 
next prove Lemmas \ref{lm:lm7} and \ref{lm:lm8}.

{\it Proof of Lemma \ref{lm:lm1}:} 
We have the following chain of inequalities:
\beqno
& &\left. h(\irg{\vc Z}_i \right| \irg{\vc Z}_{[i]}^K \irb{W}^K)
\nonumber\\
&\leq&
   h( \irg{{\vc Z}}_i -\hat{\irg{{\vc Z}}}_{i}
      \left. \right| \irg{{\vc Z}}_{[i]}^K -\hat{\irg{{\vc Z}}}_{[i]}^K)
\nonumber\\
&\leq&\sum_{t=1}^n
   h(\irg{Z}_i(t) -\hat{\irg{Z}}_{i}(t)
     \left. \right|
     \irg{Z}_{[i]}^K(t) -\hat{\irg{Z}}_{[i]}^K(t))
\nonumber\\
&\MLeq{a}&\sum_{t=1}^nF_i
\left.\left(
\Sigma_{{X}^K(t)-\hat{\irg{X}}^K(t)}\right|Q 
\right)
\nonumber\\
&\MLeq{b}& nF_i
\left.\left({\frac{1}{n}}\sum_{t=1}^n
\Sigma_{{X}^K(t)-\hat{\irg{X}}^K(t)}\right|Q 
\right)
\nonumber\\
&=& nF_i\left.\left(
     \ts{\frac{1}{n}}\Sigma_{{\lvc X}^K-\hat{\irg{\lvc X}}^K}
        \right|Q\right)
\nonumber\\
&\MEq{c}&\frac{n}{2}\log 
\left\{
(2\pi {\rm e})
\left[
Q
\left(
{\ts \frac{1}{n}}\Sigma_{{\lvc X}^K-\hat{\lvc X}^K}^{-1}
\right){}^{\rm t}Q
\right]_{ii}^{-1}
\right\}\,.
\eeqno
Step (a) follows from the definition of $F_i(\Sigma|Q)$.
Step (b) follows from Lemma \ref{lm:lm7}.
Step (c) follows from Lemma \ref{lm:lm8}.
\hfill \IEEEQED

{\it Proof of Lemma \ref{lm:lm7}:}
For given covariance matrices 
$\Sigma^{(0)}$ and $\Sigma^{(1)}$, let 
$p_{\tilde{X}^K|X^K}^{(0)}$ and 
$p_{\tilde{X}^K|X^K}^{(1)}$
be conditional densities achieving 
$F_i(\Sigma^{(0)}|Q)$ and $F_i(\Sigma^{(1)}|Q)$, respectively.
For $0\leq \alpha \leq 1$, define 
a conditional density parameterized with $\alpha$ by 
$$
p_{\tilde{X}^K|X^K}^{(\alpha)}
= (1-\alpha) p_{\tilde{X}^K|X^K}^{(0)}
+     \alpha p_{\tilde{X}^K|X^K}^{(1)}\,.
$$
Let $p_{X^K\tilde{X}^K}^{(\alpha)}$ 
be a density function of $(X^K,\tilde{X}^K)$ 
defined by 
$(p_{\tilde{X}^K|X^K}^{(\alpha)},$
$p_{X^K}^{(\alpha)})$. Let 
$\Sigma_{\tilde{X}}^{(\alpha)}$ be a covariance matrix 
computed from the density $p_{\tilde{X}^K}^{(\alpha)}$. 
Since 
$$
p_{\tilde{X}^K}^{(\alpha)}
= (1-\alpha)p_{\tilde{X}^K}^{(0)}
+    \alpha p_{\tilde{X}^K}^{(1)}\,,
$$
we have 
\beqa
\Sigma_{\tilde{X}}^{(\alpha)}
&=&
 (1-\alpha)\Sigma_{\tilde{X}}^{(0)}
   +\alpha \Sigma_{\tilde{X}}^{(1)}
\nonumber
\\
&\preceq &
 (1-\alpha)\Sigma^{(0)}
   +\alpha \Sigma^{(1)}\,.
\label{eqn:cov0}
\eeqa
Let $q_{Z^K\tilde{Z}^K}^{(\alpha)}$ be a density function 
of $(Z^K,\tilde{Z}^K)$ induced by the unitary matrix $Q$, that is, 
$$
q_{Z^K\tilde{Z}^K}^{(\alpha)}(z^K,\tilde{z}^K)
\defeq p_{{}^{\rm t}QZ^K {}^{\rm t}Q\tilde{Z}^K}^{(\alpha)}
({}^{\rm t}Qz^K,{}^{\rm t}Q\tilde{z}^K)\,.
$$
By definition it is obvious that
$$
            q_{\tilde{Z}^K}^{(\alpha)}
= (1-\alpha)q_{\tilde{Z}^K}^{(0)}
+    \alpha q_{\tilde{Z}^K}^{(1)}\,.
$$
Then we have 
\beqno
&  & (1-\alpha)F_i(\Sigma^{(0)}|Q) 
      + \alpha F_i(\Sigma^{(1)}|Q)
\\
&=& -(1-\alpha)\int q_{\tilde{Z}^K}^{(0)}(z^K)
\log \frac{q_{\tilde{Z}^K}^{(0)}(z^K)}
          {q_{\tilde{Z}_{[i]}^K}^{(0)}(z_{[i]}^K)}{\rm d}z^K
\\
& &
 \quad \quad \:\:- \alpha\int q_{\tilde{Z}^K}^{(1)}(z^K)
\log \frac{q_{\tilde{Z}^K}^{(1)}(z^K)}
          {q_{\tilde{Z}_{[i]}^K}^{(1)}(z_{[i]}^K)}{\rm d}z^K
\\
&\MLeq{a}& -\int q_{\tilde{Z}^K}^{(\alpha)}(z^K)
\log \frac{q_{\tilde{Z}^K}^{(\alpha)}(z^K)}
          {q_{\tilde{Z}_{[i]}^K}^{(\alpha)}(z_{[i]}^K)}{\rm d}z^K
\\ 
&=&-\int q_{\tilde{Z}^K}^{(\alpha)}(z^K)
    \log q_{\tilde{Z}_{i}|\tilde{Z}_{[i]}^K}^{(\alpha)}(z_i|z_{[i]}^K){\rm d}z^K
\\
&\MLeq{b}& 
F_i \left(\left.(1-\alpha)\Sigma^{(0)}+\alpha \Sigma^{(1)}
\right|Q\right)\,.
\eeqno
Step (a) follows from log sum inequality. 
Step (b) follows from the definition of 
$F_i(\Sigma|Q)$ and (\ref{eqn:cov0}).
\hfill \IEEEQED

{\it Proof of Lemma \ref{lm:lm8}:}
Let 
\beqno
q_{\tilde{Z}^K}^{(\rm G)}(z^K)
&\defeq&
\frac{1}{(2\pi{\rm e})^{\frac{K}{2}}\left|\Sigma_{\tilde{Z}^K}\right|^{\frac{1}{2}}}
{\rm e}^{\scs 
-\frac{1}{2}
{}^{\rm t}[z^K]
\Sigma_{\tilde{Z}^K }^{-1}
\mbox{\scriptsize $[z^K]$ }} 
\eeqno
and let 
$$
{q_{\tilde{Z}_{i}|\tilde{Z}_{[i]}^K}^{({\rm G})}(z_i|z_{[i]}^K)}
=\frac{q_{\tilde{Z}^K}^{(\rm G)}(z^K)}
{q_{\tilde{Z}_{[i]}^K}^{({\rm G})}(z_{[i]}^K)}
$$
be a conditional density function induced by 
$q_{\tilde{Z}^K}^{(\rm G)}(\cdot)$.
We first observe that
\beq
\int q_{\tilde{Z}^K}(z^K)
    \log 
\frac
{q_{\tilde{Z}_{i}|\tilde{Z}_{[i]}^K}(z_i|z_{[i]}^K)}
{q_{\tilde{Z}_{i}|\tilde{Z}_{[i]}^K}^{({\rm G})}(z_i|z_{[i]}^K)}
{\rm d}z^K
\geq 0 \,. \label{eqn:div}
\eeq
From (\ref{eqn:div}), we have the following chain of inequalities:
\beqa
h(\tilde{Z}_{i}|\tilde{Z}_{[i]}^K)
&=&-\int q_{\tilde{Z}^K}(z^K)
    \log q_{\tilde{Z}_{i}|\tilde{Z}_{[i]}^K}(z_i|z_{[i]}^K)
    {\rm d}z^K
\nonumber\\
&\leq &-\int q_{\tilde{Z}^K}(z^K)
    \log q_{\tilde{Z}_{i}|\tilde{Z}_{[i]}^K}^{({\rm G})}(z_i|z_{[i]}^K)
    {\rm d}z^K
\nonumber\\
&=& -\int q_{\tilde{Z}^K}(z^K)
    \log \frac{q_{\tilde{Z}^K}^{({\rm G})}(z^K)}
              {q_{\tilde{Z}_{[i]}^K}^{({\rm G})}(z_{[i]}^K)}
     {\rm d}z^K
\nonumber\\
&=& - \int q_{\tilde{Z}^K}(z^K)
      \log q_{\tilde{Z}^K}^{({\rm G})}(z^K)
      {\rm d}z^K
\nonumber\\
& & + \int q_{\tilde{Z}^K}(z^K)
      \log q_{\tilde{Z}_{[i]}^K}^{({\rm G})}(z_{[i]}^K)
      {\rm d}z^K
\nonumber\\
&\MEq{a}&
   - \int  q_{\tilde{Z}^K}^{({\rm G})}(z^K)
      \log q_{\tilde{Z}^K}^{({\rm G})}(z^K)
      {\rm d}z^K
\nonumber\\
& & + \int q_{\tilde{Z}^K}^{({\rm G})}(z^K)
      \log q_{\tilde{Z}_{[i]}^K}^{({\rm G})}(z_{[i]}^K)
      {\rm d}z^K
\nonumber\\
&=& \frac{1}{2}
    \log
    \left\{
     {(2\pi{\rm e})}
     \frac{|\Sigma_{\tilde{Z}^K     }|}
          {|\Sigma_{\tilde{Z}_{[i]}^K}|}
   \right\}
\nonumber\\
&\MEq{b}&\frac{1}{2}
    \log
    \left\{
     {(2\pi{\rm e})}
     \left[\Sigma_{\tilde{Z}^K}^{-1}\right]_{ii}^{-1}
    \right\}
\nonumber\\
&=&\frac{1}{2}
   \log
   \left\{
   {(2\pi{\rm e})}\left[Q\Sigma_{\tilde{X}^K}^{-1}{}^{\rm t}Q\right]_{ii}^{-1}
   \right\}
\nonumber\\
&\MLeq{c}&\frac{1}{2}
   \log
   \left\{
   {(2\pi{\rm e})}\left[Q\Sigma^{-1}{}^{\rm t}Q \right]_{ii}^{-1}
   \right\}\,.
\nonumber
\eeqa
Step (a) follows from the fact that 
$q_{\tilde{Z}^L}$ and 
$q_{\tilde{Z}^L}^{({\rm G})}$ 
yield the same moments of the quadratic form 
$\log q_{\tilde{Z}^L}^{({\rm G})}$.  
Step (b) is a well known formula on the determinant of matrix.
Step (c) follows from $\Sigma_{\tilde{X}^L} \preceq \Sigma$.
Thus
$$
F_i(\Sigma|Q) \leq \frac{1}{2}
\log
 \left\{
 {(2\pi{\rm e})}\left[Q\Sigma^{-1}{}^{\rm t}Q\right]_{ii}^{-1}
\right\}
$$ 
is concluded. Reverse inequality holds by letting 
$p_{\tilde{X}^K|{X}^K}$ be Gaussian with 
covariance matrix $\Sigma$. 
\hfill\IEEEQED

\subsection{
Proof of Lemma \ref{lm:lm2}
} 

In this appendix we prove Lemma \ref{lm:lm2}.
%Without loss of generality we may assume 
%that $S=\{1,2,\cdots, s\}$. 

We write a unitary matrix $Q$ as $Q=[q_{ij}]$, where $q_{ij}$ 
stands for the $(i,j)$ entry of $Q$. 
The unitary matrix ${Q}$ transforms 
$X^K$ into $\irg{Z}^K$$=QX^K$.
Set $\tilde{Q}=Q{}^{\rm t}A$ and let 
$\tilde{q}_{ij}$ be the $(i,j)$ entry of $Q{}^{\rm t}A$.
The following lemma states an important 
property on the distribution of Gaussian random vector 
$Z^K$. This lemma is a basis of the proof 
of Lemma \ref{lm:lm2}. 

\begin{lm}\label{lm:LmO}
For any $i=1,2,\cdots,K$, we have the following.
\beq 
{Z}_i=-\frac{1}{g_{ii}}\sum_{j\ne i}\nu_{ij}{Z}_j
+ \frac{1}{g_{ii}}\sum_{j=1}^{L}
\frac {\tilde{q}_{ij}}{\sigma_{N_j}^2}{Y}_j + \hat{N}_i\,,
\label{eqn:prlmaa00}
\eeq
where 
\beq
g_{ii}= \left[Q\Sigma_{X^K}^{-1}{}^{\rm t}Q\right]_{ii} 
+\sum_{j=1}^{L}\frac{\tilde{q}_{ij}^2}{\sigma_{N_j}^2}\,,
\label{eqn:defajj}
\eeq
$\nu_{ij},$ $j\in \{1,2,\cdots,K\}-\{i\}$ are suitable constants 
and $\hat{N}_i$ is a zero mean Gaussian random 
variables with variance $\frac{1}{g_{ii}}$. 
For each $i\in S$, $\hat{N}_i$ is independent 
of ${Z}_j,j\in \{1,2,\cdots,K\}-\{i\}$ 
and ${Y}_j,j\in \Lambda$. 
\end{lm}

{\it Proof:} Without loss of generality we may assume $i=1$.
%Let $\Sigma_{X^KY^L}$ be a covariance matrix
%on the pair of the Gaussian random vectors $X_S$ 
%and $Y^L$. 
Since $Y^L=AX^K+N^L$, we have 
\beqno
\Sigma_{X^KY^L}=
\left[
\ba{cc} 
\Sigma_{X^K} &\Sigma_{X^K}\\
\Sigma_{X^K} & A\Sigma_{X^K}{}^{\rm t}A+\Sigma_{N^L}
\ea
\right]\,.
\eeqno
Since $Z^K=QX^K$, we have
\beqno
\Sigma_{Z^KY^L}=
\left[
\ba{cc}
Q\Sigma_{X^K}{}^{\rm t}Q &   Q\Sigma_{X^K}\\
        \Sigma_{X^K}{}^{\rm t}Q  & A\Sigma_{X^K}{}^{\rm t}A+\Sigma_{N^L}
\ea
\right]\,.
\eeqno
The density function $p_{Z^KY^L}(z^K,y^L)$ of  
$(Z^K,Y^L)$ is given by 
\beqno
& &p_{Z^KY^L}(z^K,y^L)
\\
&=&
\frac{1}{(2\pi{\rm e})^{\frac{K+L}{2}}\left|\Sigma_{Z^KY^L}\right|^{\frac{1}{2}}}
{\rm e}^{\scs 
-\frac{1}{2}
{}^{\rm t}[z^K y^L]
\Sigma_{Z^KY^L}^{-1}
\mbox{\scriptsize 
$\left[\ba{c}z^K\\
y^L\ea\right]$}},
\eeqno
where $\Sigma_{Z^KY^L}^{-1}$ has the following form:
\beqno
\Sigma_{Z^KY^L}^{-1}=
\left[
\ba{cc}
   Q(\Sigma_{X^K}^{-1}+ {}^{\rm t}A\Sigma_{N^L}^{-1}A ){}^{\rm t}Q 
   & -Q{}^{\rm t}A\Sigma_{N^L}^{-1}\\
    -\Sigma_{N^L}^{-1}A{}^{\rm t}Q & \Sigma_{N^L}^{-1}
\ea
\right]\,.
\eeqno
Set
\beq
\left.
\ba{rcl}
\nu_{ij}  &\defeq &\ds 
   \left[Q(\Sigma_{X^K}^{-1}+{}^{\rm t}A\Sigma_{N^L}^{-1}A)
{}^{\rm t}Q\right]_{ij}
\vspace*{1mm}\\
&=&\ds\left[Q\Sigma_{X^K}^{-1}{}^{\rm t}Q\right]_{ij}
+\sum_{k=1}^L\frac{\tilde{q}_{ik}\tilde{q}_{jk}}{\sigma_{N_k}^2}\,,
\vspace*{1mm}\\
\beta_{ij}&\defeq &\ds 
-\left[Q{}^{\rm t}A\Sigma_{N^L}^{-1}\right]_{ij}=
-\frac{\tilde{q}_{ij}}{\sigma_{N_j}^2}\,.
\ea
\right\}
\label{eqn:prlmz2z} 
\eeq
Now, we consider the following partition of 
$\Sigma_{Z^KY^L}^{-1}$: 
\beqno
\Sigma_{Z^KY^L}^{-1}
&=&
\left[
\ba{cc}
   Q(\Sigma_{X^K}^{-1}+{}^{\rm t}A\Sigma_{N^L}^{-1}A){}^{\rm t}Q 
& -Q{}^{\rm t}A\Sigma_{N^L}^{-1}\\
 -\Sigma_{N^L}^{-1}A{}^{\rm t}Q & \Sigma_{N^L}^{-1}
\ea
\right]
\nonumber\\
&=&
\left[
\ba{c|c}    
      g_{11} & {}^{\rm t}g_{12} \\\hline
      g_{12} & G_{22}    
\ea
\right]\,,
\label{eqn:prlmz3} 
\eeqno 
where $g_{11}$, $g_{12}$, and $G_{22}$ are scalar, 
$K+L-1$ dimensional vector, and $(K+L-1)$
$\times(K+L-1)$ 
matrix, respectively. It is obvious from 
the above partition of $\Sigma_{Z^KY^L}^{-1}$ 
that we have 
\beq
\left.
\ba{rcl}
g_{11}&=&\ds\nu_{11}=
   \left[Q\Sigma_{X^K}^{-1}{}^{\rm t}Q\right]_{11}
   +\sum_{k=1}^L\frac{\tilde{q}_{1k}^2}{\sigma_{N_k}^2}\,,
\vspace{1mm}\\
g_{12}&=&{}^{\rm t}\left[\nu_{12}\cdots\nu_{1K}
               \beta_{11}\beta_{12}\cdots\beta_{1L}\right]\,.
\ea
\right\}
\label{eqn:prlmz5} 
\eeq
It is well known that 
$\Sigma_{Z^KY^L}^{-1}$ has the following expression:
\beqno
\Sigma_{Z^KY^L}^{-1}
&=&\left[\ba{c|c}    
      g_{11} & {}^{\rm t}g_{12} \\\hline
      g_{12} & G_{22}    
      \ea\right]
\nonumber\\
&=&\left[
\ba{c|c}    
      1 & {}^{\rm t}0_{12} \\\hline
      \frac{1}{g_{11}}g_{12} & I_{L-1}    
\ea
\right]
\left[
\ba{c|c}    
      g_{11} & {}^{\rm t}0_{12} \\\hline
      0_{12}&G_{22}-\frac{1}{g_{11}}{}^{\rm t}g_{12}g_{12}    
\ea
\right]
\nonumber\\
& &\qquad\qquad \qquad \times
\left[
\ba{c|c}    
      1 & \frac{1}{g_{11}}{}^{\rm t}g_{12} \\\hline
      0_{12} & I_{L-1}    
\ea
\right]\,.
%\label{eqn:prlmz6} 
\eeqno
Set
\beq
%\left.\ba{rcl}
\hat{n}_1
\defeq 
\left[z_1|z_{[1]}^Ky^L\right]
\left[
      \ba{c}
      1\\
      \hline 
      \frac{1}{g_{11}}g_{12}
      \ea
\right]
%\vspace{1mm}\\
=z_1+\frac{1}{g_{11}}\left[z_{[1]}^Ky^L\right]g_{12}\,.
%\ea\right\}
\label{eqn:prlmz64} 
\eeq
Then, we have
\beqa
\hspace*{-6mm}& &{}^{\rm t}[z^Ky^L]\Sigma_{Z^KY^L}
    \left[\ba{c}
          z^K\\
          y^L
          \ea
    \right]
\nonumber\\
\hspace*{-6mm}&=&{}^{\rm t}[ z_1 | z_{[1]}^Ky^L]
   \left[\ba{c|c}    
      g_{11} & {}^{\rm t}g_{12} \\\hline
      g_{12} & G_{22}    
      \ea\right]
 \left[\ba{c}
        z_1\\
       \hline
       \vspace*{-3.5mm}\\
       z_{[1]}^K\\
        y^L
        \ea
\right]  
\nonumber\\
\hspace*{-6mm}&=&[\hat{n}_1|z^K_{[1]}y^L]
\left[
\ba{c|c}    
      g_{11} & {}^{\rm t}0_{12}\\\hline
      0_{12}&G_{22}-\frac{1}{g_{11}}g_{12}{}^{\rm t}g_{12}    
\ea
\right]
\left[\ba{c}\hat{n}_1\\
      \hline 
      \vspace*{-3.5mm}\\
      z_{[1]}^K\\
      y^L
      \ea\right]\,.  
\label{eqn:zasds}
\eeqa
From (\ref{eqn:prlmz2z})-(\ref{eqn:prlmz64}), we have 
\beqa 
\hat{n}_1&=&z_1
+\frac{1}{g_{11}}\sum_{j=2}^L\nu_{1j}z_j
+\frac{1}{g_{11}}\sum_{j=1}^L\beta_{1j}y_j
\nonumber\\
&=&z_1
+\frac{1}{g_{11}}\sum_{j=2}^L\nu_{1j}z_j
-\frac{1}{g_{11}}\sum_{j=1}^L
\frac{\tilde{q}_{1j}}{\sigma_{N_j}^2}y_j\,.
\label{eqn:prlmz66} 
\eeqa
It can be seen from (\ref{eqn:zasds}) and (\ref{eqn:prlmz66}) 
that the random variable $\hat{N}_1$ defined by
$$
\hat{N}_1\defeq 
Z_1+\frac{1}{g_{11}}\sum_{j=2}^L\nu_{1j}Z_j
   -\frac{1}{g_{11}}\sum_{j=1}^L\frac{\tilde{q}_{1j}}{\sigma_{N_j}^2}Y_j
$$
is a zero mean Gaussian random variable with variance 
$\frac{1}{{g}_{11}}$ and is independent 
of $Z_{[1]}^K$ and $Y^L$. This completes the proof 
of Lemma \ref{lm:LmO}.
\hfill\IEEEQED

The followings are two variants of the entropy power 
inequality.

\begin{lm}\label{lm:lm5zz} Let ${\vc U}_i,i=1,2,3$ 
be $n$ dimensional 
random vectors with densities and let 
$T$ be a random variable taking values in a finite set. 
We assume that ${\vc U}_3$ is independent of ${\vc U}_1$, 
${\vc U}_2$, and $T$. Then,  we have 
\beqno
\EP{{\lvc U}_2+{\lvc U}_3|{\lvc U}_1T}
\geq 
\EP{{\lvc U}_2|{\lvc U}_1T}+\EP{{\lvc U}_3}\,.
\eeqno
\end{lm}
\begin{lm}\label{lm:lm5zzb} Let ${\vc U}_i$, $i=1,2,3$ be 
$n$ random vectors with densities. Let $T_1, T_2$ 
be random variables taking values in finite sets. 
We assume that those five random variables
form a Markov chain 
$
(T_1,{\vc U}_1) \to {\vc U}_3 \to (T_2,{\vc U}_2) 
$
in this order. Then, we have    
\beqno
& &\EP{{\lvc U}_1+{\lvc U}_2|{\lvc U}_3T_1T_2}
\\
&\geq& \EP{{\lvc U}_1|{\lvc U}_3T_1}
      +\EP{{\lvc U}_2|{\lvc U}_3T_2}\,.
\eeqno
\end{lm}

{\it Proof of Lemma \ref{lm:lm2}:}
%Without loss of generality we may assume 
%that $S=\{1,2,\cdots, s\}$. 
%We write unitary matrix $Q_S$ as $Q_S=[q_{ij}]$, where $q_{ij}$ 
%stands for the $(i,j)$ element of $Q_S$. By an elementary 
%computation, for any $i\in S$, we have the following.
By Lemma \ref{lm:LmO}, we have
\beq 
{\vc Z}_i=-\frac{1}{g_{ii}}\sum_{j\ne i}\nu_{ij}{\vc Z}_j
+ \frac{1}{g_{ii}}\sum_{j=1}^{L}
\frac {\tilde{q}_{ij}}{\sigma_{N_j}^2}{\vc Y}_j + \hat{\vc N}_i\,,
\label{eqn:prlmaa}
\eeq
where 
%\beq
%g_{ii}= \left[Q\Sigma_{X^K}^{-1}Q_S\right]_{ii} 
%+ \sum_{j=1}^{s}\frac{q_{ji}^2}{\sigma_{N_j}^2}\,,
%\label{eqn:defajj}
%\eeq
%$\nu_{ij},$ $j\in S-\{i\}, $ are suitable constants 
%and 
$\hat{\vc N}_i$ is a vector of $n$ independent 
copies of zero mean Gaussian random 
variables with variance $\frac{1}{g_{ii}}$. 
For each $i\in \Lambda$, $\hat{\vc N}_i$ is independent of 
${\vc Z}_j, j \in \{1,2,\cdots,$ $K\}-\{i\}$
and ${\vc Y}_j, j \in \Lambda$. 
Set 
\beqno
h^{(n)}&\defeq& \frac{1}{n}h({\vc Z}_i|{\vc Z}_{[i]}^K,W^L)\,.
\eeqno
Furthermore, for $k\in \Lambda$, define
\beqno
&&S_k \defeq \{k,k+1,\cdots,L\}\,,
%\\ 
%& &
\Psi_k=\Psi_k({\vc Y}_{S_k}) 
\defeq \sum_{j=k}^{L} \frac {\tilde{q}_{ij}}{\sigma_{N_j}^2}{\vc Y}_j\,.
\eeqno
Applying Lemma \ref{lm:lm5zz} to (\ref{eqn:prlmaa}), we have
\beq  
\frac{{\baseN}^{2h^{(n)}}}{2\pi{\rm e}}
\geq \frac{1}{(g_{ii})^2}
\frac{1}{2\pi{\rm e}}{\baseN}^{\frac{2}{n}h(\Psi_1|{\svc Z}_{[i]}^K,W^L)} 
+\frac{1}{g_{ii}}\,.
\label{eqn:PrConvLm1}
\eeq
On the quantity $h(\Psi_1|{\svc Z}_{[i]}^K,W^L)$ 
in the right member of (\ref{eqn:PrConvLm1}), we have 
the following chain of equalities:
\beqa
& &h(\Psi_1|{\vc Z}_{[i]}^K,W^L)
\nonumber\\
&=&I(\Psi_1;{\vc X}^K|{\vc Z}_{[i]}^K,W^L)
   +h(\Psi_1|{\vc X}^K,{\vc Z}_{[i]}^K,W^L)
 \nonumber\\
&\MEq{a}&
   I(\Psi_1;{\vc Z}^K|{\vc Z}_{[i]}^K,W^L)  
   +h(\Psi_1|{\vc X}^K,W^L)
\nonumber\\
&=& I(\Psi_1;{\vc Z}_i|{\vc Z}_{[i]}^K,W^L) 
   +h(\Psi_1|{\vc X}^K,W^L)
\nonumber\\
&=&h({\vc Z}_i|{\vc Z}_{[i]}^K,W^L)
    -h({\vc Z}_i|\Psi_1,{\vc Z}_{[i]}^K,W^L) 
\nonumber\\
& &+h(\Psi_1|{\vc X}^K,W^L)
\nonumber\\
&\MEq{b}& nh^{(n)}-h({\vc Z}_i|\Psi_1,{\vc Z}_{[i]}^K) 
           +h(\Psi_1|{\vc X}^K,W^L)
\nonumber\\
&=&nh^{(n)}-\frac{n}{2}\log\left[{2\pi{\rm e}}(g_{ii})^{-1}\right]
   +h(\Psi_1|{\vc X}^K,W^L)\,. 
\label{eqn:PrConvLm2}
\eeqa
Step (a) follows from that ${\vc Z}^K$ can be obtained from 
${\vc X}^K$ by the invertible matrix $Q$. 
Step (b) follows from the Markov chain 
$${\vc Z}_i\to (\Psi_1,{\vc Z}_{[i]}^K)\to {\vc Y}^L\to W^L.$$ 
From (\ref{eqn:PrConvLm2}), we have
\beq
\frac{1}{2\pi{\rm e}}{\baseN}^{\frac{2}{n}h(\Psi_1|{\svc Z}_{[i]}^K,W^L)}
=\frac{{\baseN}^{2h^{(n)}}}{2\pi{\rm e}}g_{ii}\cdot
 \frac{1}{2\pi{\rm e}}{\baseN}^{\frac{2}{n}h(\Psi_1|{\svc X}^K,W^L)}.
\label{eqn:PrConvLm3}
\eeq
Substituting (\ref{eqn:PrConvLm3}) into (\ref{eqn:PrConvLm1}), 
we obtain  
\beq
\frac{{\baseN}^{2h^{(n)}}}{2\pi{\rm e}}
\geq \frac{{\baseN}^{2h^{(n)}}}{2\pi{\rm e}}\frac{1}{g_{ii}}
     \cdot\frac{1}{2\pi{\rm e}}{\baseN}^{\frac{2}{n}h(\Psi_1|{\svc X}^K,W^L)} 
     +\frac{1}{g_{ii}}\,.
\label{eqn:PrConvLm4}
\eeq
Solving (\ref{eqn:PrConvLm4}) with respect to 
$\frac{{\baseN}^{2h^{(n)}}}{2\pi{\rm e}}$, we obtain
%and the technique that Oohama \cite{oh4}, \cite{oh2} developed 
%to prove the converse coding theorem, we obtain 
\beq
\frac{{\baseN}^{2h^{(n)}}}{2\pi{\rm e}}
\geq
  \left[g_{ii} 
 -\frac{1}{2\pi{\rm e}}{\baseN}^{\frac{2}{n}h(\Psi_1|{\svc X}^K,W^L)} 
  \right]^{-1}\,.
\label{eqn:prlmz}
\eeq
Next, we evaluate a lower bound of  
$
{\baseN}^{\frac{2}{n}h(\Psi_1|{\svc X}^K,W^L)}\,.
$ 
Note that for $j=1,2,\cdots,s-1$ we have 
the following Markov chain:
\beq
\left(W_{S_{j+1}},\Psi_{j+1}({\vc Y}_{S_{j+1}})\right)\to {\vc X}^K 
\to \left(W_j,\ts \frac{\tilde{q}_{ij}}{\sigma_{N_j}^2}{\vc Y}_j\right)\,. 
\label{eqn:Markov}
\eeq
Based on (\ref{eqn:Markov}), we apply 
Lemma \ref{lm:lm5zzb} to 
$
\frac{1}{2\pi{\rm e}}  
{\baseN}^{\frac{2}{n}h(\Psi_j|{\svc X}^K,W^L)}
$
for $j=1,2,\cdots,s-1$. Then, for $j=1,2,$ $\cdots,s-1$, 
we have the following chains of inequalities : 
\beqa
& &\frac{1}{2\pi{\rm e}} %\cdot 
  {\baseN}^{\frac{2}{n}h(\Psi_j|{\svc X}^K,W^L)}
\nonumber\\
&=&\frac{1}{2\pi{\rm e}} %\cdot 
  {\baseN}^{\frac{2}{n}
  h\left(\left.\Psi_{j+1}+\frac{\tilde{q}_{ij}}{\sigma_{N_1}^2}{\svc Y}_j
         \right|{\svc X}^K,W_{S_{j+1}},W_j\right)}
\nonumber\\
&\geq & \frac{1}{2\pi{\rm e}} 
        {\baseN}^{\frac{2}{n}
        h\left(\left.\Psi_{j+1}\right|{\svc X}^K,W_{S_{j+1}}\right)}
        +\frac{1}{2\pi{\rm e}} 
         {\baseN}^{
         \frac{2}{n}
         h\left(\left.\frac{\tilde{q}_{ij}}{\sigma_{N_j}^2}{\svc Y}_j\right|
         {\svc X}^K,W_j\right)
         }
\nonumber\\
&=& \frac{1}{2\pi{\rm e}} 
        {\baseN}^{\frac{2}{n}
         h\left(\left.
         \Psi_{j+1}\right|{\svc X}^K,W_{S_{j+1}}\right)}
        +\tilde{q}_{ij}^2\frac{{\baseN}^{-2r_j^{(n)}}}{\sigma_{N_j}^2}\,.
\label{eqn:prlmz1}
\eeqa
Using (\ref{eqn:prlmz1}) iteratively for
$j=1,2,\cdots, s-1$, we have
\beq
\frac{1}{2\pi{\rm e}} %\cdot 
  {\baseN}^{\frac{2}{n}h(\Psi_1|{\svc X}^K,W^L)}
\nonumber\\
\geq 
\sum_{j=1}^{s}\tilde{q}_{ij}^2\frac{{\baseN}^{-2r_j^{(n)}}}{\sigma_{N_j}^2}\,.
\label{eqn:prlmz2}
\eeq
Combining (\ref{eqn:defajj}), (\ref{eqn:prlmz}), 
and (\ref{eqn:prlmz2}), we have 
\beqa
\frac{{\baseN}^{2h^{(n)}}}{2\pi{\rm e}}
&\geq&
  \left\{
  \left[Q\Sigma_{X^K}^{-1}{}^{\rm t}Q\right]_{ii} 
  +\sum_{j=1}^{s}\tilde{q}_{ij}^2
   \frac{1-{\baseN}^{-2r_j^{(n)}}}{\sigma_{N_j}^2}
   \right\}^{-1}
\nonumber\\
&=&\left[Q
         \left(
         \Sigma_{X^K}^{-1}
         +{}^{\rm t}A\Sigma_{N_{\Lambda}(r_{\Lambda}^{(n)})}^{-1}A
         \right)
         {}^{\rm t}Q\right]_{ii}^{-1}\,,
\nonumber
%\label{eqn:prlmz4}
\eeqa
completing the proof. 
\hfill \IEEEQED

\newcommand{\Skip}{}
{%\footnotesize

}

\end{document}